\documentclass[11pt,a4paper]{article}
\pdfoutput=1   
\usepackage{amsmath,amsfonts,amssymb,amsbsy}
\usepackage{mathrsfs,latexsym}
\usepackage{graphicx}
\usepackage{color}
\usepackage{booktabs}
\usepackage{multirow}
\usepackage{multicol}
\usepackage{subfig}
\usepackage{makecell}
\usepackage{alpha}
\hypersetup{%
  pdftitle = {The determination of potential scales in 2+1 flavor QCD},
  pdfauthor = {Tom~M.~B.~Asmussen, Roman~H\"ollwieser, Francesco Knechtli, Tomasz Korzec},
  pdfkeywords = {Lattice QCD, Static Potential, Scale Setting, Lambda Parameter, Sommer Scale}
}%
\usepackage{macros_alpha}
\usepackage{textcomp}

\usepackage{array}
\usepackage{booktabs}
\usepackage{multirow}
\usepackage{subfig}
\usepackage{placeins}
\usepackage{blindtext}
\usepackage{appendix}
\usepackage{bm}
\usepackage{times}
\usepackage{float}
\usepackage{wrapfig}
\usepackage[utf8]{inputenc} 
\usepackage[T1]{fontenc} 
\usepackage{amsmath,amssymb}

\usepackage{hyperref}

\begin{document}

\title{The determination of potential scales in 2+1 flavor QCD\\
}
\preprintno{WUB/24-01}


\author{Tom~M.~B.~Asmussen} 
\author{Roman~H\"ollwieser} 
\author{Francesco~Knechtli}
\author{Tomasz~Korzec}
\address[buw]{Department of Physics, University of Wuppertal, Gau{\ss}stra{\ss}e 20, 42119 Germany}


\begin{abstract}
We calculate the hadronic scales $r_0$, $r_1$ and their ratio $r_0/r_1$ on $\Nf=2+1$ flavor QCD ensembles generated by the CLS consortium. These scales are determined from a tree-level improved definition of the static force on the lattice, which we measure using Wilson loops. Our analysis involves various continuum and chiral extrapolations of data that cover pion masses between 134 MeV and 420 MeV and five lattice spacings down to 0.039 fm. We compare the potential scales to gradient flow scales by forming corresponding ratios. We find $r_0=0.4757(64)$ fm at the physical point. As a byproduct of our analysis we express the $\Nf=3$ QCD Lambda parameter determined by the ALPHA Collaboration in units of the scale $r_0$ and obtain $r_0 \Lambda^{(3)}_{\overline{\rm{MS}}} = 0.820(28)$. Furthermore we present results for the second derivative of the potential to study its shape and compare it to phenomenological potential models.
\end{abstract}
  



\maketitle


\newpage

\section{Introduction}
One of the essential requirements in lattice QCD calculations is the determination of the scale, i.e. of the lattice spacing in physical units. Among various scales used in lattice Quantum Chromodynamics (QCD), the $r_0$ scale~\cite{Sommer:1993ce} has been used in practical computations for a long time. The $r_0$ scale is convenient due to its simple definition in terms of the static quark anti-quark potential $V_{0}(r)$, which can be computed via Wilson loops~\cite{Wilson:1974sk}. On the other hand, the determination of $r_0$ from experimental input is not straightforward and is only indirect, since it involves phenomenological potential models.

Computations of Wilson loops established an understanding of confinement and its interplay with asymptotic freedom, a central problem of particle physics, via the formation of a flux tube between quark-anti-quark static charges~\cite{DiGiacomo:1989yp, DiGiacomo:1990hc, Singh:1993jj, Bali:1994de, Bali:2000gf, Luscher:2002qv, Greensite:2005yu, Andreev:2020pqy}. Confinement manifests itself in the asymptotic linear rise of $V_{0}(r)$ at large $r$; the corresponding slope is known as the string tension. In quenched calculations, the scale has been set using the string tension, but in full QCD the string breaks at the pair-production threshold, making a precise definition difficult \cite{Bulava:2019iut,Bulava:2024jpj}. Instead of the string tension, one can consider the force $F(r) \equiv dV_{0}(r)/dr$ and form the dimensionless product $r^{2}F(r)$. The latter can be used to set the scale at distances defined by $r_{i}^{2} F(r_{i})= c_{i}$, with $c_{0}=1.65$~\cite{Sommer:1993ce}, $c_{1}=1$~\cite{Bernard:2000gd}, or $c_{2}=\frac{1}{2}$~\cite{Bazavov:2017dsy}. The definition of $r_0$ amounts to $r_0\approx0.5$ fm in phenomenological potential models.

In this work, we determine the $r_0$ and $r_1$ scales on the ensembles of gauge configurations generated by the Coordinated Lattice Simulations (CLS) initiative~\cite{Bruno:2014jqa,Mohler:2017wnb}. These ensembles include the dynamics of the three light (up, down and strange) quarks. The CLS consortium has produced these ensembles with an improved lattice action, high statistics, and multiple lattice spacings, making them particularly suitable for precision calculations. Incorporating the $r_0$ scale into the analysis of CLS ensembles offers valuable insights into the continuum and physical quark mass limits. 

The article is organized as follows: First, we introduce the potential between a quark and anti-quark in the static limit and discuss its properties in~\sect{sec:pot}. Then  in~\sect{sec:sim} we describe the ensembles used in this work which were generated by the CLS consortium. Next, we present our determination of the scales $r_0$ and $r_1$ in~\sect{sec:r0}. We perform various continuum and chiral extrapolations of the scales $r_0$ and $r_1$ in units of the gradient flow scale $\sqrt{t_0}$ \cite{Luscher:2010iy} as well as the ratio $r_0/r_1$ in~\sect{sec:ex}. We draw our conclusions by quoting our final results for the potential scales and the $\Nf=3$ QCD $\Lambda$ parameter in units of $r_0$ in~\sect{sec:co}. There we also discuss the shape of the static potential at distances smaller than the string breaking distance and compare it to potential models.

\section{The Static Potential}\label{sec:pot}

The potential $V(r)$ between a static quark and anti-quark separated by a distance $r$ from each other is one of the most fundamental and conceptually simplest quantities computable in QCD. 
At short distances it is very well described by perturbation theory and calculations to two and three loop order have been 
completed~\cite{Peter:1997me,Schroder:1998vy,Schroder:1998vy,Melles:2000dq}~\cite{Anzai:2009tm,Brambilla:2009bi} both in pure gauge theory and in full QCD. Renormalized couplings can be defined both from the 
potential directly~\cite{Brodsky:1999fr} (the $V-$scheme),
\begin{equation}
   \alpha_V(\mu) = -\frac{1}{\CF}r V(r)\Bigr|_{\mu=r^{-1}} \,,
\end{equation}
but also from the static force $F(r) \equiv V'(r)$~\cite{Sommer:1993ce} or even from $V''(r)$~\cite{Donnellan:2010mx},
\begin{eqnarray}
   \alpha_{qq}(\mu) &=& \frac{1}{\CF} r^2 V'(r)\Bigr|_{\mu=r^{-1}} \, , \\
   \alpha_c(\mu)    &=& - \frac{1}{2\CF} r^3 V''(r)\Bigr|_{\mu=r^{-1}}\, . \label{e:ccoupling}
\end{eqnarray}
The shape of the potential, or the force, at short distances is related to the $\beta$-functions of these couplings,
which are known to 4 loops, derived from the 3-loop calculations of the potential together with the 4-loop $\beta$-function~\cite{vanRitbergen:1997va, Czakon:2004bu}
of the $\overline{\rm MS}$-coupling.

At long distances without dynamical fermions, the potential 
is expected to follow predictions from effective string theories~\cite{Nambu:1978bd,Luscher:2002qv}
\begin{equation}\label{e:effstring}
   V(r) = \sigma\, r + \mu + \frac{\gamma}{r} + O(1/r^2),\qquad \gamma=-\frac{\pi}{24}(d-2)\, ,
\end{equation}
where $\sigma$ is the string tension, $\mu$ is a regularization dependent mass and $\gamma$ is a coefficient
that depends on the space-time dimensionality $d$ and which is universal across a wide range of effective string theories.
Potential models~\cite{Eichten:1978tg} assume a certain shape of the static potential to predict heavy meson spectra. 
Over the years different phenomenological potentials have emerged, one being the Cornell potential~\cite{Eichten:1979ms}
\begin{equation}\label{e:Cornell}
   V_{\rm Cornell}(r) = - \frac{0.52}{r} + \frac{r}{(2.34\ {\rm GeV}^{-1})^2}\, ,
\end{equation}
another the Richardson potential~\cite{Richardson:1978bt}
\begin{eqnarray}
   V_{\rm Richardson}(r) &=& \frac{8\pi}{33-2\Nf}\Lambda\left(\Lambda r - \frac{f(\Lambda r)}{\Lambda r}\right)\, ,\label{e:Richardson} \\
   f(t) &=& \frac{4}{\pi}\int\limits_0^\infty \! {\rm d}q\, \frac{\sin(qt)}{q}\left[\frac{1}{\ln(1+q^2)}-\frac{1}{q^2} \right]\, ,
\end{eqnarray}
where $\Nf=3$, $\Lambda=398$ MeV was used in the original publication.
In full QCD the string picture is only approximate, because the string ``breaks'' at a distance around 
$r\approx 1.2$ fm\cite{Bali:2005fu,Bulava:2019iut,Bulava:2024jpj}, and $V(r)$
smoothly transitions from being a static quark potential to being a (very flat) static-light molecule potential.
At intermediate distances and in full QCD, the potential is best
computed numerically in the framework of lattice QCD. When computed on a lattice with lattice spacing $a$, $V(r)$ contains a linearly divergent constant $E_{\rm self}(a) \sim \frac{1}{a}$~\cite{DellaMorte:2005nwx,Grimbach:2008uy}, hence the static force, in which this is absent, is
more widely used. An important application is the definition of hadronic scales~\cite{Sommer:1993ce, Bernard:2000gd}
\begin{equation}
   r_1^2 F(r_1) = 1 \qquad {\rm or} \qquad r_0^2 F(r_0) = 1.65\, .
   \label{eq:r1_r0}
\end{equation}
They play an important role when lattice calculations are compared to each other and a dependence on experimental inputs
is undesired. One example is the $\Lambda$ parameter of QCD, which often is published in $r_0$-units~\cite{FlavourLatticeAveragingGroupFLAG:2021npn}.
In more recent times the $r_0$ scale is increasingly being replaced by gradient flow scales like $t_0$~\cite{Luscher:2010iy} 
or $w_0$~\cite{BMW:2012hcm} which are maybe more abstract, but easier to compute to a high precision. However the deep theoretical understanding  
and the mild lattice artifacts of $r_0$ make it still an appealing choice.

\section{Simulations}\label{sec:sim}

For our calculation we use a subset of the gauge field ensembles generated by the CLS 
consortium~\cite{Bruno:2014jqa,Mohler:2017wnb}.
These were generated using a L\"uscher-Weisz-gauge action~\cite{Luscher:1984xn} with $\Nf=2+1$ flavors of dynamical 
$O(a)$ improved Wilson quarks~\cite{Sheikholeslami:1985ij,Bulava:2013cta}. The ensembles cover pion masses from the $SU(3)$-symmetric 
mass point with $m_\pi\approx 420$ MeV down to the physical pion mass, along trajectories where
the sum of bare quark masses stays constant. Five different lattice spacings between 0.039 fm and 0.085 fm 
are available. The lattices have periodic boundary conditions for both gauge and fermion fields, except 
for the temporal direction which has either anti-periodic fermionic boundaries or, especially on all fine lattices,
open boundaries~\cite{Luscher:2011kk}. In case of open boundaries, tree-level values for the fermionic and gluonic 
boundary improvement coefficients were used. 

An even-odd preconditioned hybrid Monte-Carlo~\cite{Duane:1987de} with frequency splitting~\cite{Hasenbusch:2001ne} of the quark determinant was used for the light quark doublet. The strange quark was treated with an RHMC~\cite{Kennedy:1998cu}. For further
details concerning the simulation algorithm we refer the reader to~\cite{Luscher:2012av} and the documentation of
the \verb+openQCD+ package~\cite{openQCD}.

The simulated gauge field distribution deviates slightly from the desired target distribution and all observables 
are corrected by the inclusion of a reweighting factor. This accounts for 
twisted-mass-reweighting~\cite{Luscher:2008tw,Kuberski:2023zky},
inperfections of the rational approximation and an occasionally negative strange quark determinant~\cite{Mohler:2020txx}.

Keeping the sum of bare quark masses constant along a chiral trajectory is almost equivalent to keeping
a suitable combination of pion and kaon masses constant. In nature the dimensionless combination $\phi_4 := 8t_0 (m_K^2 + m_\pi^2/2)$ has a value very close to 1.098~\cite{Strassberger:2021tsu}. 
All our ensembles deviate only slightly from this value. The physical mass point has $\phi_2 := 8t_0 m_\pi^2 = 0.0779(7)$~\cite{Strassberger:2021tsu},
which at the flavor symmetrical point should be $2\phi_4/3$. Slight mistunings of the masses are inevitable, and therefore
our symmetric point ensembles do not all have exactly the same $\phi_2$ values, and the chiral trajectories do not have
exactly constant $\phi_4$ values. In the past such mistunings were corrected by computing quark mass derivatives of
all observables and using these to shift the results~\cite{Bruno:2016plf}.
In this work the main focus is on purely gluonic observables and
on dimensionless ratios of gluonic observables. In such cases the quark mass dependence is mild enough to simply 
ignore the mistunings. This has been observed in~\cite{Bruno:2016plf}, where the necessary shifts in 
the gluonic quantity $t_0$ were insignificant, but it is also what we observe for potential scales on
the ensembles where we computed the quark mass derivatives. 
For example on ensemble D450, $\phi_4$ needs to be reduced by 0.51 \%. This changes
$r_0$ by only 0.46 \textperthousand, which is negligible, given that its statistical error is 0.45 \%.
In the end, we avoid using the rather noisy quark mass derivatives, and instead take the mistuning into account by including terms proportional to $(\phi_4-1.098)$ in some of our
fits.

The data analysis of auto-correlated Monte-Carlo time series is treated with the $\Gamma$-method~\cite{Wolff:2003sm}.
An estimate of the exponential autocorrelation time $\tau_{\rm exp}$ is used to take slowly decaying modes of the 
auto-correlation functions into accounts, as proposed in~\cite{Schaefer:2010hu}. ``Derived observables'', i.e. non-linear functions of Monte-Carlo estimates, are conviniently treated as described e.g. 
in~\cite{dobs,Ramos:2018vgu,Joswig:2022qfe}, i.e. by storing their mean values and
``projected fluctuations'' in suitable structures. This greatly simplifies linear error propagation with full control over all correlations and auto-correlations.

Table~\ref{tab:sim} summerizes the simulations used for this study.

\begin{table}[!h]
  \centering
  \scriptsize{
   \begin{tabular}{c c c c c c c c c}
   \toprule
       $\beta$& id & $\frac{T}{a}\times \frac{L^3}{a^3}$ & b.c. & $(\kappa_{u,d}, \kappa_s)$ & $m_\pi$[MeV] & stat. [MDU] & $\tau_{\rm exp}$ [MDU] \\
   \midrule
       \multirow{5}{*}{\rotatebox[origin=c]{90}{3.40}}
       &H101 & $96\times32^3$& o & (0.13675962, 0.13675962) & 420  & 8064 & 26.6  \\
       &H102-0 & $96\times32^3$ & o & (0.136865, 0.136549339) & 353  & 3988 & 26.6 \\
       &H102-1 & $96\times32^3$ & o &                         & 356  & 4032 & 26.6 \\
       &H105 & $96\times32^3$& o & (0.13697, 0.13634079)    & 281  & 7880 & 26.6 \\
       &C101 & $96\times48^3$& o & (0.137055, 0.136172656)  & 221  & 8000 & 26.6 \\
   \midrule
       \multirow{4}{*}{\rotatebox[origin=c]{90}{3.46}}
       &B450 & $ 64\times32^3$& p & (0.13689, 0.13689)      & 424  & 6448 & 32.9 \\
       &S400 & $128\times32^3$& o & (0.136984, 0.136702387) & 357  & 11492 & 32.9 \\
       &D450 & $128\times64^3$& p & (0.137126, 0.136420428639937) & 220 & 2000 & 32.9 \\
       &D452 & $128\times64^3$& p & (0.137163675, 0.136345904546) & 157  & 3996 & 32.9 \\
   \midrule
       \multirow{5}{*}{\rotatebox[origin=c]{90}{3.55}}
       &N202 & $128\times48^3$& o & (0.137, 0.137)          & 418  & 7608 & 47.3 \\
       &N203 & $128\times48^3$& o & (0.13708, 0.136840284)  & 350  & 6172 & 47.3 \\
       &N200 & $128\times48^3$& o & (0.13714, 0.13672086)   & 288  & 6848 & 47.3 \\
       &D200 & $128\times64^3$& o & (0.1372, 0.136601748)   & 200  & 8004 & 47.3 \\
       &E250 & $192\times96^3$& p & (0.137232867, 0.136536633) & 131  & 3800 & 47.3 \\
   \midrule
       \multirow{4}{*}{\rotatebox[origin=c]{90}{3.70}}
       &N300-0 & $128\times48^3$& o & (0.137, 0.137)        & 427  & 2028 & 77.9 \\
       &N300-1 & $128\times48^3$& o &                       & 427 & 5916 & 77.9 \\
       &J303 & $192\times64^3$& o & (0.137123, 0.1367546608)& 260  & 8584 & 77.9 \\
       &E300 & $192\times96^3$& o & (0.137163, 0.1366751636177327) & 177  & 4548 & 77.9 \\
   \midrule
       \multirow{2}{*}{\rotatebox[origin=c]{90}{3.85}}
       &J500 & $192\times64^3$& o & (0.136852, 0.136852)   & 416  & 11552 & 126.5\\
       &J501 & $192\times64^3$& o & (0.1369032,0.136749715)& 339  & 14236 & 126.5\\
   \bottomrule
   \end{tabular}
   }
   \caption{CLS ensembles used in this work. The ensembles are labeled by and id. For the H102 and
	    N300 ensembles, runs with the same physical but different simulation parameters exist. The table
	    furthermore lists the inverse bare coupling $\beta$, the lattice sizes, the boundary
	    conditions in time direction (b.c.), where ``o'' stands for open and ``p'' for (anti-)periodic,
	    the bare quark masses parametrized by the hopping parameters for up/down and 
	    strange quarks, the pion mass in MeV taken from~\cite{Strassberger:2021tsu} and the statistics 
	    in molecular dynamic units.
	    The final entry is a rough estimate of the exponential autocorrelation time, which is 
	    used during error analysis to attach ``tails'' to auto-correlation functions.}
   \label{tab:sim}
\end{table}

\section{Determination of $r_0/a$ and $r_1/a$}\label{sec:r0}

The potential $V(r)$ between a static (infinitely massive) quark and anti-quark separated by distance $r$ can be extracted from the expectation values of Wilson loops, which are traces of products of links along rectangular paths extending in Euclidean time and one spatial direction. In this article we consider only $r/a \times T/a$ on-axis Wilson loops, but off-axis (non-planar) Wilson loops can also be used. More precisely, we compute
\begin{eqnarray}
 W^{(l,k)}(x_0,r,T)&=& \frac{a^3}{2 L^3}\sum\limits_{\mu=1}^3 \sum\limits_{\vec x} \biggl( \nonumber \\
 &&\hspace{-3.5cm}\frac{\langle \mathcal{W}^{(l)}(x,x+r\hat \mu)\ \mathcal{W}^{(0)}(x+r\hat \mu, x+r\hat \mu +T\hat 0) \
              {\mathcal{W}^{(k)}}^\dagger(x+T\hat 0,x+r\hat \mu+T \hat 0)\ {\mathcal{W}^{(0)}}^\dagger (x, x+T\hat 0) 
              w \rangle}{\langle w \rangle}\nonumber \\
 &+& T\to -T \biggr) \, ,
\end{eqnarray}
where $\mathcal{W}^{(l)}(x,y)$ denotes a smeared straight Wilson line from $x$ to $y$. We follow the method of~\cite{Donnellan:2010mx} for the measurement of the of Wilson loops and use the {\tt wloop} package~\cite{wloop}.
All gauge links have been HYP2 smeared using parameters $\alpha_1 = 1.0, \alpha_2 = 1.0, \alpha_3 = 0.5$ 
\cite{Hasenfratz:2001hp, DellaMorte:2005nwx}, which amounts to a particular modification of the 
Eichten-Hill static action \cite{Eichten:1989zv}. In addition the links for the spatial lines in
$\mathcal{W}^{(l)}$ have undergone $l$ iterations of 3D HYP smearing with $\alpha_2 = 0.6, \alpha_3 = 0.3$,
which allows us to construct a variational basis of creation/anihilation operators. 
After each iteration we project the gauge fields onto $SU(3)$ as described in \cite{DellaMorte:2005nwx} and always use 
Eq. (2.24) and four iterations of Eq. (2.25) of \cite{DellaMorte:2005nwx} for the projection. 
The numbers of smearing iterations are scaled with the lattice spacing and are tabulated in \tab{tab:smearing} of appendix~\ref{s:apploop}.
Finally $w$ denotes 
the overall reweighting factor, as described in the previous section.

Assuming that $x_0$ and $x_0+T$ are sufficiently far away from the temporal lattice boundaries, each of the measured correlation functions posseses a spectral decomposition
\begin{equation}
   W^{(l,k)}(x_0,r,T) = \sum_{n=0}^\infty c^{(l,k)}_n(r) e^{-V_n(r) T}\, ,
\end{equation}
with complex coefficients $c^{l,k}_n = {c^{k,l}_n}^*$. At short to intermediate 
separations $r$, the lowest lying energy level $V_0(r)\equiv V(r)$ is the static potential. 
The level above, $V_1(r)$, can be an excitation of the static potential, a static potential with hadrons that have
vacuum quantum numbers (e.g. two pions) or a pair of static-light mesons. Its nature can change depending on 
the distance. 

The correlation matrix is Hermitian, $ W^{(l,k)} =  {W^{(k,l)}}^*$, but with finite statistics, this symmetry has to be
enforced by symmetrization, before $W$ is used in a GEVP procedure to extract the energy levels~\cite{Campbell:1987nv,Guagnelli:1998ud,Luscher:1990ck,Blossier:2009kd,Fischer_2020}.

In the statistical average of standard Monte Carlo lattice simulations, the signal of Wilson loops is the result of strong cancellations between positive and negative contributions. This leads to an exponentially growing noise-to-signal ratio which prevents the calculation of the potential at large distances. In pure gauge theory an exponential suppression of the statistical noise of Wilson loops can be achieved by the multi-hit (or one-link) method \cite{Parisi:1983hm} and much further by the multilevel algorithm \cite{Luscher:2001up}. These algorithms are not applicable in presence of dynamical fermions due to the non-locality of the effective gauge action when the logarithm of the fermion determinant is included. HYP smearing is known to reduce the noise problem both in pure gauge theory and in 
full QCD. In pure gauge theory it was demonstrated in \cite{Alexandrou:2001ip} that the use of HYP smeared links leads to a determination of the static potential comparable in precision to the multi-hit method. In \cite{Donnellan:2010mx} it was shown that this procedure leads to a determination of the potential between quark and anti-quark sources that agrees with the continuum potential up to $\mathcal{O}(a^2)$ effects (after renormalization).

\subsection{Effects of open temporal boundaries}

\begin{figure}[!h]
  \centering 
  \includegraphics[width=0.48\textwidth,height=.4\linewidth]{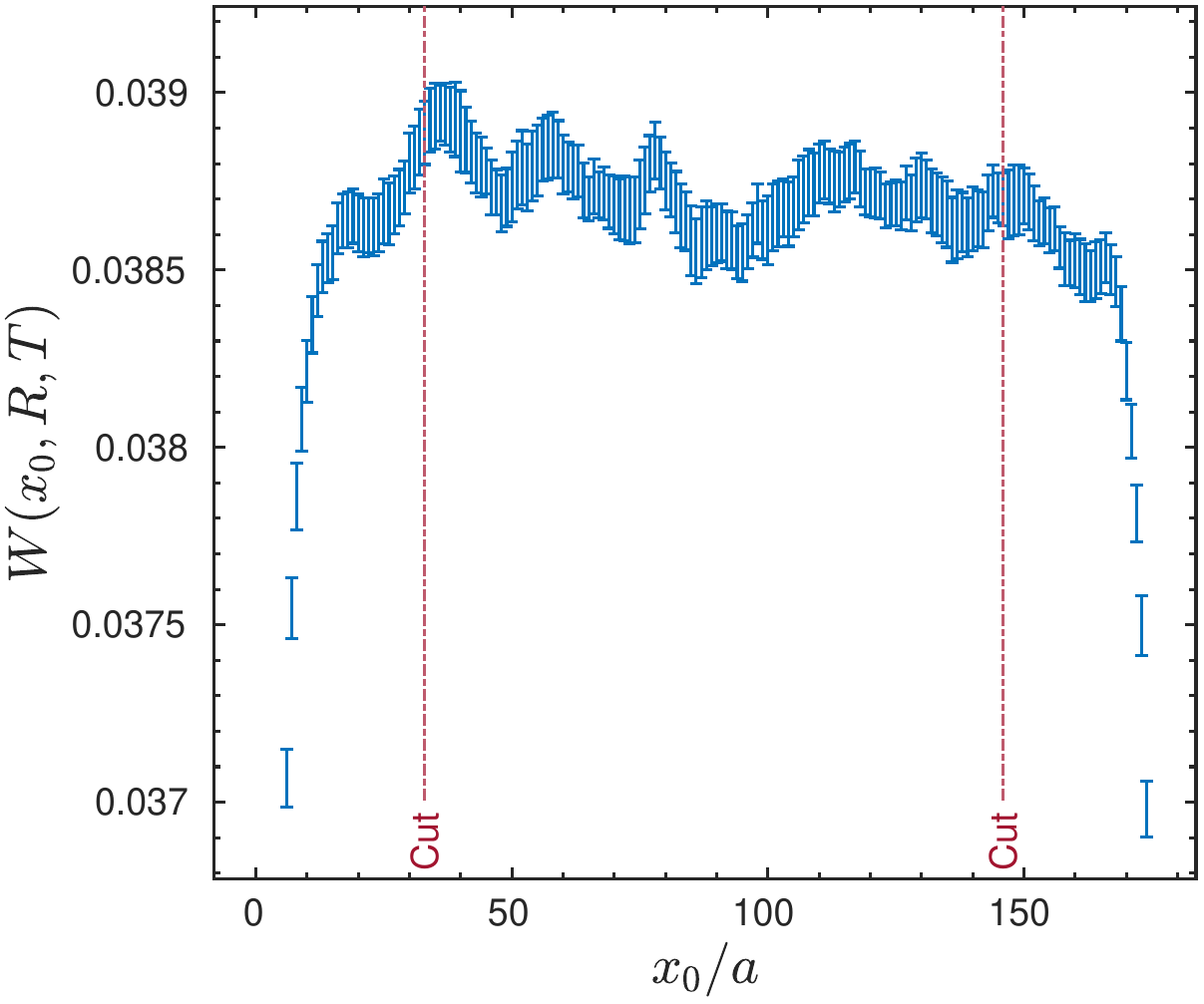}  \hfill
  \includegraphics[width=0.48\textwidth,height=.4\linewidth]{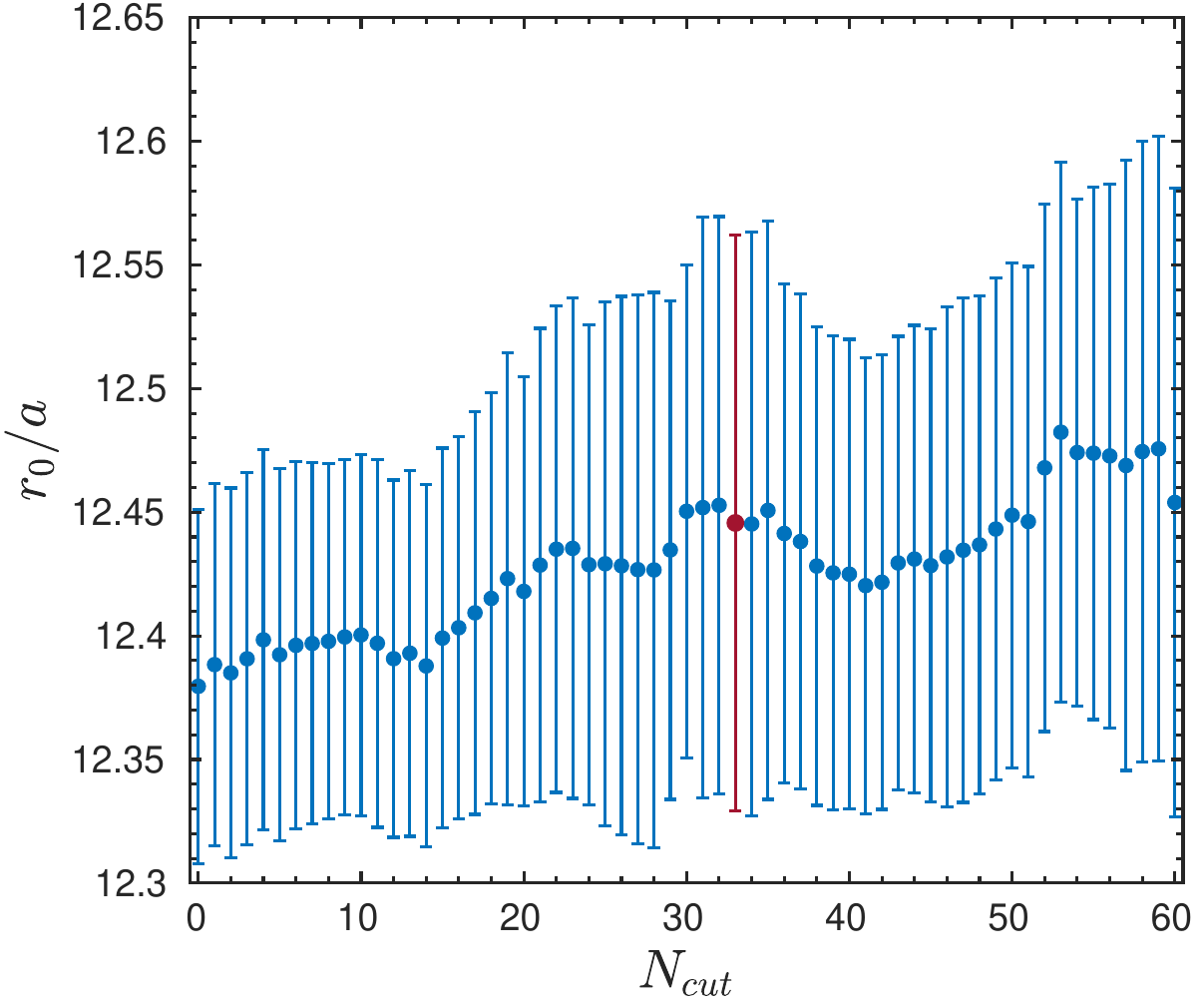}
  \caption{On the left $W(x_0,r,T)$  as a function of $x_0$ is shown for ensemble J500 with loopsize $T=13a$ $\&$ $r=6a$ together with the cut of $N_{\rm cut}=33$ timeslices from the open boundaries in time. On the right we plot the dependence of $r_0$ on $N_{\rm cut}$, with the red datapoint being a cut of 33.} 
  \label{fig:TimeCut}
\end{figure}

The observables can be averaged in the lattice time direction, but to avoid effects of the open boundaries only Wilson loops sufficiently far away from the boundary are taken into account in the average. To evaluate which cut to use, several values of the number $N_{\rm cut}$ of timeslices cut from the temporal boundaries have been analysed. 
The left side of \fig{fig:TimeCut} shows how the result for a particular Wilson loop $W(x_0,r,T)$ depends on the temporal position $x_0$,
and our preferred cut position is indicated. On the right one can see how the final result for $r_0$, whose determination will be explained in a later chapter, varies with the position of the cut. Surprisingly, the effect of including data that is clearly affected 
by boundary effects, does not lead to large systematic errors in the final result. Even the extreme case, where all data is included
is comfortably compatible with our preferred value (in red), that is free of boundary contaminations. The procedure to obtain $r_0$
from Wilson loops involves a GEVP and a subsequent plateau average of an effective energy. In this process differently sized 
Wilson loops enter, and effects of the open boundaries may cancel to some extent. In some ensembles 
the dependence of the final result on the cut is stronger than in the ensemble J500 displayed here. The main observation 
here is that the error increases when more data is discarded, while the central value remains almost constant.

In the end we set $N_{\rm cut}=15$ for the coarsest lattice and scale this number with the lattice spacing up to $N_{\rm cut}=33$ on the finest lattices.

The time-averaged and symmetrized Wilson loops $\overline W^{(l,k)}(r,T)$ enter a generalized eigenvalue problem
\begin{equation}
   \overline W(r,T) \vec v_n = \lambda_n \overline W(r,T_0) \vec v_n\, ,
\end{equation}
with eigenvalues and eigenvectors that depend on $T,T_0$ and $r$. The $n$'th eigenvalue has the asymptotical
form $\lambda_n(T,T_0,r) \propto e^{-V_n(r) (T-T_0)}$. The value for $T_0 = 5a$ is kept constant for all ensembles and 
distances $r$.

\subsection{Improved distance for $V$, $V'$, $V''$}

Following \cite{Sommer:1993ce} we eliminate tree level lattice artifacts by using the improved distance $r_I$ 
\begin{align}\label{e:latforce}
    F(r_I) &= [V(r) - V(r-a)]/a \, .
\end{align}
The tree level expression for the static potential extracted from Wilson loops using HYP smearing for the temporal lines has been derived in~\cite{Hasenfratz:2001tw,thesisHoffmann}.
The improved distance $r_I$ is defined such that the continuum tree level expression: 
\begin{equation}
    r_I ^2 F_{\rm  tree}(r_I) = C_F \frac{g_0 ^2}{4\pi} \, , 
    \label{eq:r2ftree}
\end{equation}
where, for the gauge group SU(3) $C_F= 4/3$, holds exactly. On the lattice the tree level force is given by
\begin{equation}
 F_{\rm tree}(r_I)= - \CF \frac{g_0 ^2}{a} \int\limits_{-\pi} ^\pi  \frac{d^3k}{(2\pi)^3} \frac{\left[\cos(rk_1/a)-\cos((r-a)k_1/a)\right]\times f_{\rm sm}(k)}{4\left( \sum_{j=1}^3 \sin^2(k_j/2)-4c_1\sum_{j=1}^3 \sin^4(k_j/2)\right)} \, ,
 \label{eq:ftree}
\end{equation}
where the smearing factor $f_{\rm sm}$ is given in Eqs. (A.3) and (A.4) of~\cite{Donnellan:2010mx}.
With $c_1=0$ one would get the result for the plaquette action. In our case $c_1 = -1/12$ is used, which corresponds to the L\"uscher-Weisz-gauge  action.
Similarly for the shape parameter $c(r)$ an improved definition can be found
\begin{align}
    c(\Tilde{r}) =\frac{1}{2} \Tilde{r}^3 [V(r+a)+V(r-a)-2V(r)]/a^2 \, ,
    \label{eq:ImprovedShape}
\end{align}
where $\Tilde{r}$ is chosen such that
\begin{equation}\label{eq:cTree}
    c_{\rm  tree}(\Tilde{r}) = -C_F \frac{g_0 ^2}{4\pi} \, .
\end{equation}
The values of the improved distances for the force $F(r_I)$ and for $c(\Tilde{r})$ for our choice of HYP2 smearing can be found in table \ref{tab:r_I} of appendix~\ref{s:apprI}.

\subsection{Evaluation of $V(r)$ and $F(r)$} \label{Ch:eval-v0}

\begin{figure}
    \centering
    \includegraphics[width=0.32\textwidth, height=4cm]{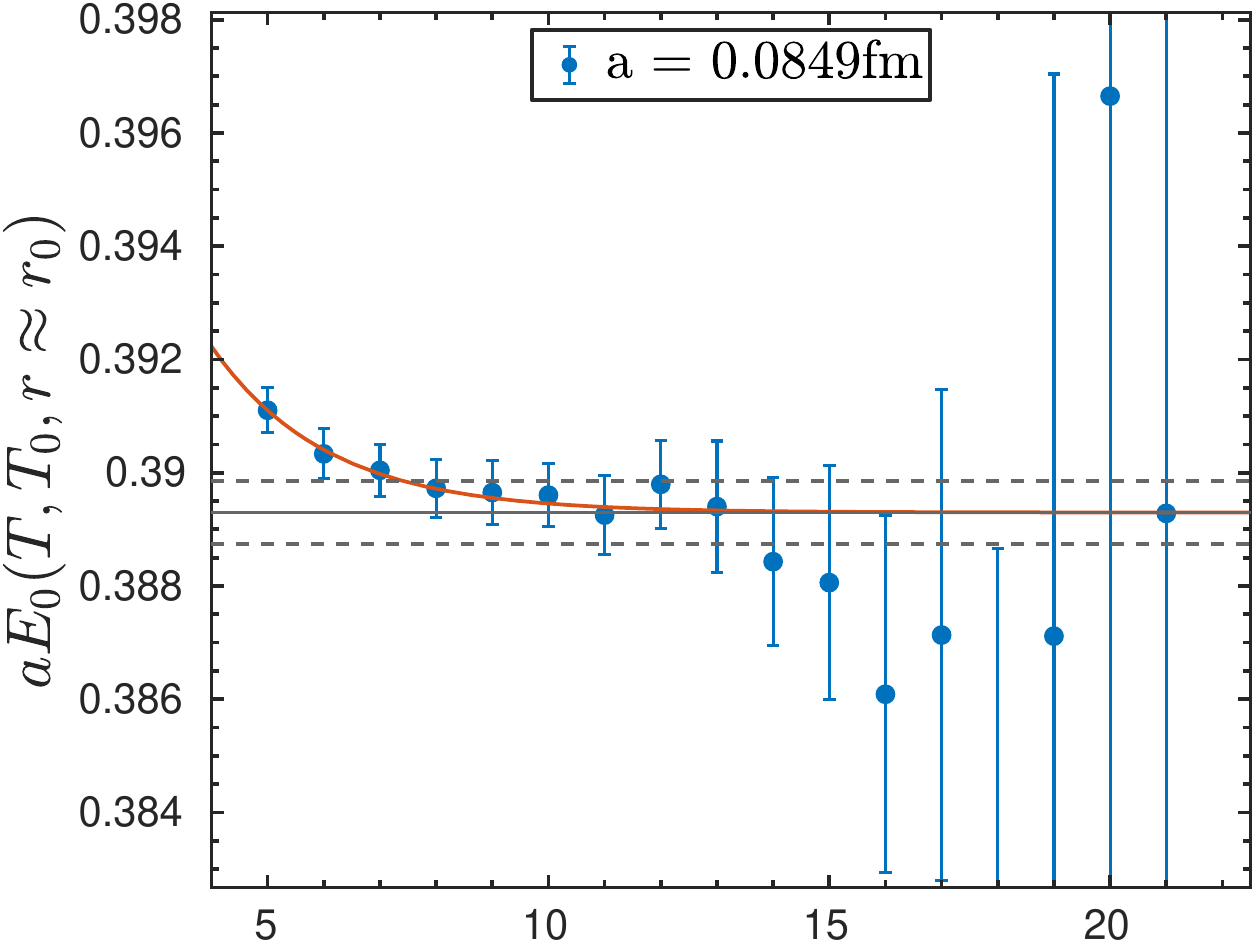}\hfill
    \includegraphics[width=0.31\textwidth, height=4cm]{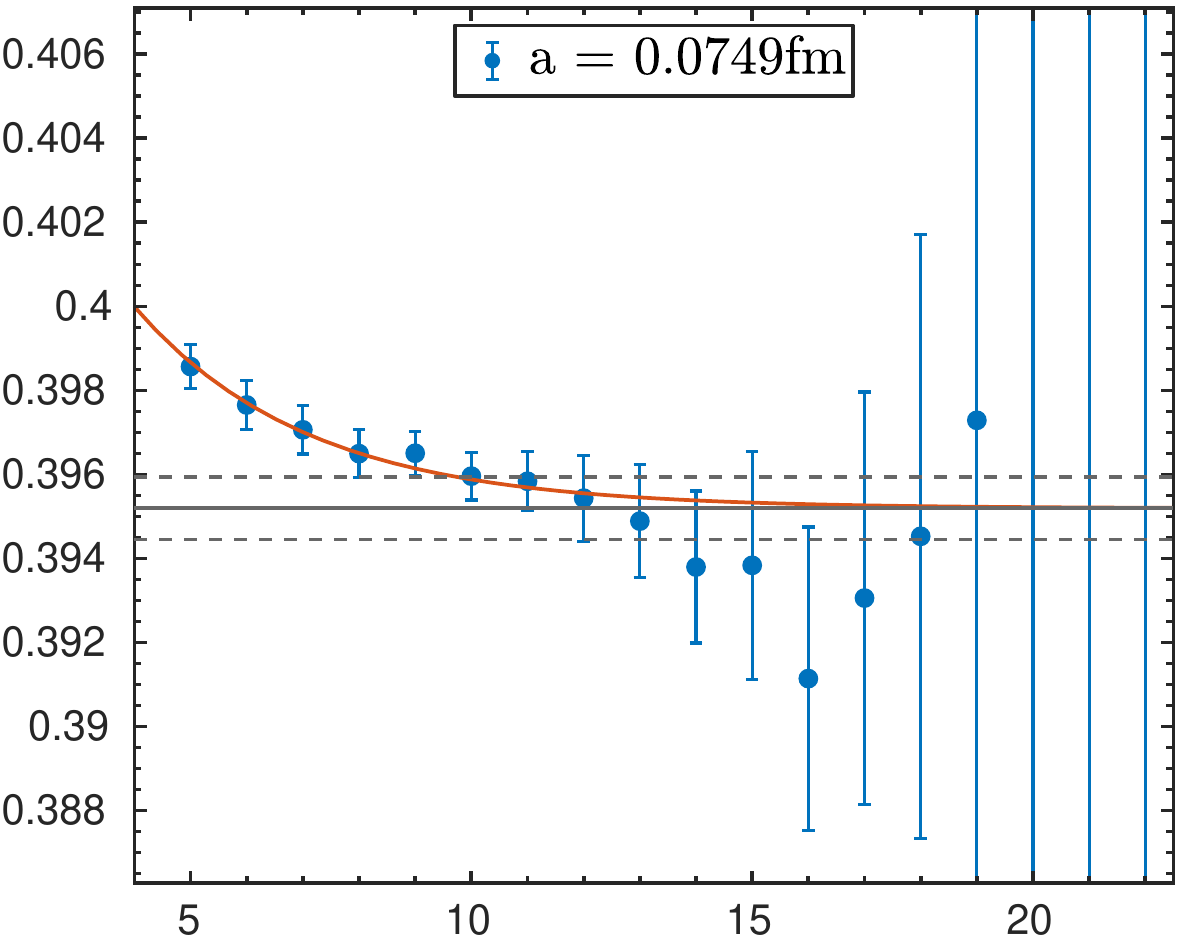}\hfill
    \includegraphics[width=0.32\textwidth, height=4cm]{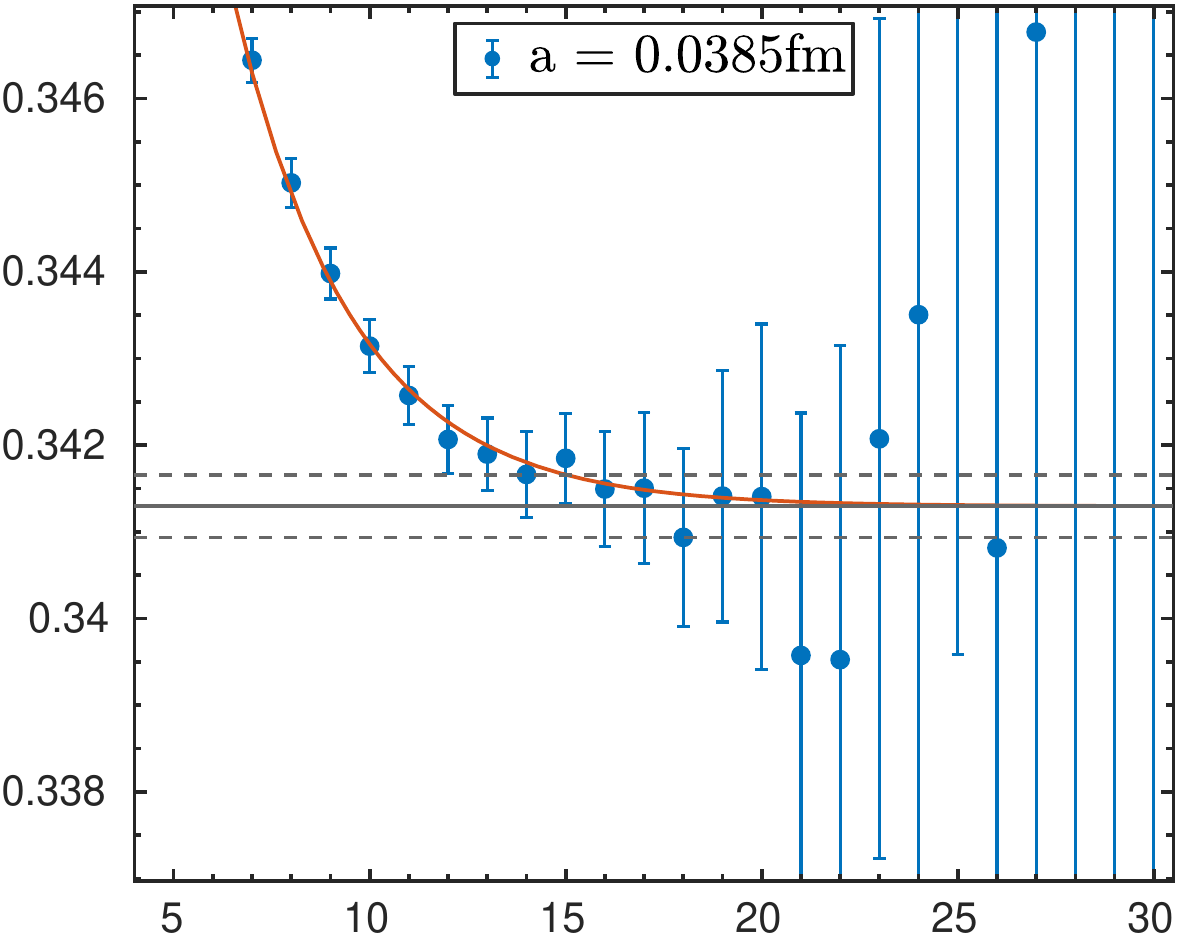}
    \\
    \includegraphics[width=0.32\textwidth, height=4cm]{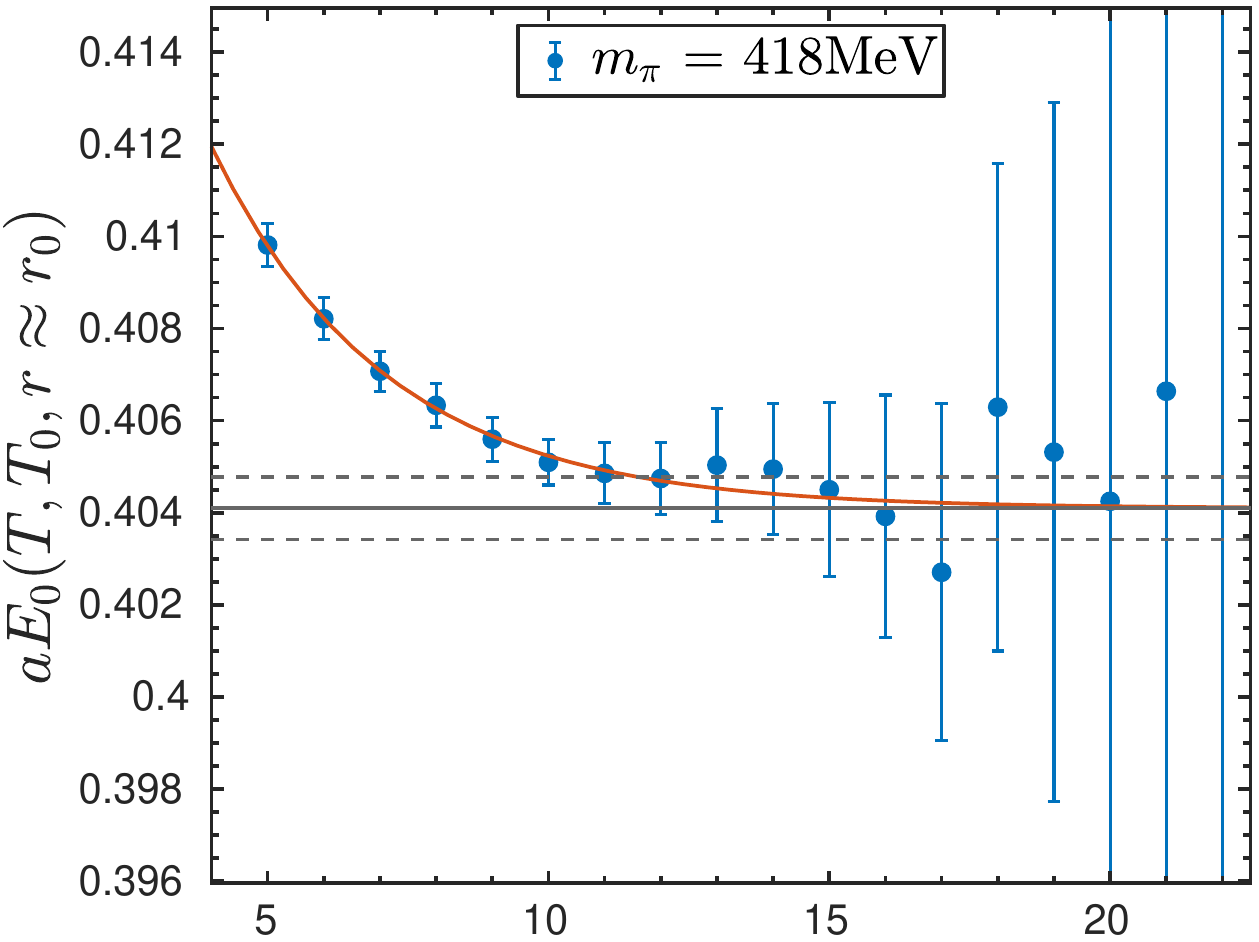}\hfill
    \includegraphics[width=0.31\textwidth, height=4cm]{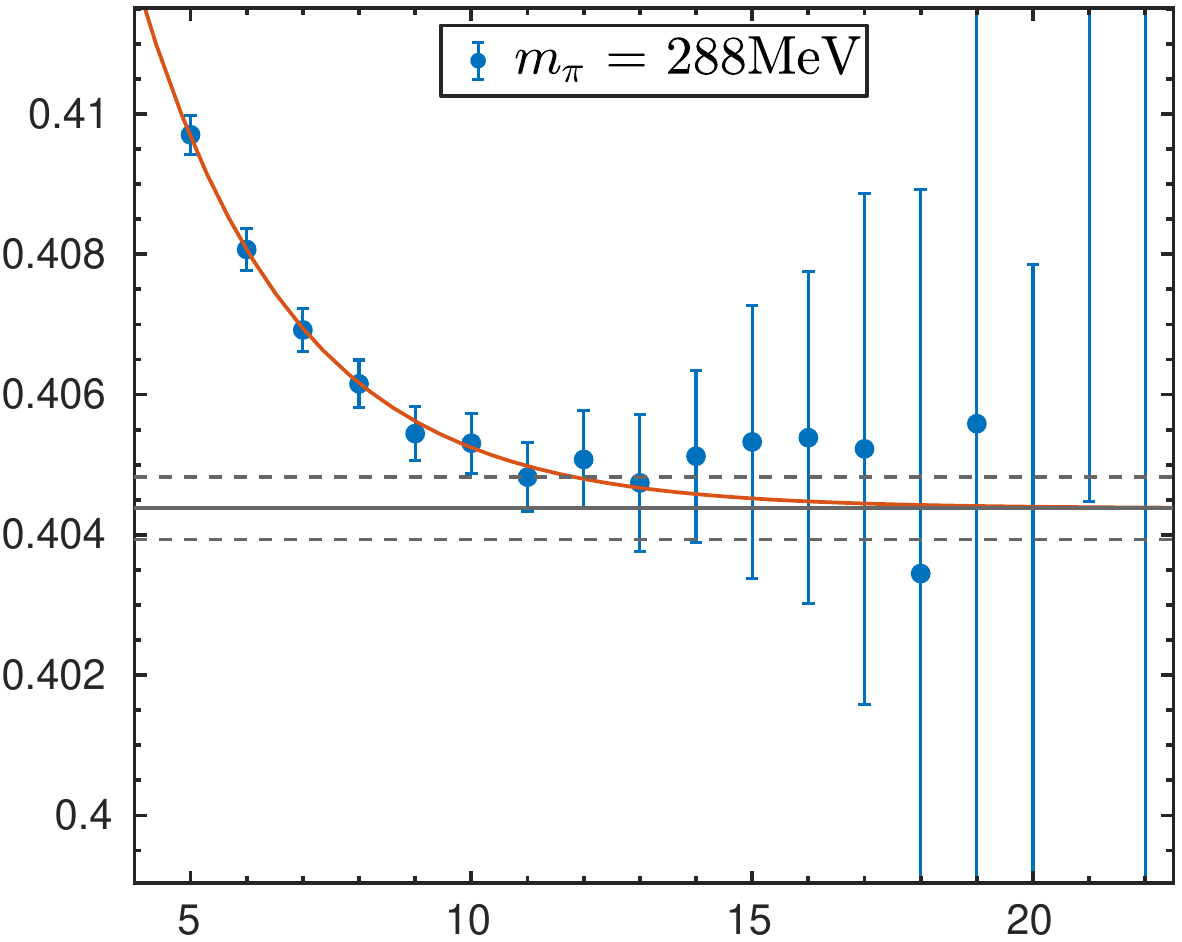}\hfill
    \includegraphics[width=0.32\textwidth, height=4cm]{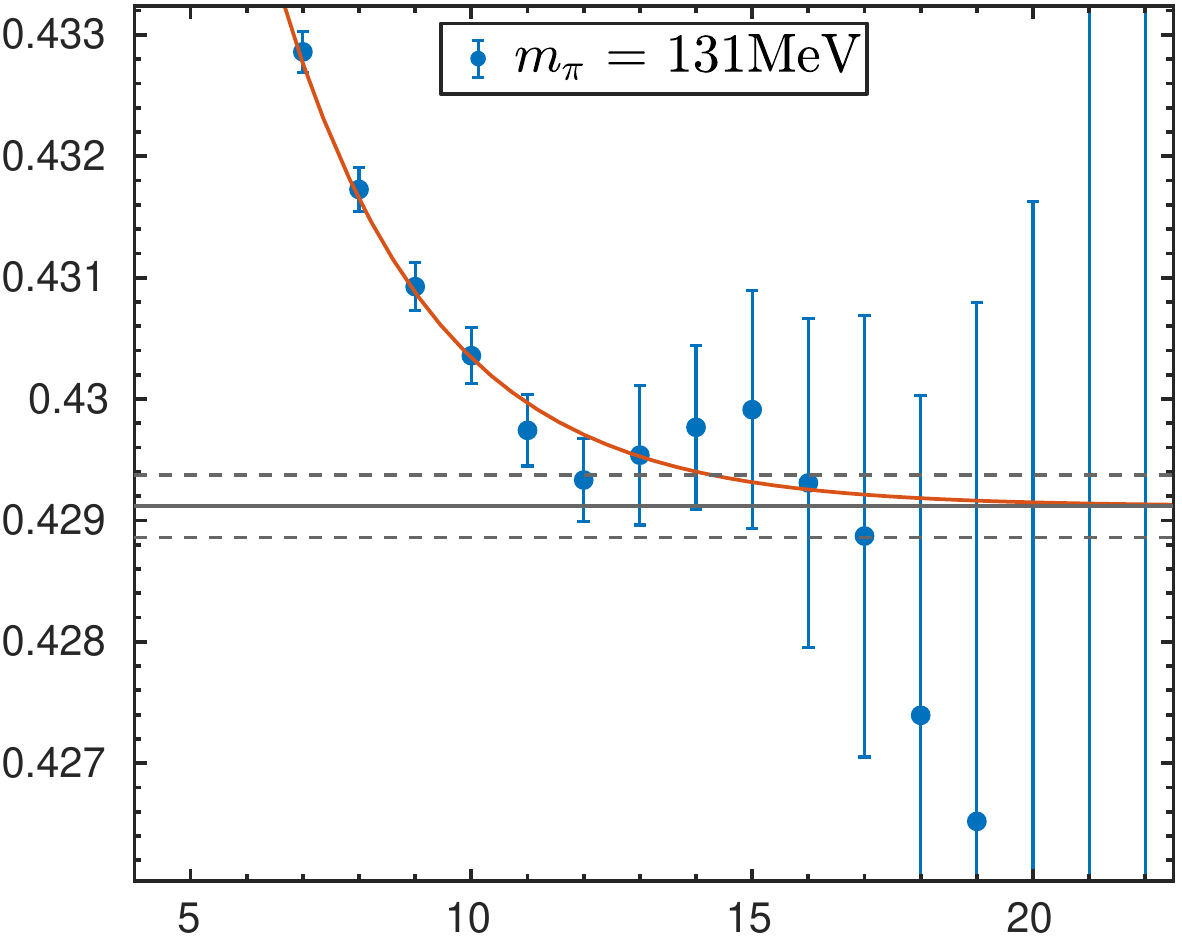}
    \\
    \includegraphics[width=0.32\textwidth, height=4.cm]{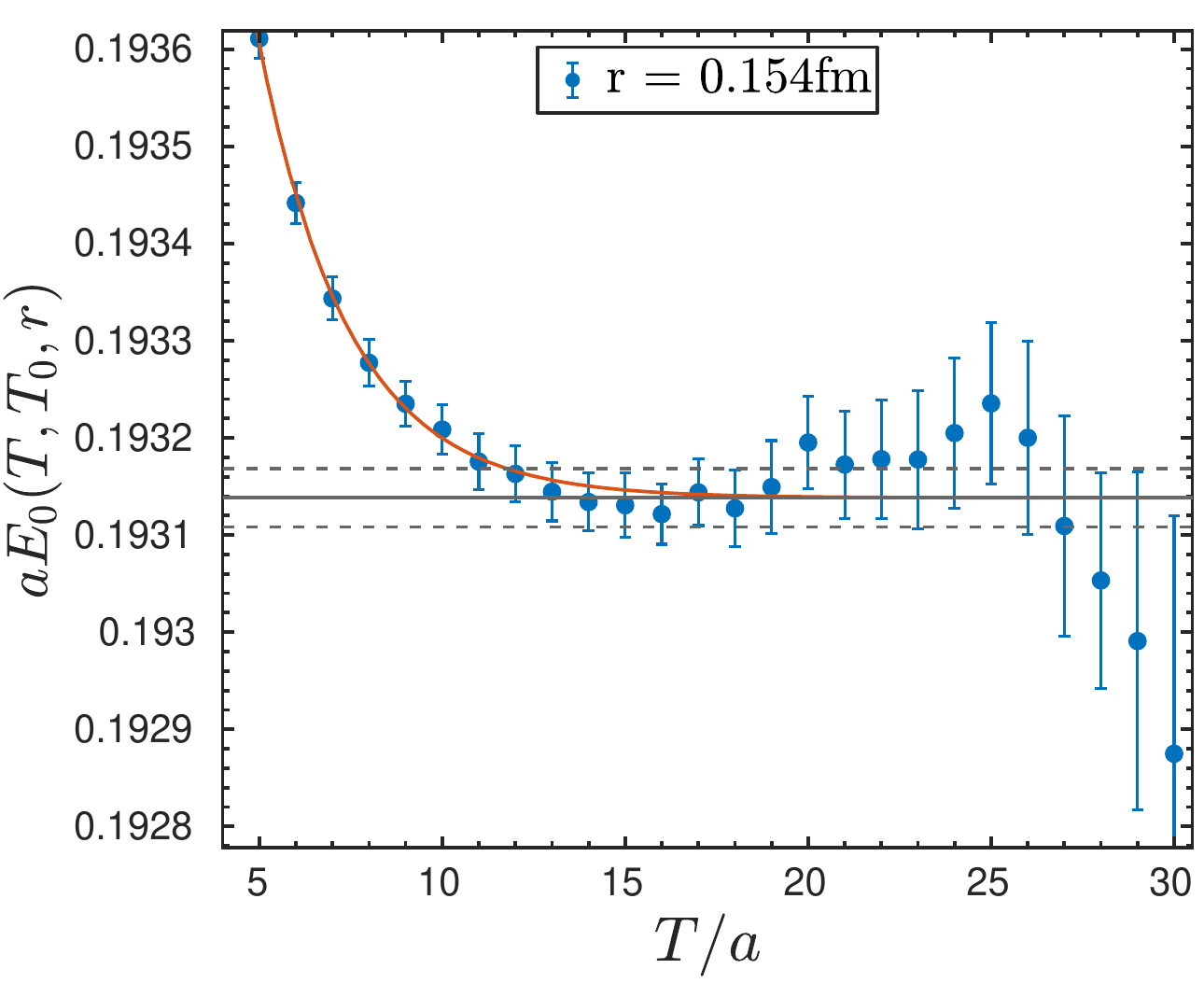}\hfill
    \includegraphics[width=0.33\textwidth, height=4.cm]{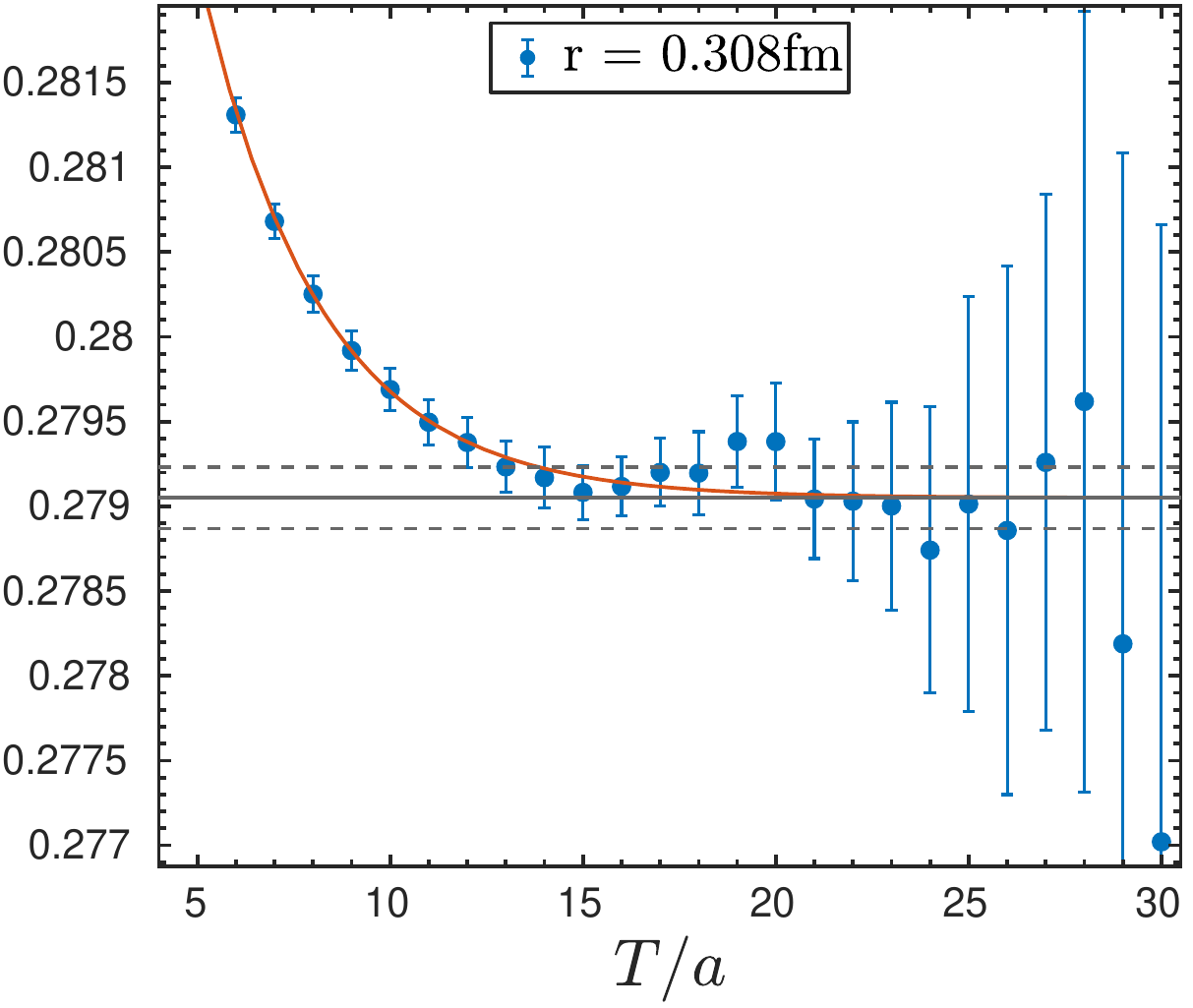}\hfill
    \includegraphics[width=0.33\textwidth, height=4.cm]{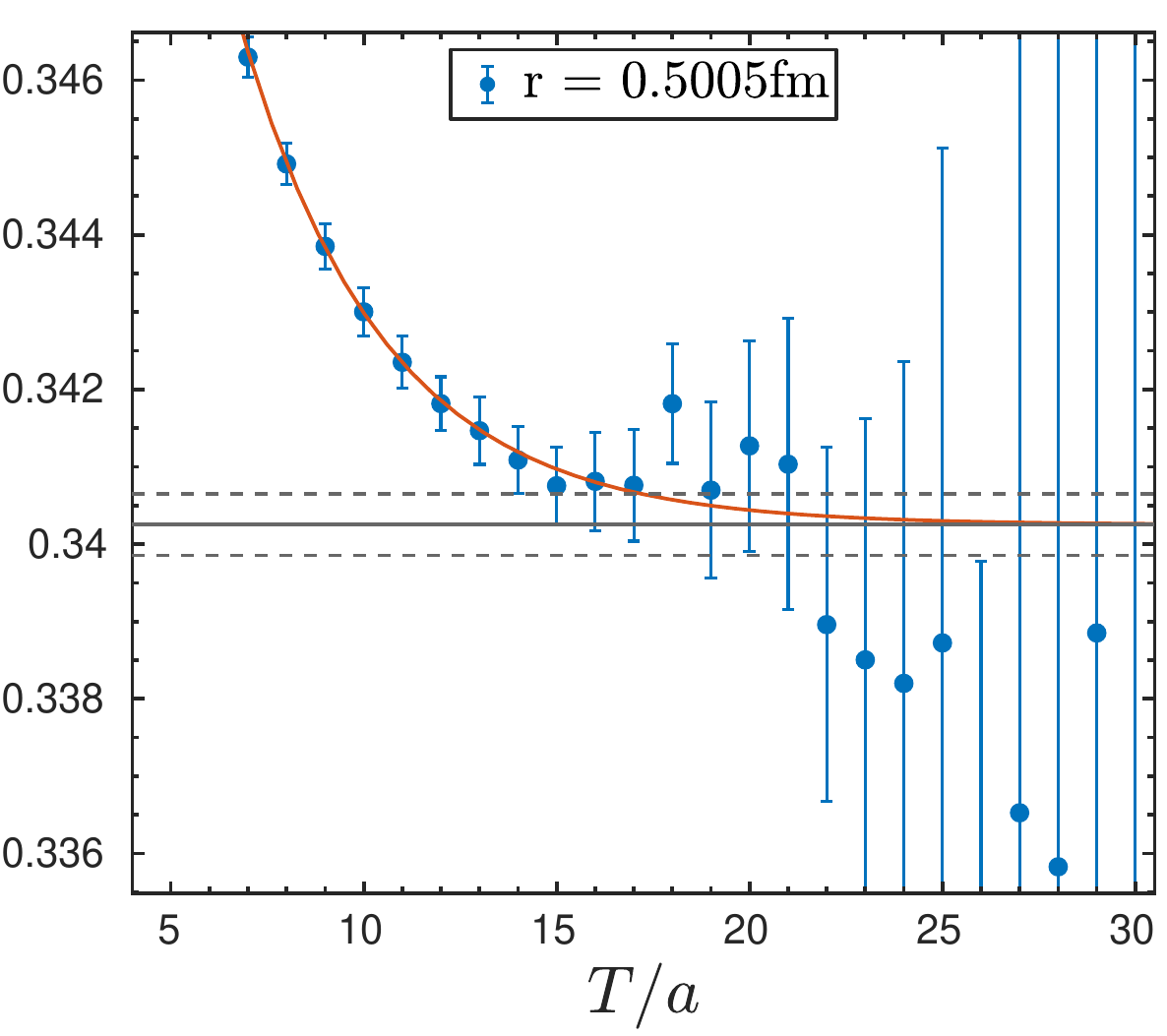}
    
    \caption{Several effective mass plots. The first row shows the symmetric mass ensembles H101, B450, and J500 with decreasing lattice spacing at $r\approx r_0$. In the second row we display ensembles N202, N200, and E250 with decreasing pion mass from the symmetrical to physical value at $r\approx r_0$. The last row shows the finest lattice J501 at $m_\pi\approx340\,$MeV with different values of $r$ from $\frac{r_1}{2}$ up to $r_0$}
    \label{f:effmasses}
\end{figure}

The ground state potential $V(r)$ can be found by performing a weighted 
average in a plateau region of the effective masses 
$E_0(T,T_0,r)=\frac{1}{a}\ln(\lambda_0(T,T_0,r)/\lambda_0(T+a,T_0,r)$ extracted from the GEVP.
Examples of effective masses are shown in figure~\ref{f:effmasses}. From top to bottom, the first row of plots show the dependence on the lattice spacing $a$, the second row the dependence on the pion mass $m_\pi$ and the third row the dependence on the distance $r$.

On some  ensembles it has been difficult to determine the crucial starting point of the plateau, 
especially for potentials around $r\approx r_1 $. Therefore in contrast to previous work with these ensembles \cite{Asmussen:2023pia} we performed a fit to the effective masses of the form
\begin{equation}
 E_0(T,T_0,r) = E_0 + B e^{-\Delta \cdot T}\, .
 \label{eq:E_0Fit}
\end{equation}
Where plateau averages were unproblematic, such fits yielded compatible values, but they have the advantage to be able to fit 
data at earlier times, in practice all data starting at $T_0$. A precise value of $E_0$ can be obtained even for cases where the 
traditional plateau starts very late or is non existent.

\begin{figure}[h!]
  \centering 
  \includegraphics[height=.45\linewidth]{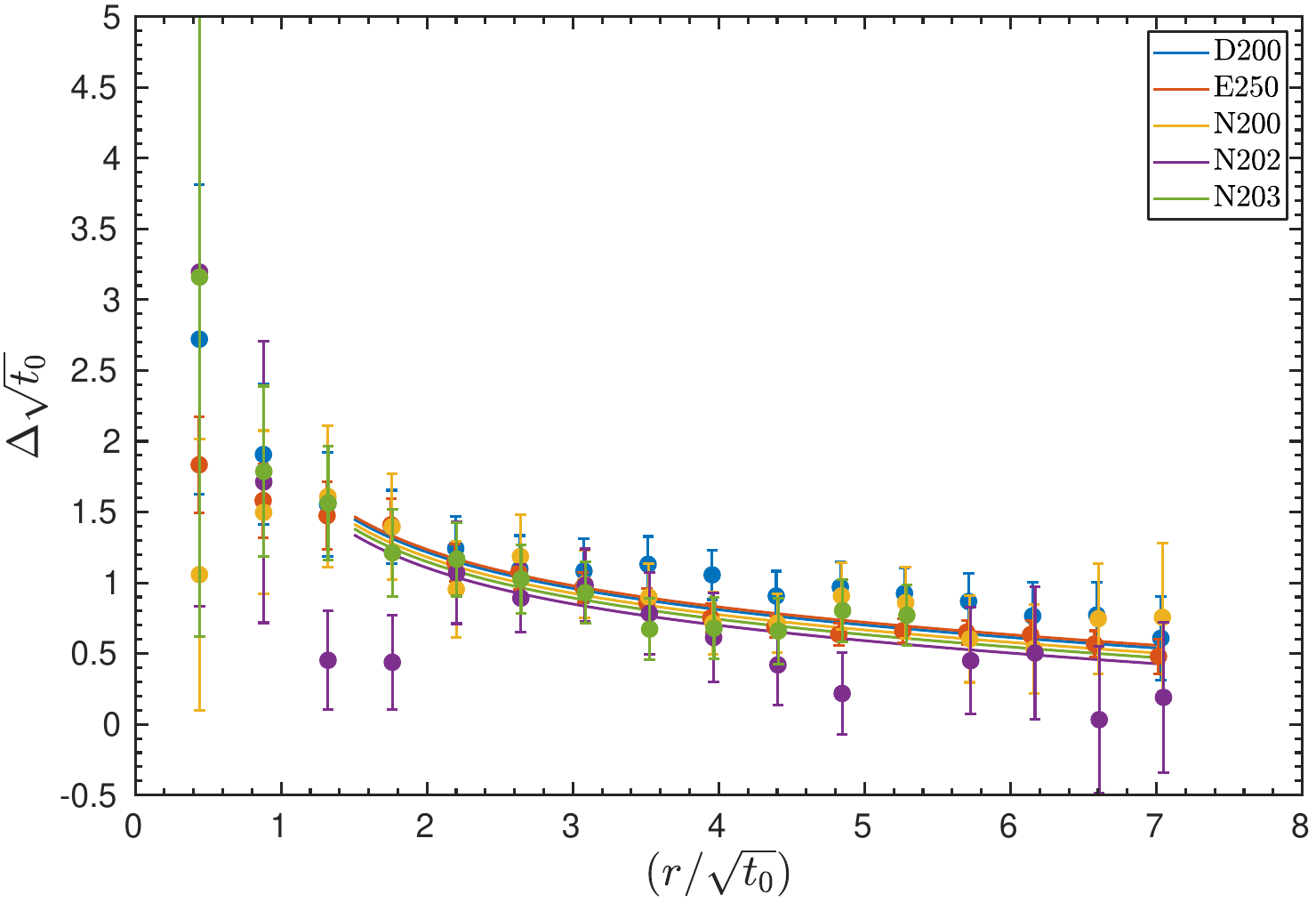}
  \caption{Global fit of the gap $\Delta$ extracted from fits given in eq. \ref{eq:E_0Fit} for $\beta$=3.55. }
  \label{fig:deltafit}
\end{figure}

Unfortunately this nonlinear three parameter fit by itself was not stable enough for all effective masses on all ensembles to confidently describe the data and find the ground state everywhere. To fix this we 
invoked a two step process. First, the ensembles where this fit is unproblematic are used to extract the $\Delta$ parameter 
as a function of $r$, $m_\pi$ and the lattice spacing. One global fit per lattice spacing is performed to establish a model for the $\Delta$ values.
The ansatz for the global fit assumes that the ``effective'' excited state with energy $E_0+\Delta$ has an approximately 
constant energy as a function of $r$, and therefore the difference to $V(r)$ behaves like $V(r)$ itself,
\begin{equation}
    \sqrt{t_0} \Delta = \hat{\Delta}_0 - \hat{\delta}_0 \frac{r}{\sqrt{t_0}} - \hat{\gamma}_0 \frac{\sqrt{t_0}}{r} + \hat{b}_0 t_0 \left( m^2_\pi-\left(m_\pi^{\rm phys} \right)^2\right) \,.
    \label{eq:DeltaFit}
\end{equation}
Once the four parameters $\hat{\Delta}_0$, $\hat{\delta}_0$, $\hat{\gamma}_0$, and $\hat{b}_0$ of this model are determined, it can be used to read off $\Delta$ on all ensembles, including those
where fit \eq{eq:E_0Fit} was unstable. That value is then used
in fit~\eq{eq:E_0Fit}, turning it into a well behaved linear fit with two parameters.
In figure~\ref{fig:deltafit} one can see how well the model~(\eq{eq:DeltaFit}) describes the $\Delta$ values for $\beta$=3.55. 
The first datapoints, which have large lattice artifacts, have not been used in the fit eq.~(\ref{eq:DeltaFit}), but have been included visually in the figure. In the interesting range $r_1 \leq r \leq r_0$ the fit describes the data very well.

\begin{figure}[h!]
  \centering 
  \includegraphics[width=0.48\textwidth,height=.4\linewidth]{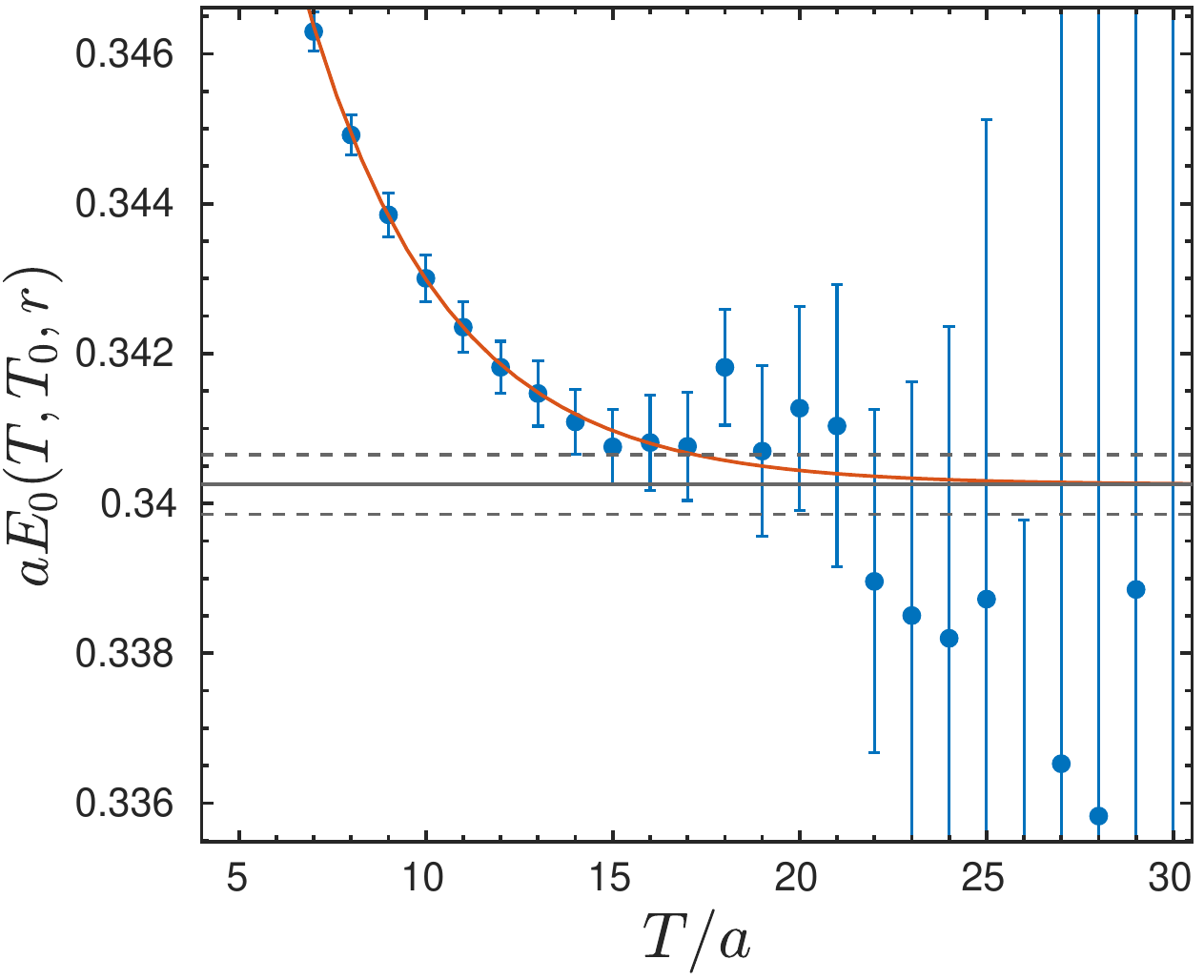}  \hfill
  \includegraphics[width=0.48\textwidth,height=.41\linewidth]{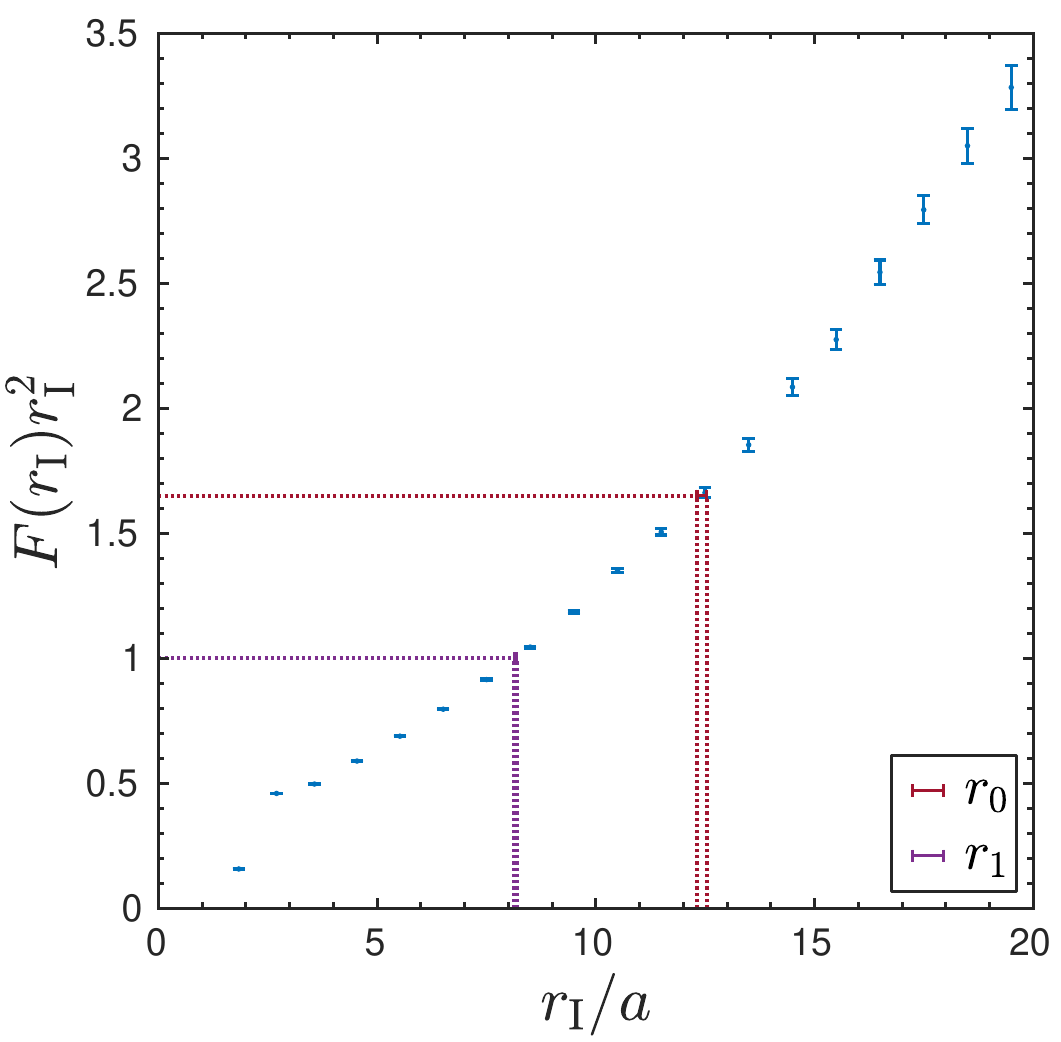}
  \caption{The effective masses for ensemble J500 at $r/a = 13$ with $t_0/a = 5$ on the left together with the fit eq. \ref{eq:E_0Fit} in red and the ground state potential as the grey line. On the right is the static force $F(r)$ of the same ensemble together with the values $r_0$ and $r_1$ of the 3-point interpolation.}
  \label{fig:EmForce}
\end{figure}

 A fit of the effective mass using the gap from the global fit~\eq{eq:DeltaFit} can be seen on the left of figure \ref{fig:EmForce}, where the red line shows the fit and the darkgrey line the resulting ground state with the associated uncertainty as dashed lines.

\begin{figure}[h!]
  \centering 
  \includegraphics[width=0.48\textwidth,height=.4\linewidth]{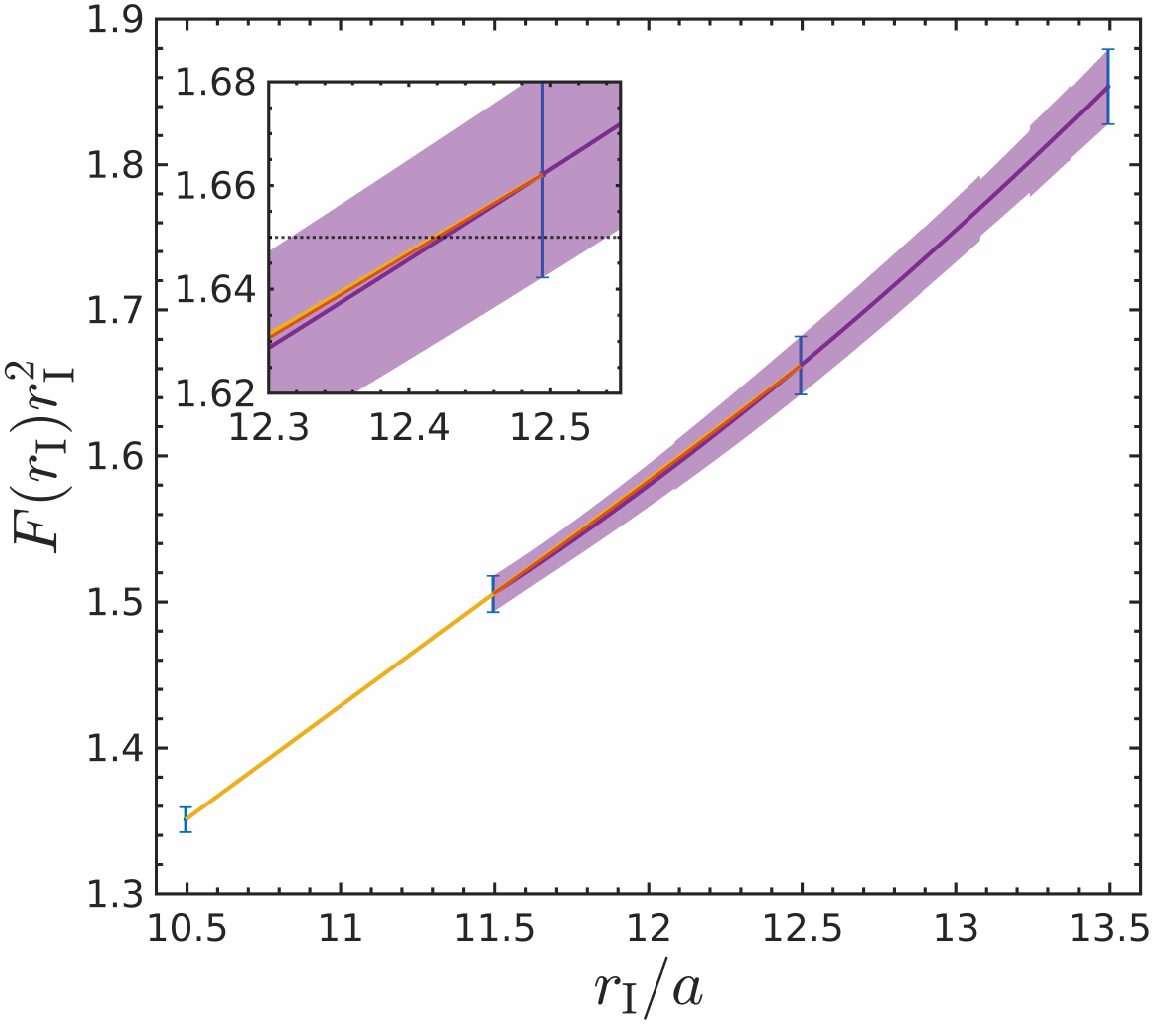} \hfill 
  \includegraphics[width=0.48\textwidth,height=.4\linewidth]{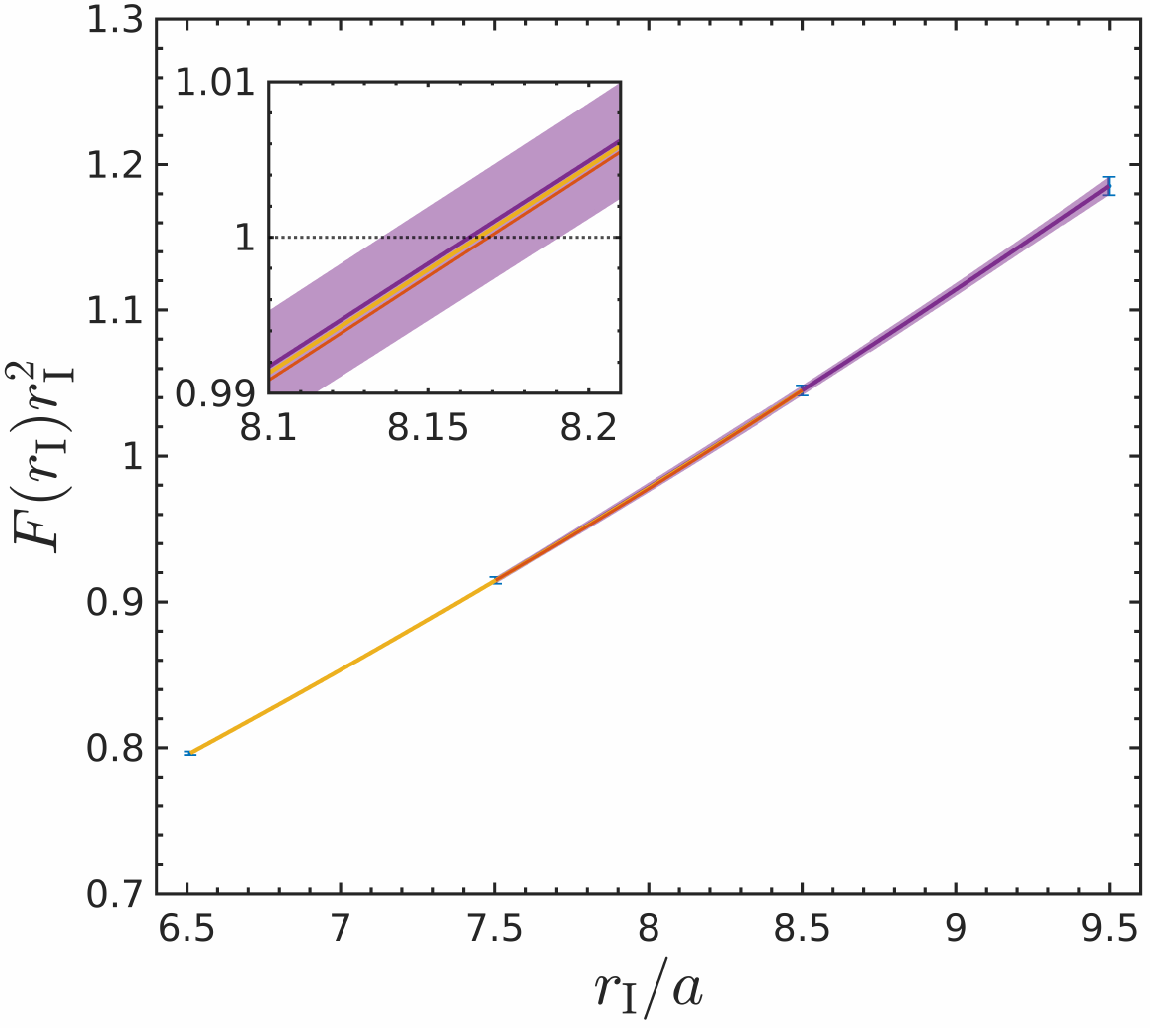}
  \caption{We show the three different interpolations for $r_0$ (left) and $r_1$ (right). The red line corresponds to the 2-point interpolation using the nearest points to the target $r$-value. The yellow line uses the same points and one point at a shorter distance. The purple line is with one point at a larger distance. The purple line is the one used for the final result and has an included errorband. On the top left of the figure is a zoomed in view of the point of interest. }
  \label{fig:Interpol}
\end{figure}

Computing $F(r)$ from the potential $V(r)$ using the improved distance, see \eq{e:latforce} one can find the values of $r_1$ and $r_0$ by~\eq{eq:r1_r0} which is visualised in the right plot figure \ref{fig:EmForce}. Different interpolation ranges can be used to determine the values of $r_1$ and $r_0$. On the one hand a 2-point interpolation is used with the two points enclosing the target value. For a 3-point interpolation a third point is included, either corresponding to a smaller or to a larger distance than the points used in the 2-point interpolation. The interpolation function used for the 2- or 3-point interpolation is
\begin{align}
    F(r_I)r_I^2 &= \hat{f}_1 r_I^2 + \hat{f}_0 \, , \quad\mbox{(2-point)} \qquad \mbox{or} \label{eq:2point}\\
    F(r_I)r_I^2 &= \hat{f}_1 r_I^2 + \hat{f}_0 + \hat{f}_2/r_I^2 \,,\quad\mbox{(3-point)} \, ,\label{eq:3point}
\end{align}
where $\hat{f}_0$, $\hat{f}_1$, and $\hat{f}_2$ are the coefficients of the interpolations. In \fig{fig:Interpol} the three interpolations are shown by the lines with different colors. The purple line, which has an errorband, is the interpolation used for the final result. It uses three points with two points at smaller distances and one point at larger distance than the targeted value of $r$. One can see that the difference between the different interpolations is minuscule and the results are compatible with each other. For coarser lattice spacings lattice artifact might be enhanced when very small Wilson loops are used in the interpolation. 
The differences of the interpolations in $r_1$ is larger, as the forces have smaller errors, and the $1/r^2$ terms are larger in this region.

\FloatBarrier

\subsection{Evalution of exceptionally problematic ensembles \label{s:E300}}

\begin{figure}[h!]
  \centering 
  \includegraphics[width=0.48\textwidth,height=.4\linewidth]{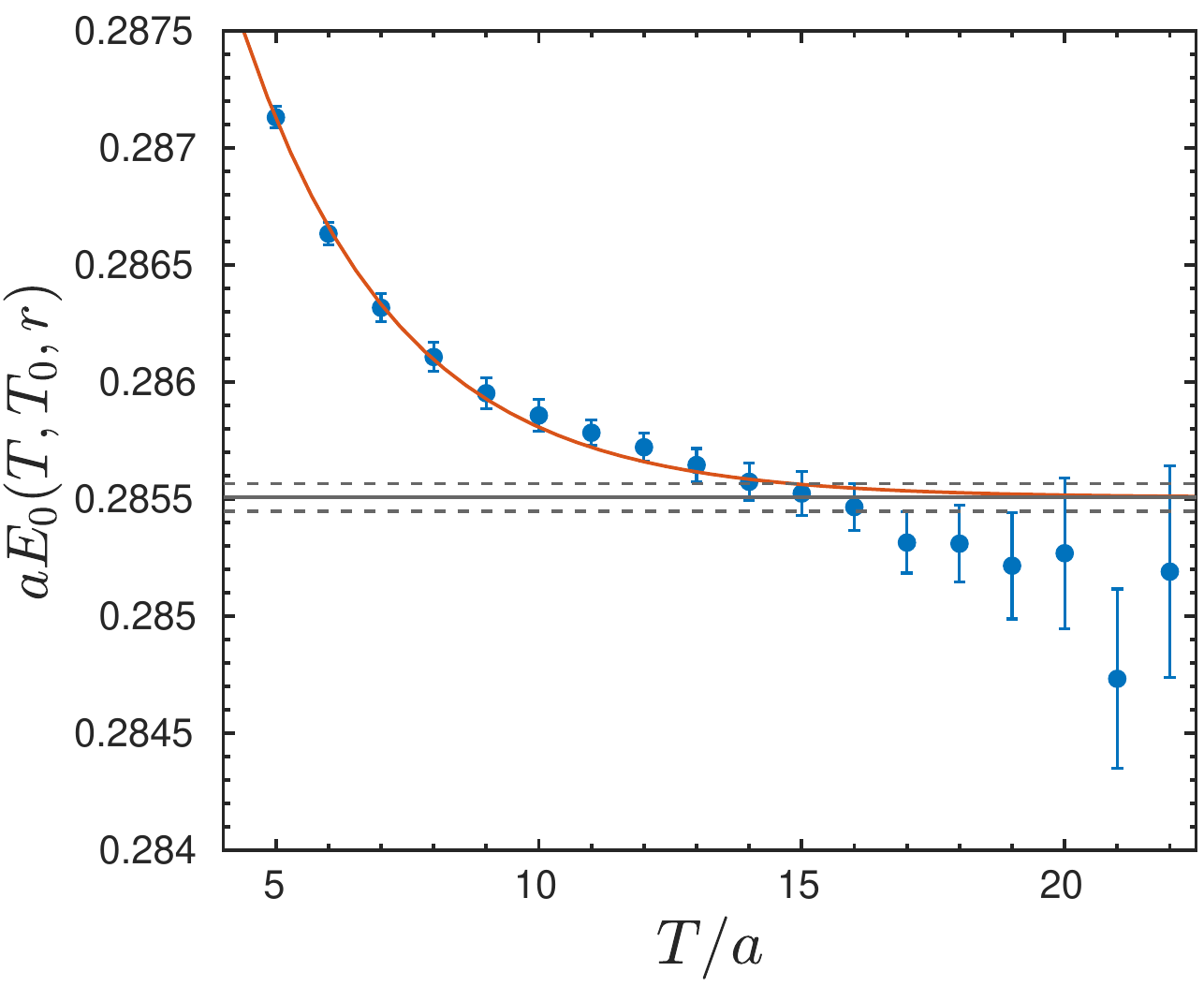}  \hfill 
  \includegraphics[width=0.48\textwidth,height=.4\linewidth]{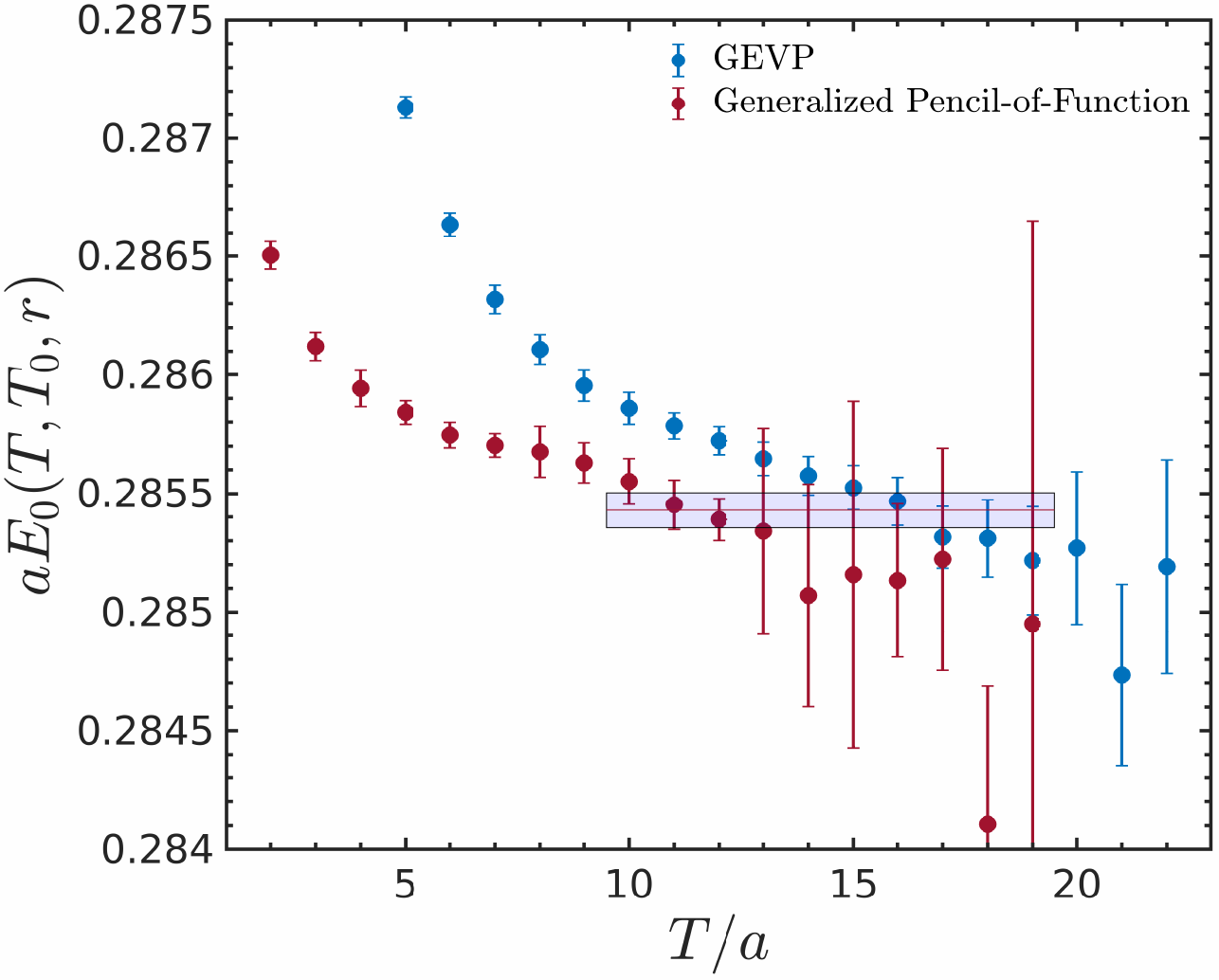}
  \caption{On the left ensemble E300 with a loopsize of $r/a=6$, which is in the range of $r_1$, showing the plateau calculation using the fit method described in section \ref{Ch:eval-v0}. The right figure shows the difference between using the GEVP method in blue and the effective mass calculated through the pencil of function method in red for E300.}
  \label{fig:E300_pencil}
\end{figure}

For a very small number of ensembles and distances,
the determination of the potential from effective energies has been particularly difficult.
In particular, we observed problems where $r$ was close to $r_1$ on ensembles with finer lattice spacings. 
A plot of one of the most problematic cases, the effective mass of ensemble E300, is shown in figure \ref{fig:E300_pencil}.
E300 has the second finest lattice spacing with a pion mass close to the physical one.
A dip of the effective mass can be observed here at large $T/a$. Several methods have been used to analyze and possibly find a better way to determine the potential for this case. We mostly focussed on E300, as the pion mass was close to the physical mass. At first we doubled the amount of different smearing levels. The hope was to be able to better suppress excited states, with a larger basis for the GEVP. Unfortunately this did not bring the desired improvement of the effective energy.

Then we looked at different more complicated fit functions for the effective masses, for example by adding an additional excited state contribution to \eq{eq:E_0Fit} that is either fixed to be two pion masses above $E_0$ or left free. Adding this second exponential term proved to make the fit less stable for a lot of ensembles, so that the fit form in \eq{eq:DeltaFit}, did not work any more. 

The method used in the end for this difficult case is the pencil of functions method 
\cite{Fischer_2020,Aubin_2011,Aubin:2012NY,PhysRevD.92.034512,Ottnad_2018}, which seems to produce a more reliable plateau albeit with a larger error. 
The method expands the original $4\times4$ correlation matrix to an $8\times 8$ in which every element $\overline W^{(k,l)}(r,T)$
of the original matrix is replaced by 
$  \begin{pmatrix}   
  \overline W^{(k,l)}(r,T)   & \overline W^{(k,l)}(r,T+\tau) \\
  \overline W^{(k,l)}(r,T+\tau)  & \overline W^{(k,l)}(r,T+2\tau)
  \end{pmatrix}\, .$
We use $\tau=2a$ here. This larger matrix is ``pruned''~\cite{Niedermayer:2000yx} down to $6 \times 6$, before the GEVP problem is 
solved as before.
A difference between the ``normal'' effective energies and the ones calculated with this new method can be seen on the right of figure \ref{fig:E300_pencil} for the E300 ensemble.

Other cases in which the plateau was not well defined were ensembles B450 and J501. In the latter only in the range of $r_1$. The problems seem to be unrelated to the quark mass or resolution of the lattices, as it did not appear for E250 at the physical pion mass nor for other ensembles at the fine lattice spacing.

\FloatBarrier

\section{Continuum and Chiral Extrapolations}\label{sec:ex}
\begin{figure}[h]
  \centering 
  \includegraphics[width=0.47\textwidth,height=.46\linewidth]{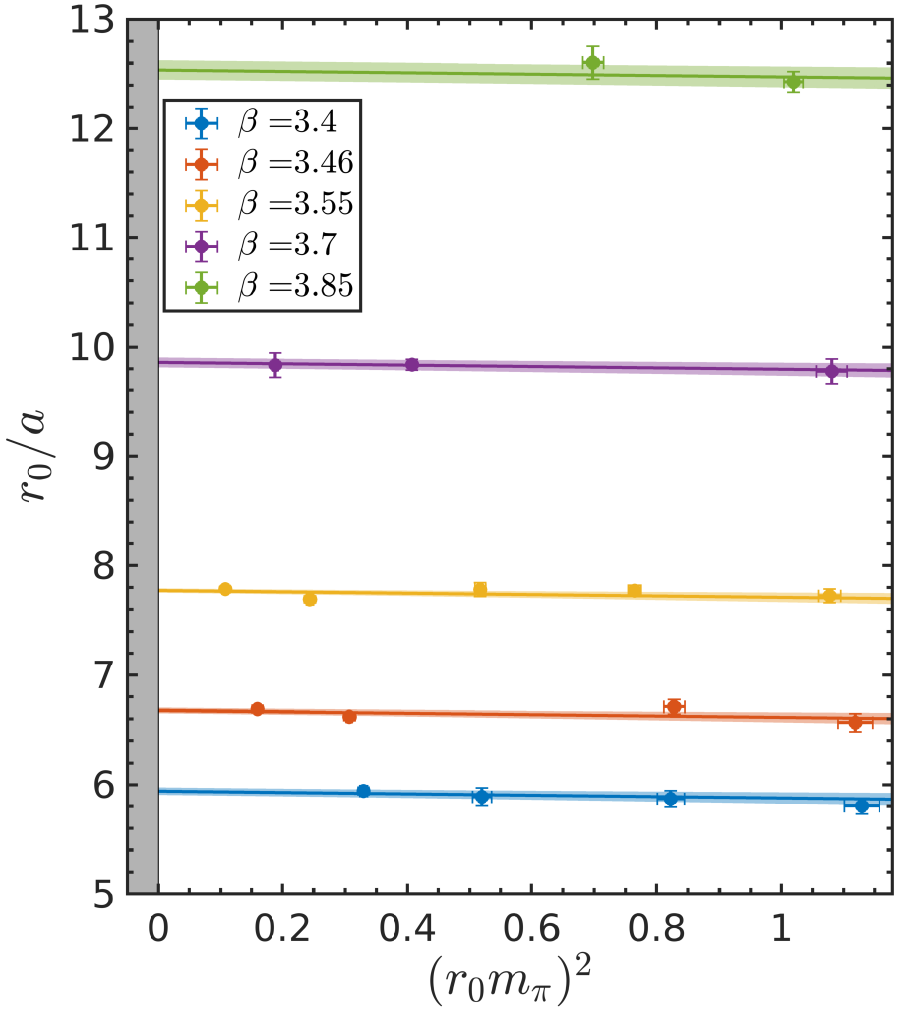}  \hfill 
  \includegraphics[width=0.47\textwidth,height=.46\linewidth]{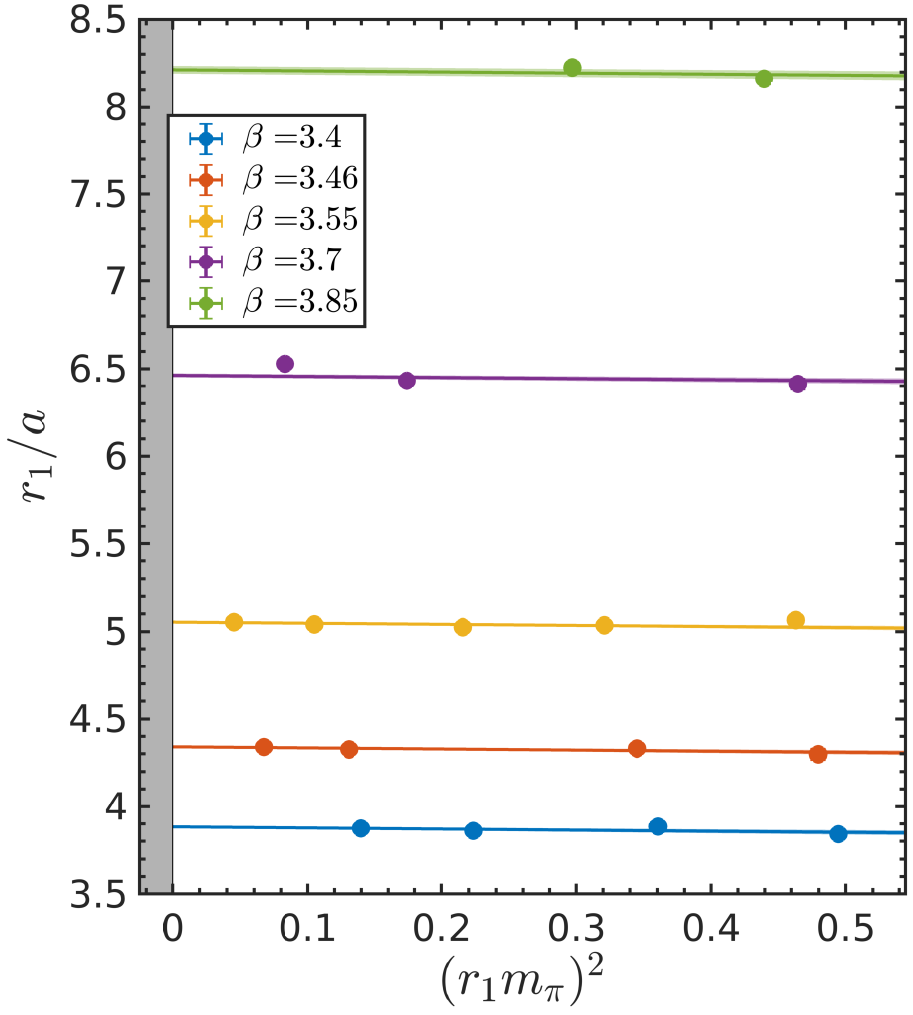}
  \caption{$r_0/a$ and $r_1/a$ for all ensembles as a function of $(r_0m_\pi)^2$ (left) and $(r_1m_\pi)^2$ (right). A global fit~\eq{eq:gfitr0} is performed on the data.}
  \label{fig:r0r1data}
\end{figure}

Using the methods explained in the previous chapter a value for $r_0/a$ and $r_1/a$ is determined and collected in \tab{tab:r0_res} of appendix~\ref{s:appr0} for each of the 18 ensembles . A plot with the values together with a global fit of the form
\begin{equation}\label{eq:gfitr0}
    \frac{r_0}{a} =c_1\rvert_{\beta}  + c_2 (r_0 m_\pi )^2
\end{equation}
is shown in figure \ref{fig:r0r1data}. 
For both $r_0$ and $r_1$ the pion mass dependence is very mild, leading to nearly horizontal trends of the fit functions.
Note that the data point for E300 (purple point farthest to the left) is more than one sigma away from the 
fit which we also
observe for $r_0/r_1$ and $r_{1}/\sqrt{t_0}$. The results of the scales extrapolated to the physical value of $m_\pi$ and to chiral limit ($m_\pi=0$) can be found in \tab{tab:r0_res_lim}. These results will be used in chapter \ref{Cha:Lambda} to calculate $\Lambda \times r_0$.

\begin{table}[!h]
   \centering
   \begin{tabular}{c c c c c c c c c}
   \toprule
       $\beta$&  $r_0^{\chi}/a$ & $r_0/a$ & $r_1^{\chi}/a$ & $r_1/a$\\
   \midrule
       3.4  & 5.941(35)  & 5.934(35)  & 3.883(8)  & 3.880(8)  \\
       3.46 & 6.676(26)  & 6.670(26)  & 4.341(6)  & 4.338(6)  \\
       3.55 & 7.773(21)  & 7.766(21)  & 5.055(5)  & 5.052(5)  \\
       3.7  & 9.860(46)  & 9.854(46)  & 6.462(12)  & 6.460(12)  \\
       3.85 & 12.540(88)  & 12.533(88)  & 8.214(23)  & 8.211(23)  \\
   \bottomrule
   \end{tabular}
   \caption{Results of $\frac{r_0}{a}$,  $\frac{r_1}{a}$ in the chiral limit ($m_\pi =0$, superscript ``$\chi$'') and at the physical point (no superscript).}
   \label{tab:r0_res_lim}
\end{table}

\begin{figure}[h]
  \centering
  \includegraphics[width=0.48\textwidth,height=.4\linewidth]{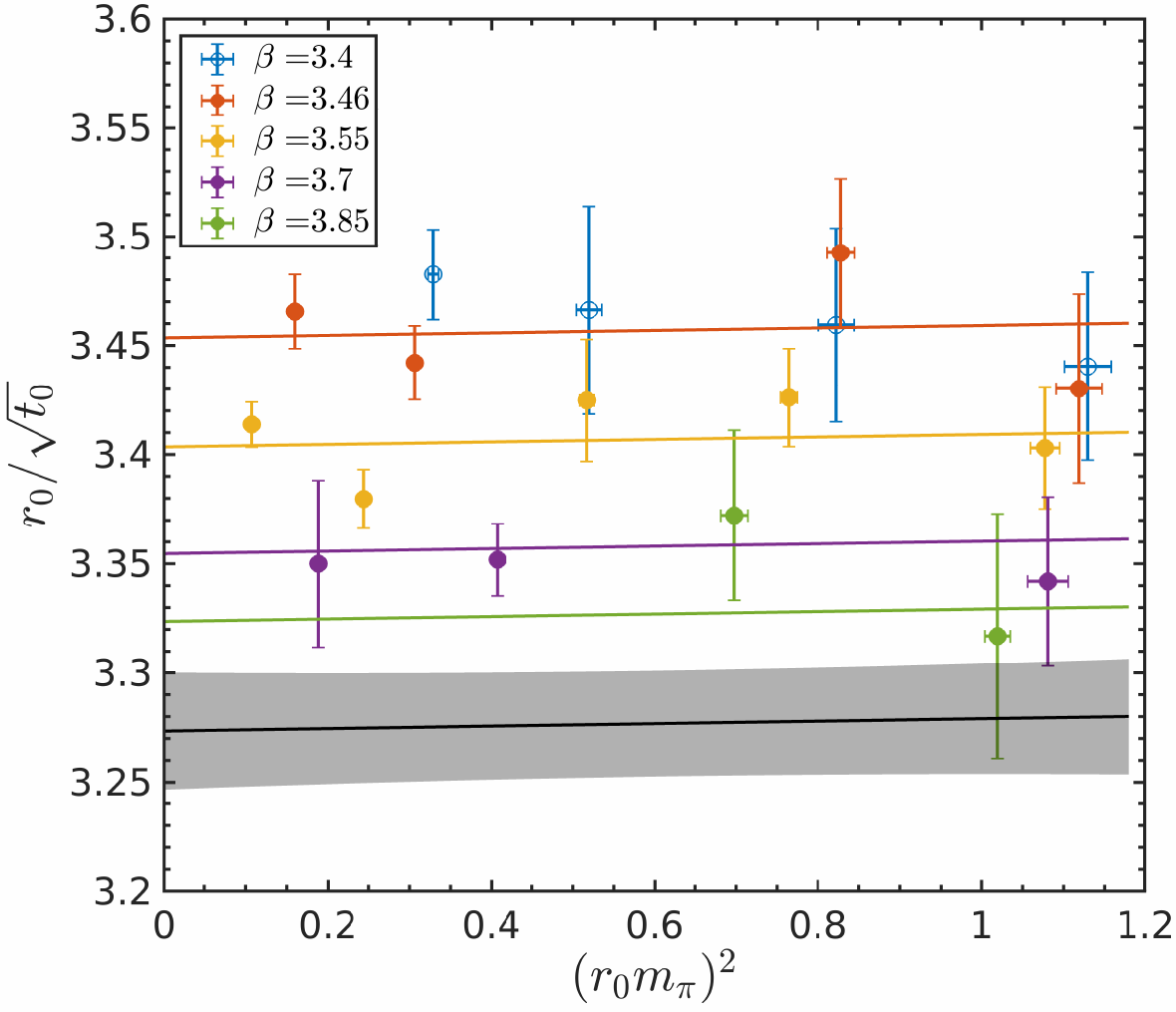}\hfill 
  \includegraphics[width=0.48\textwidth,height=.4\linewidth]{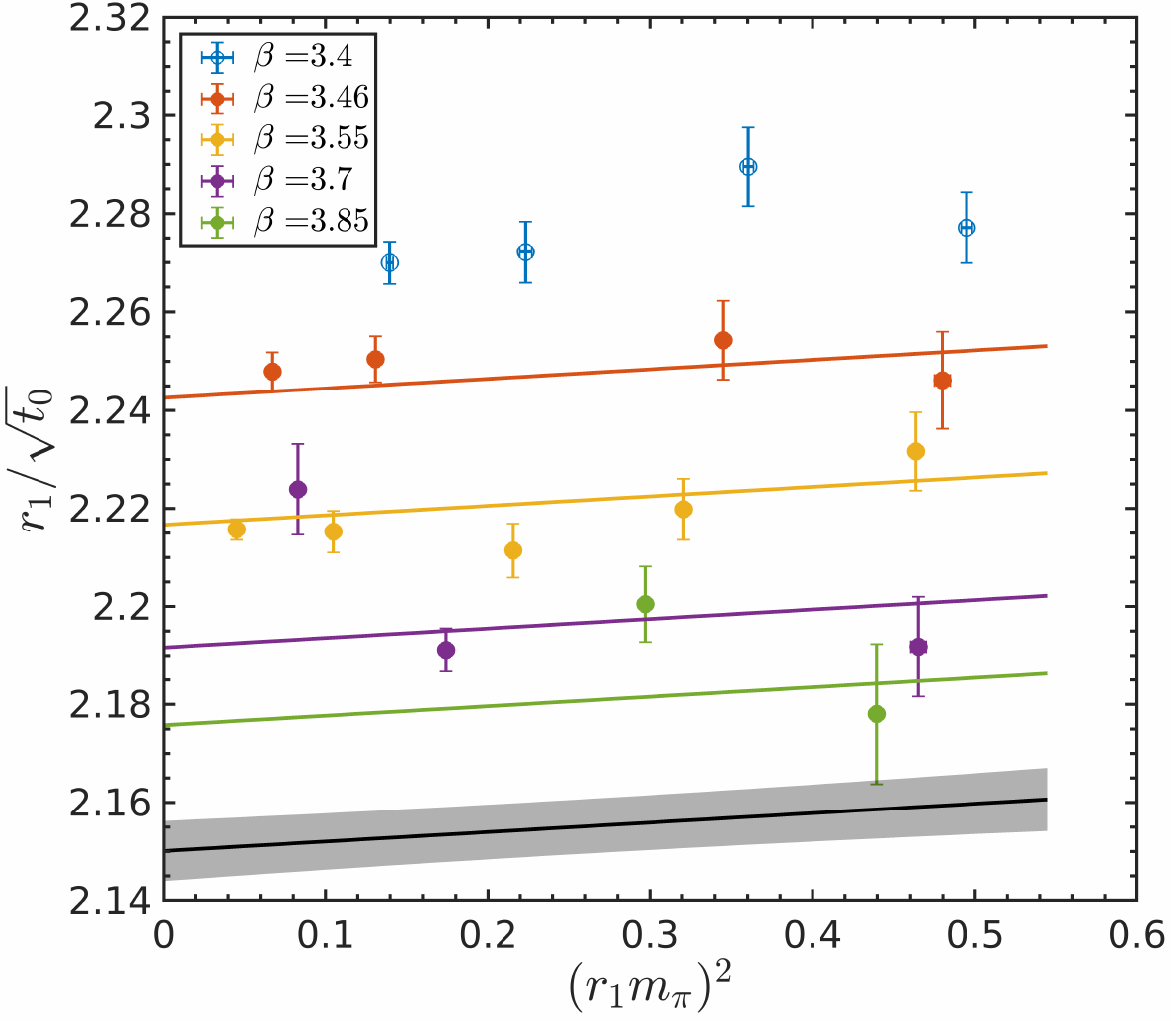}
  \caption{On the left the values $r_0/\sqrt{t_0}$ and the right $r_0/\sqrt{t_0}$ are ploted together with a global fit corresponding to Fit 1 in eq. \ref{eq:Fit1}. The grey band on the bottom corresponds to the continuum extrapolation of said fit.}
  \label{fig:r0r1globaldata}
\end{figure}

For the next step we combine our values of $r_0$ and $r_1$ with $\sqrt{t_0}$ into a dimensionless ratio. Statistical correlations between the quantities entering the ratios are propagated into our analysis. Such ratios are independent of 
the overall scale and data from different lattice spacings fall onto a universal curve (a function of $m_\pi$), up to 
lattice artifacts. A combined continuum and chiral (inter\footnote{One ensemble is generated at the physical pion mass}-)extrapolation
requires an ansatz that models the lattice artifacts and the mass dependence. The global fit used in figure \ref{fig:r0r1globaldata}
is of the form
\begin{align}
    {\rm Fit\ 1}:\quad&\frac{r_0}{\sqrt{t_0}}=c_1^{(1)} + c_2^{(1)} \left( \frac{a}{r_{0}^{\rm sym}}\right) ^2 + c_3^{(1)} (r_0 m_\pi )^2\,, \label{eq:Fit1}
\end{align}
where $r_{0}^{\rm sym}$ is the value for the ensembles calculated at the symmetric point with degenerate quark masses. 
This particular fit assumes that lattice artifacts are purely $O(a^2)$ and do not depend on the mass. The physical 
mass dependence is assumed to be linear in $m_\pi^2$.
The symmetric ensembles correspond to the point farthest to the right for each lattice spacing. Ideally these points would be on top of each other, but scatter slightly because of the mistunings, as discussed in section \ref{sec:sim}. The black/grey band on the bottom of the figure shows the mass dependence in the continuum limit by setting $c_2$ in \eq{eq:Fit1} to zero. Using the physical mass of the pion $m_\pi=134.8\,$MeV (after correcting for isospin breaking effects \cite{Bruno:2016plf}) and the physical value of $\sqrt{t_0}=0.1443(7)(13)$ fm \cite{Strassberger:2021tsu} the physical value for $r_0$ can be extrapolated.

\begin{figure}[h!]
  \centering \includegraphics[width=0.48\textwidth,height=.4\linewidth]{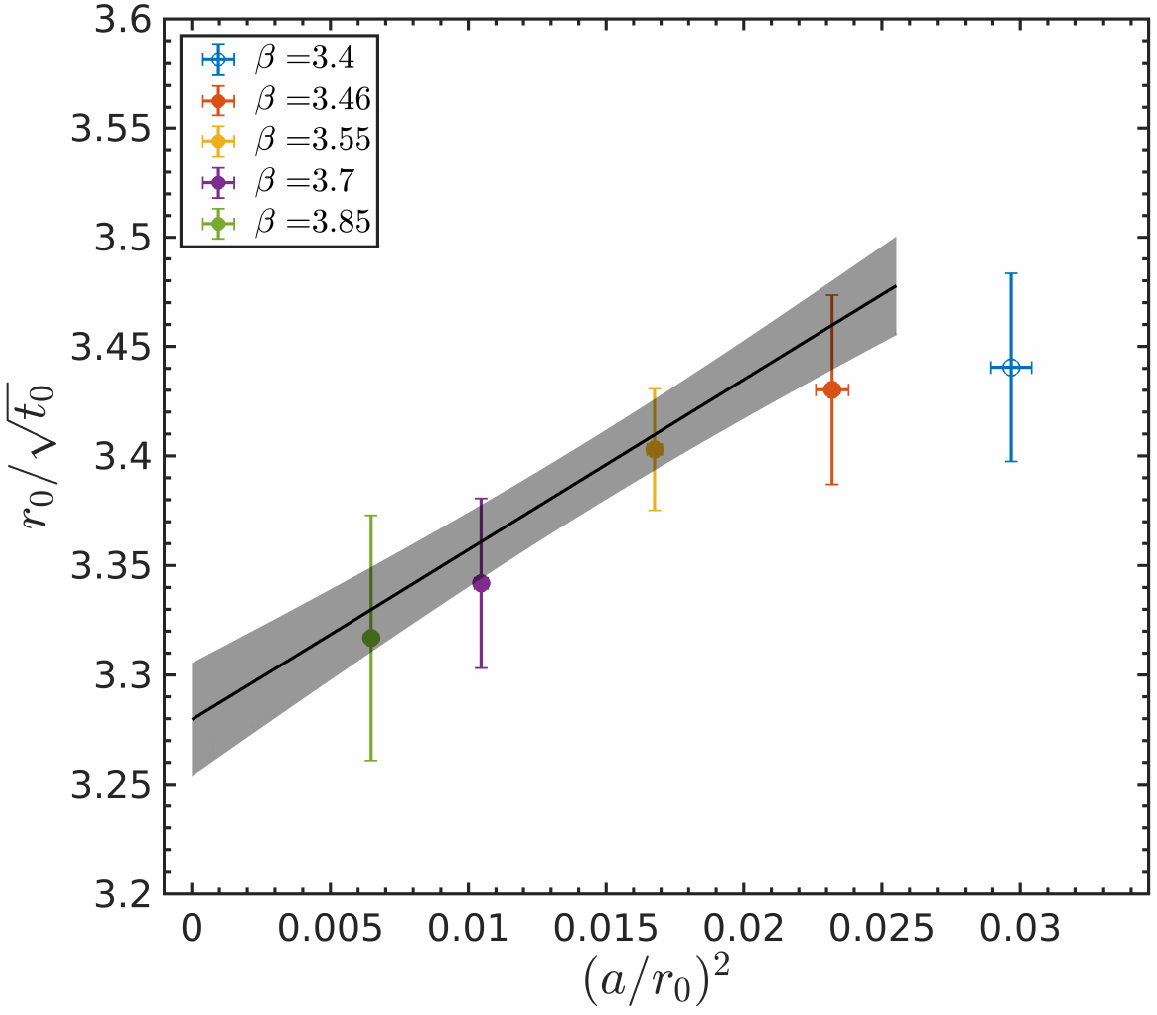}  \hfill 
  \includegraphics[width=0.48\textwidth,height=.4\linewidth]{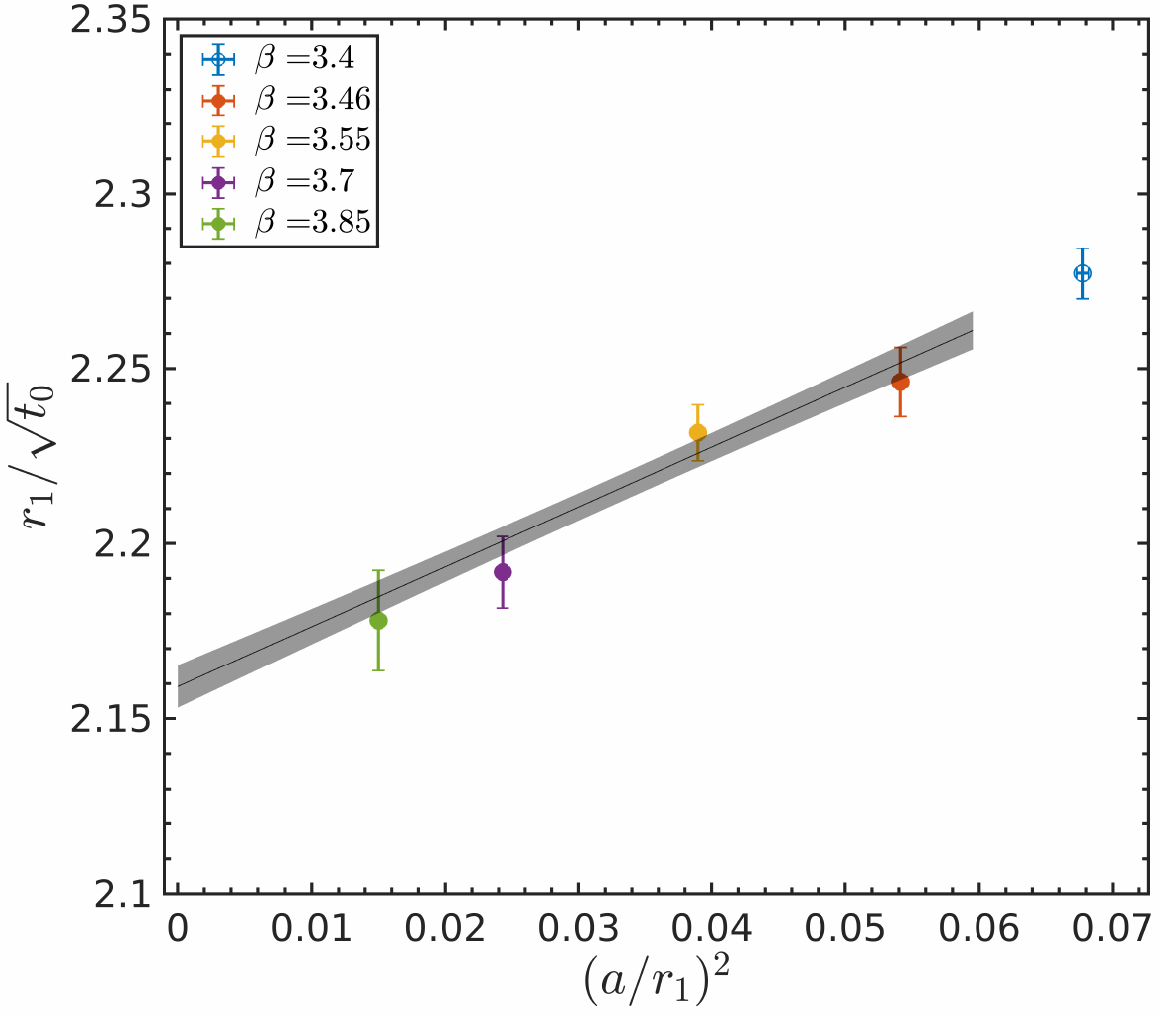} 
  \caption{The same fits as in figure~\ref{fig:r0r1globaldata} displayed as a function of the lattice spacing at the symmetric point.}
  \label{fig:sym}
\end{figure}

In addition to looking at all data, in \fig{fig:r0r1globaldata}, we plot the data at the symmetric mass point as a function
of $a^2$ to visualize the continuum extrapolation entailed in the fit. The solid lines in figure \ref{fig:sym} are the fit functions
evaluated setting $(r_0 m_\pi)^2 = 1.1$ and $(r_1m_\pi)^2=0.47$ in \eq{eq:Fit1} respectively.  The fit curve follows the displayed data nicely, even if it comes from a global fit including all data.

Complementing the fit defined in \eq{eq:Fit1}, which is labeled as $F_1$, three other fits are performed on the $r_0/\sqrt{t_0}$ data with changed or added terms,

\begin{align}
    {\rm Fit\ 2}:\quad& \frac{r_0}{\sqrt{t_0}} = c^{(2)}_1+c^{(2)}_2 \left( \frac{a}{r_{0}^{\rm sym}} \right)^2 + c^{(2)}_3(r_0 m_\pi)^2 + c^{(2)}_4(m_\pi a)^2\,,
    \label{eq:Fit2}\\
    {\rm Fit\ 3}:\quad& \frac{r_0}{\sqrt{t_0}} = c^{(3)}_1+c^{(3)}_2 \left( \frac{a}{r_{0}^{\rm sym}} \right)^2 + c^{(3)}_3 \phi_2 + c^{(3)}_4(1.098-\phi_4)\,,
    \label{eq:Fit3} \\
    {\rm Fit\ 4}:\quad& \frac{r_0}{\sqrt{t_0}} = c^{(4)}_1+c^{(4)}_2 \left( \frac{a}{t_{0}^{\rm sym}} \right)^2 + c^{(4)}_3\phi_2+ c^{(4)}_4(1.098-\phi_4)\,.
    \label{eq:Fit4}
\end{align}

The second fit labeled $F_2$ defined in \eq{eq:Fit2} is similar to Fit 1, but with an added 
term that parameterizes mass dependent lattice artifacts.
In Fits 3 \eq{eq:Fit3} and 4 \eq{eq:Fit4} we allow for a mistuning of data with respect to the desired value $\phi_4=1.098$. In Fit 3 the lattice artifact 
is $\propto a^2/r_0^{\rm sym}$, while Fit 4 uses $a^2/t_0^{\rm sym}$.

\begin{figure}[h]
  \centering 
  \includegraphics[width=0.8\textwidth,height=.5\linewidth]{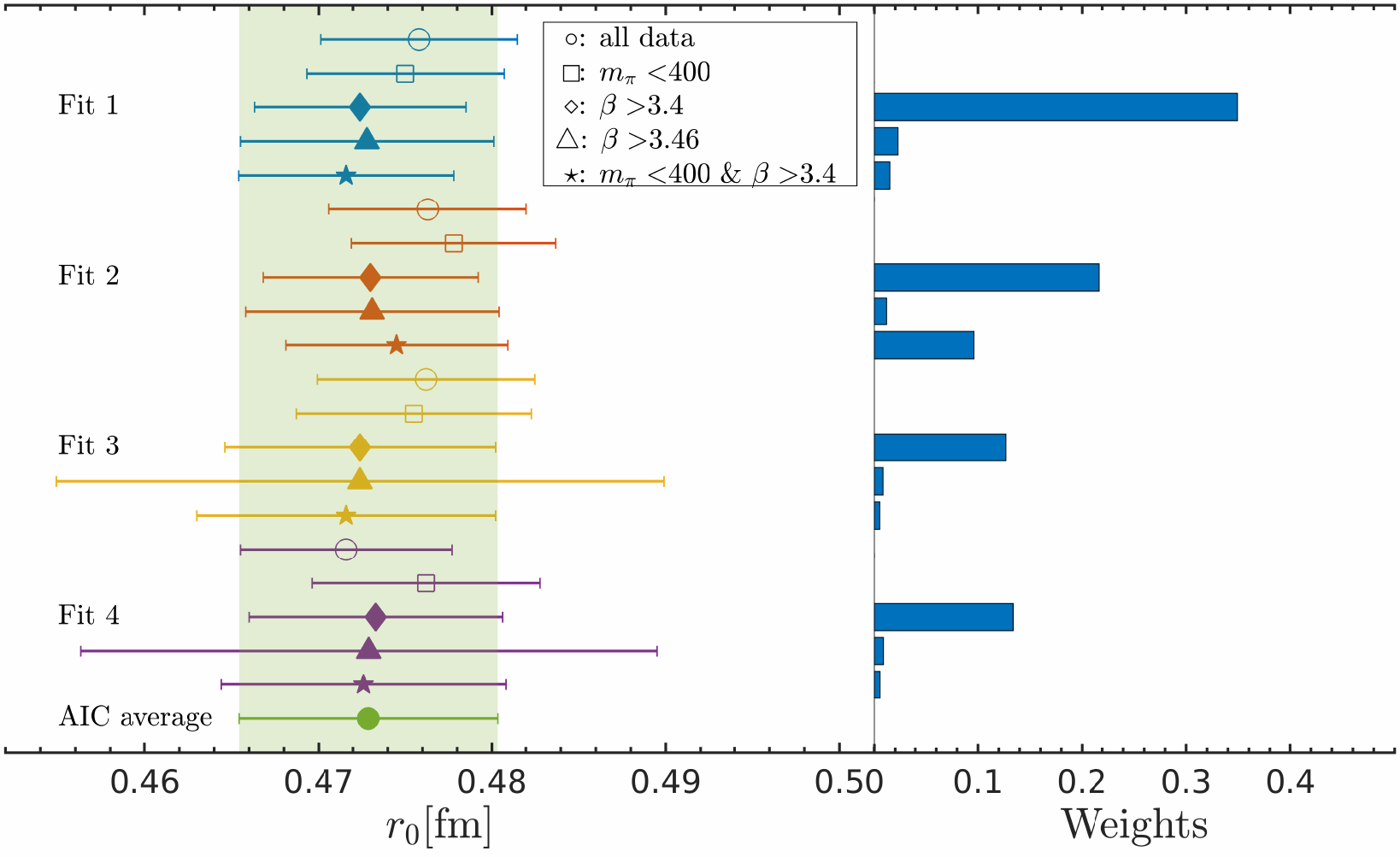} \\
  \includegraphics[width=0.8\textwidth,height=.5\linewidth]{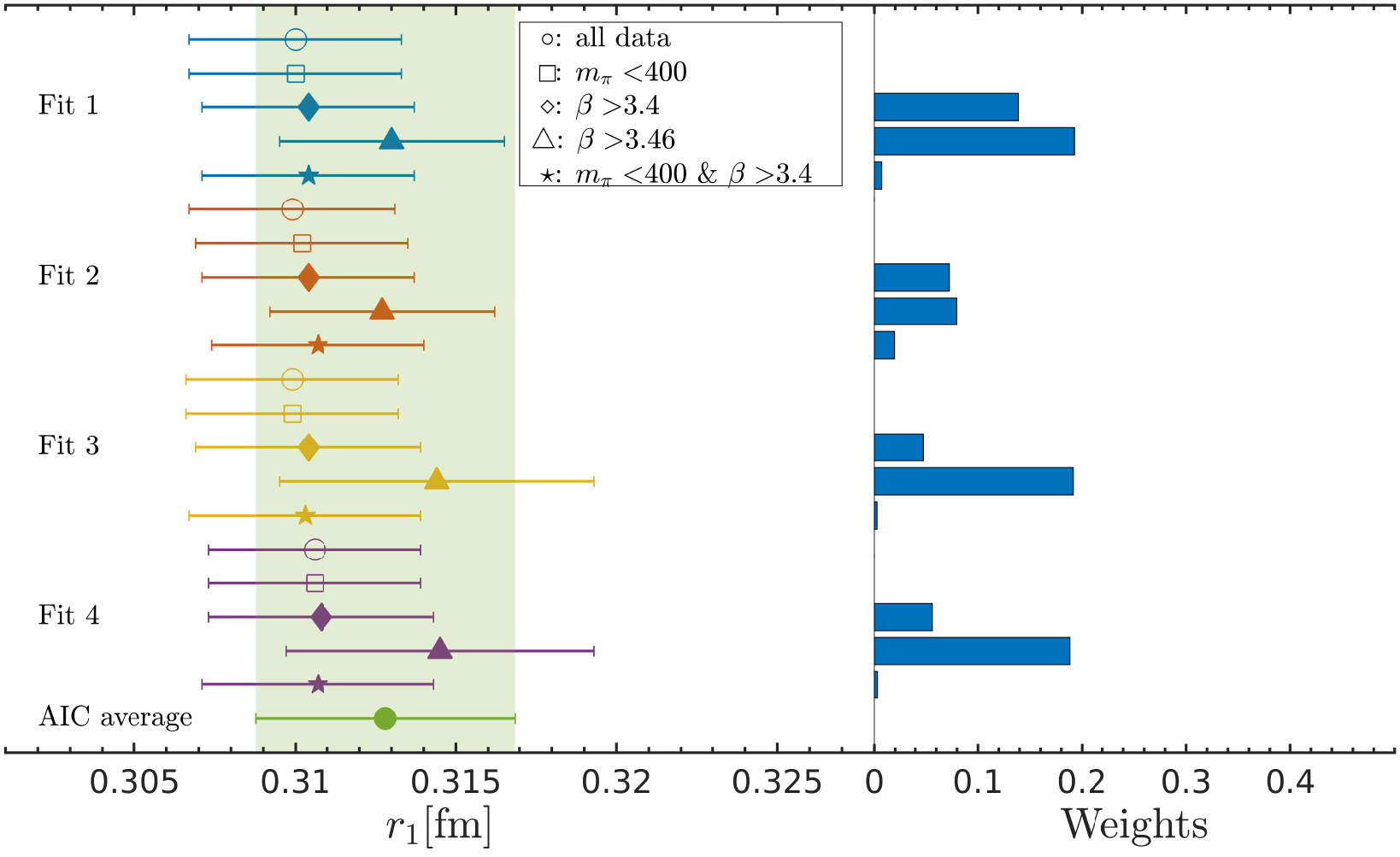}
	\caption{The top panel shows the values of $r_0$ at the physical point obtained from the different fits listed in table \ref{tab:FitResults} of appendix~\ref{s:appfit} together with different cuts, as well as the weights~(\ref{eq:weights}). At the bottom the same is shown for $r_1$. The green point corresponds to the AIC average excluding the coarsest lattice spacing shown as empty markers.}
  \label{fig:Fit}
\end{figure}

 To further investigate possible systematic errors, several cuts on the data are performed. 
 Either one or two of the coarsest lattices are neglected. In addition data too far away from the physical mass point can 
 be excluded. The mass cut, when in place,  eliminates all data at the symmetrical mass point.
 Results of $r_0$ depending on the different fits and cuts together with their associated $\chi^2 / d.o.f.$ can be found in table \ref{tab:FitResults} in appendix~\ref{s:appfit} and visualized in a plot in figure \ref{fig:Fit}. All in all the extrapolated values are in agreement with each other. The main effect of the cuts is a slight increase in error. The biggest disparity is observed for $r_1$ when discarding the coarsest 2 lattice spacings. An explanation might be that with the cuts in place the slightly out-lying E300 value gets a relatively higher statistical weight. In the initial phase of the analysis we considered computing the derivative of all observables with 
 respect to $\phi_4$ and applying a mistuning correction on all ensembles, but since Fit 3 and Fit 4 yield very similar results to the fits without a parameterization of the mistunings, we did not deem the correction to be necessary. Especially since these derivatives 
 are quite noisy for purely gluonic observables.
 The final values are given by an average according to the Akaike Infomation Criterion \cite{1100705,Jay:2020jkz,Neil:2022joj,CamposPlasencia:2023cir}, short AIC.
\begin{align}
  \langle \mathcal{O} \rangle = \sum^{N_{\rm fits}} _{n=1} w^{\textrm{AIC}}_n \langle O \rangle _n,
 \end{align}
where the AIC weight for a given fit and cut labelled by $n$ is given by
\begin{align}
	w^{\textrm{AIC}}_n = N \exp \left[ -\frac{1}{2} ( \chi^2 _n +2N^{\rm parameters}_n +2N^{\rm cut}_n \right].\label{eq:weights}
\end{align}

$N$ is a normalization factor, $N^{\rm parameters}_n$ the number of fit parameters for a given fit, and $N^{\rm cut}_n$ the number of excluded datapoints. The points with the coarsest lattice spacing are excluded in the overall average, which is visualized by the empty markers in the figure.

\begin{figure}[h]
  \centering 
  \includegraphics[width=0.44\textwidth,height=.45\linewidth]{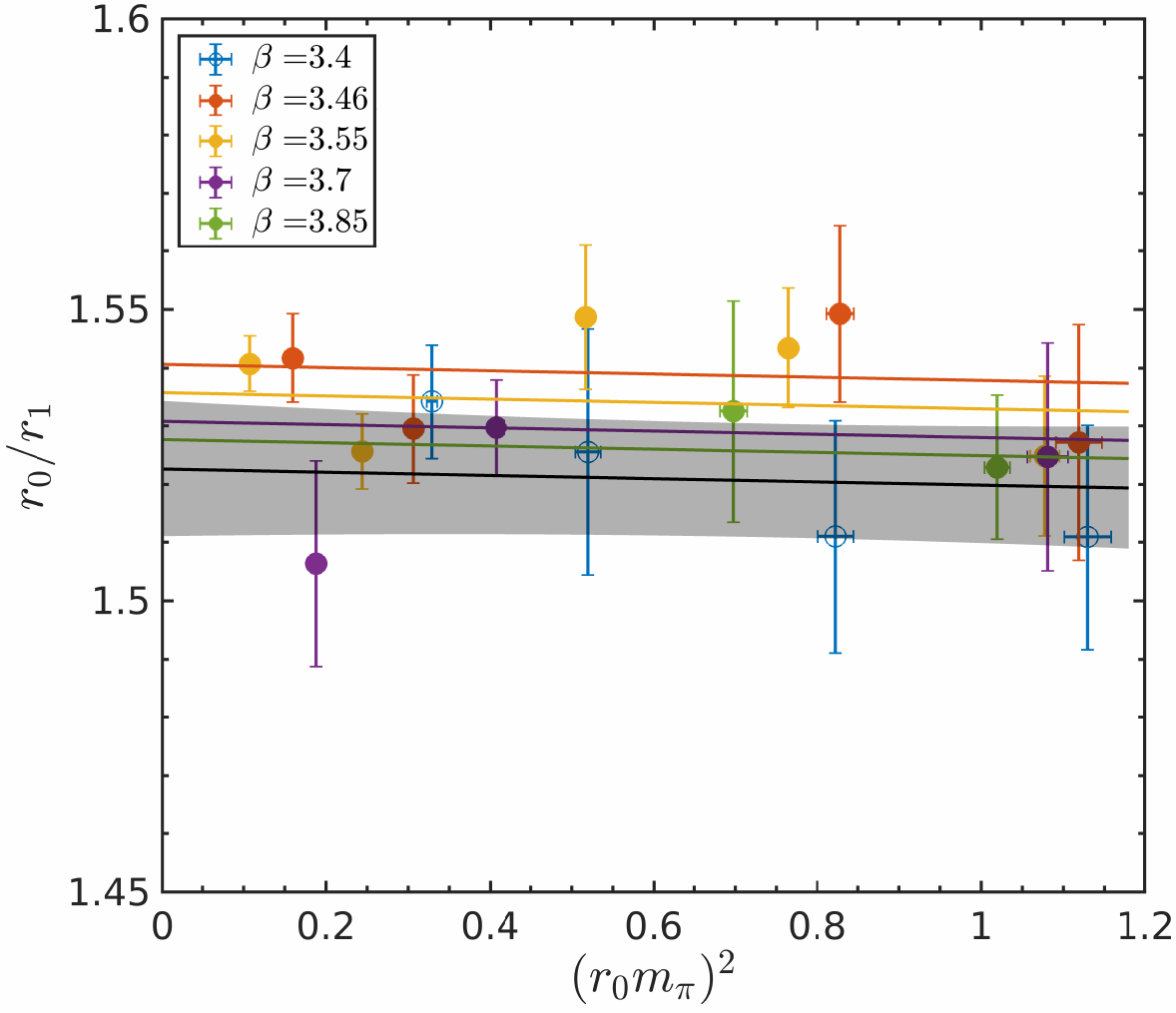} \hfill 
  \includegraphics[width=0.55\textwidth,height=.45\linewidth]{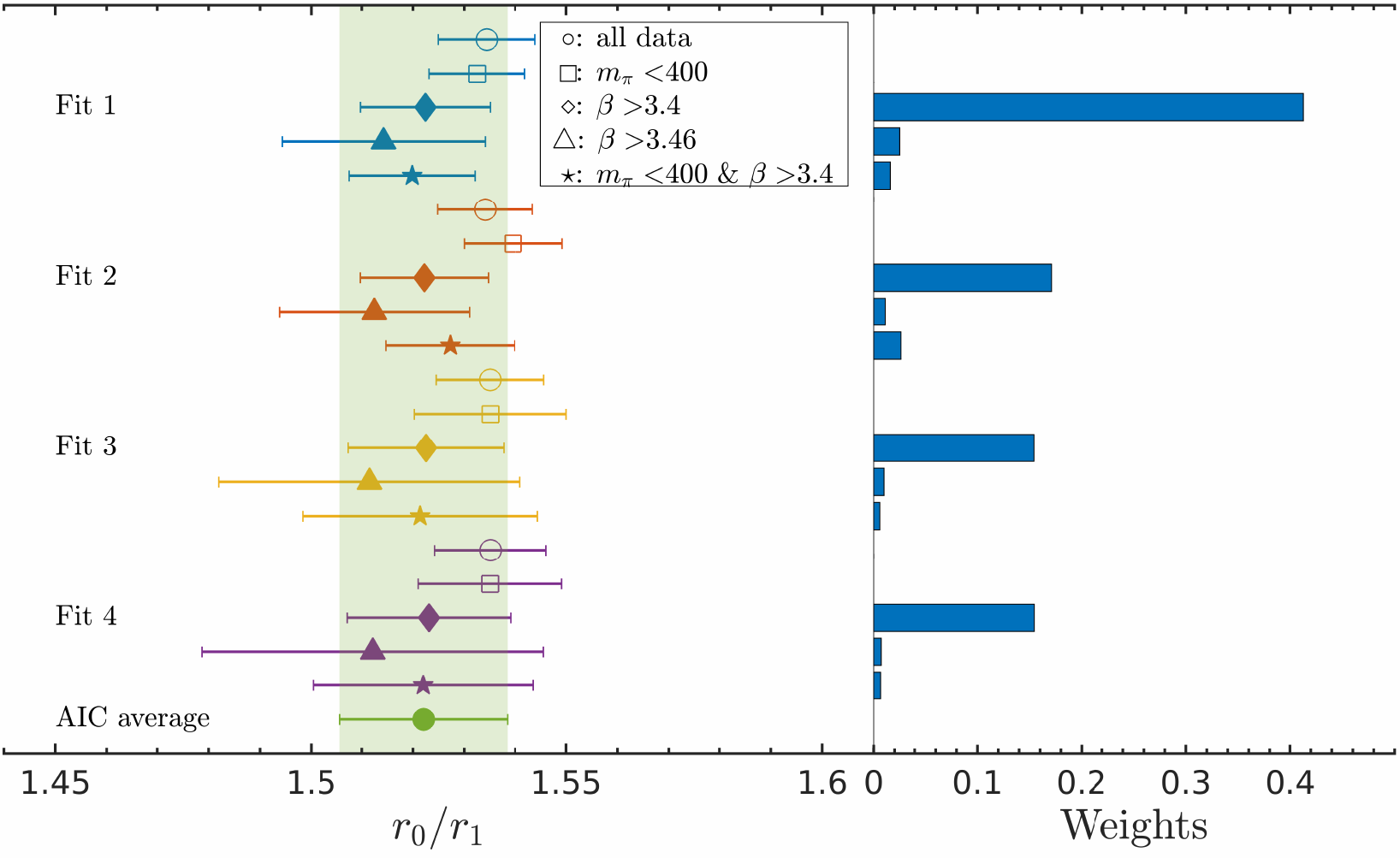}
  \caption{On the left: the values of $r_0/r_1$  of the different ensembles together with a fit like \eq{eq:Fit1}. The grey
band is the fit result in the continuum limit. On the right: the results of the different fits and cuts performed on the data together with the AIC average excluding the coarsest lattice spacing.}
  \label{fig:FitRatio}
\end{figure}

The same analysis is repeated for the values of $r_0/r_1$, seen on the left of figure \ref{fig:FitRatio}, where the fit corresponding to the grey band in the plot is done using a linear fit similar to \eq{eq:Fit1}. On the right of the same figure the different fits and cuts are seen together with the AIC average. Compared to $r_0/\sqrt{t_0}$ and $r_1/\sqrt{t_0}$, the effect of the cuts is larger for 
$r_0/r_1$.

\FloatBarrier

\subsection{Corrections to Leading Scaling Violations}
\begin{figure}
   \includegraphics[width=0.33\linewidth,height=.4\linewidth]{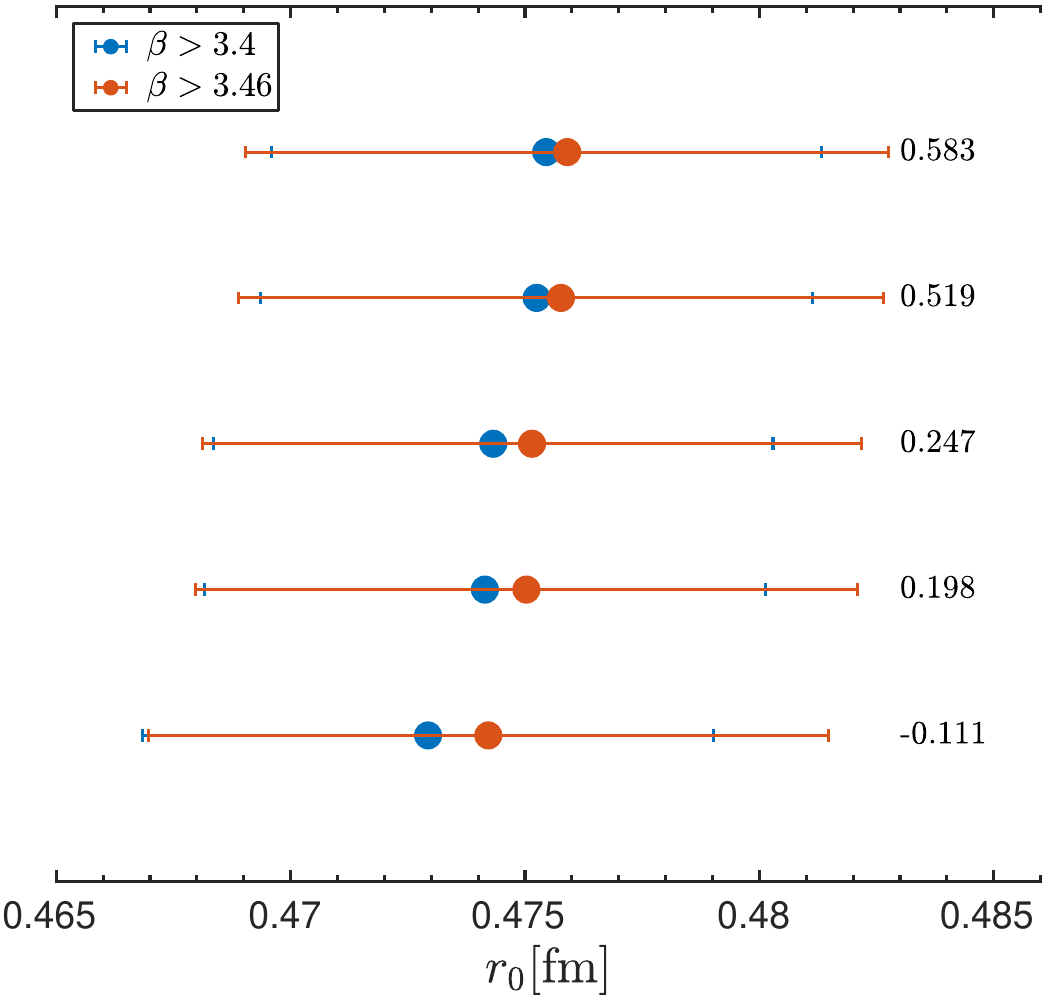} \hfill \includegraphics[width=0.33\linewidth,height=.4\linewidth]{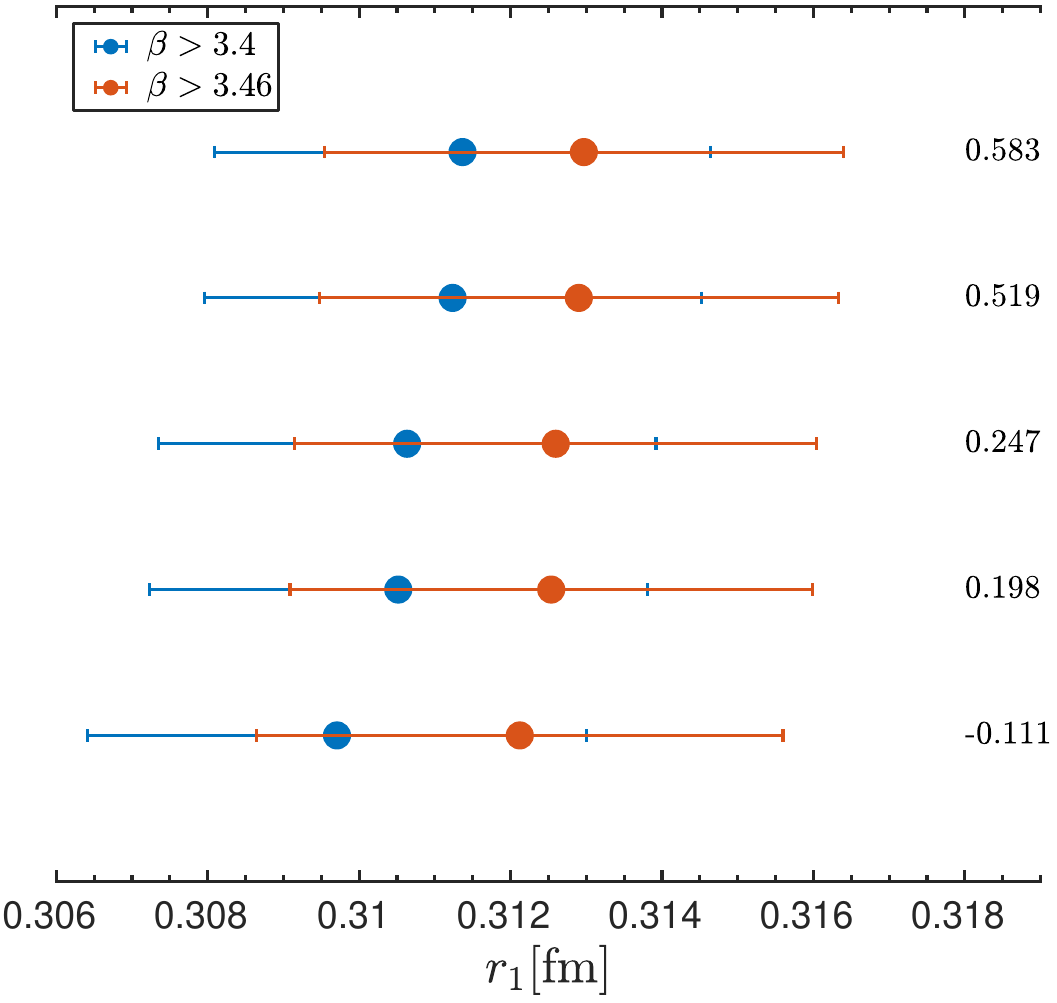} \hfill \includegraphics[width=0.33\linewidth,height=.4\linewidth]{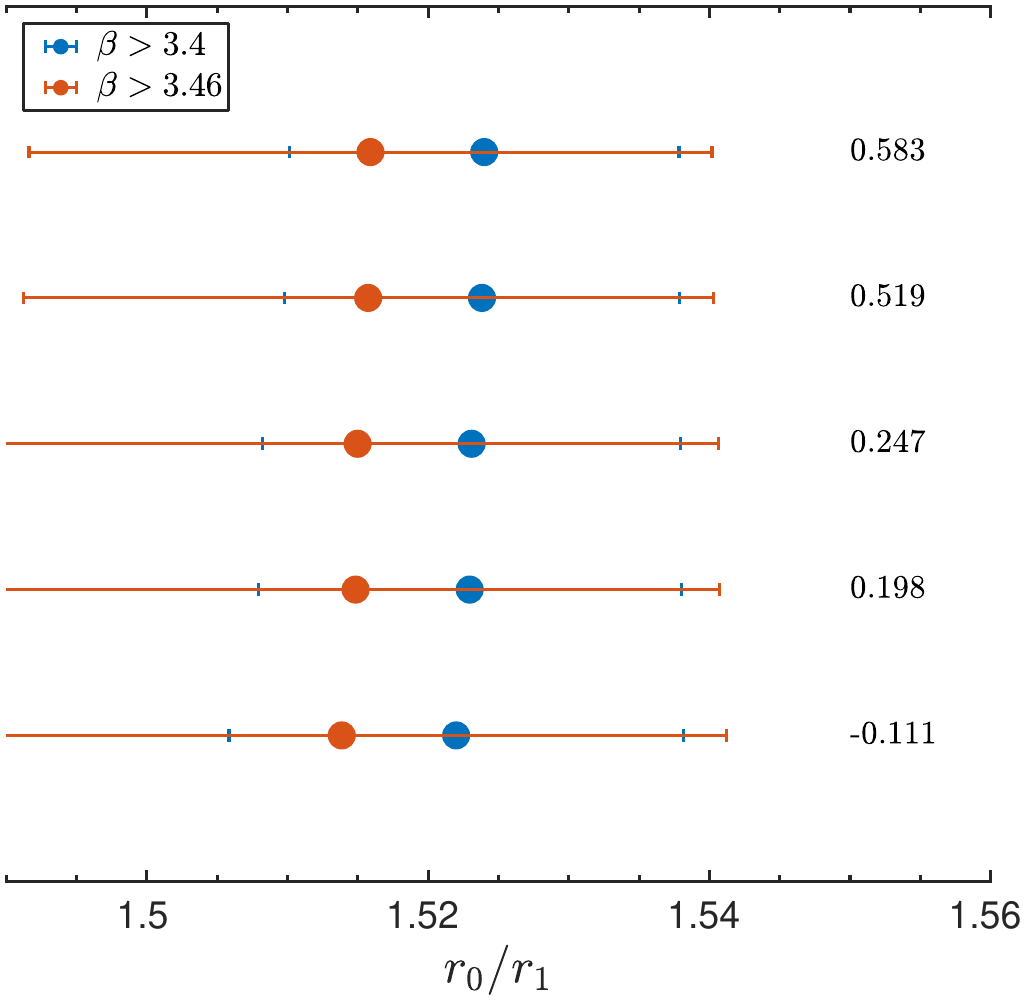} \\
	\caption{The impact of logarithmic corrections on the continuum values of $r_0$ (left), $r_1$ (middle), and $r_0/r_1$ (right). Different colors refer to 
	data sets with different cuts on the coarsest lattice spacing. The $\hat \Gamma$ used in the fit is printed to the right of the 
	corresponding error bars.
	}\label{fig:logcorr}
\end{figure}

\begin{figure}
   \centering
	\includegraphics[width=0.6\linewidth]{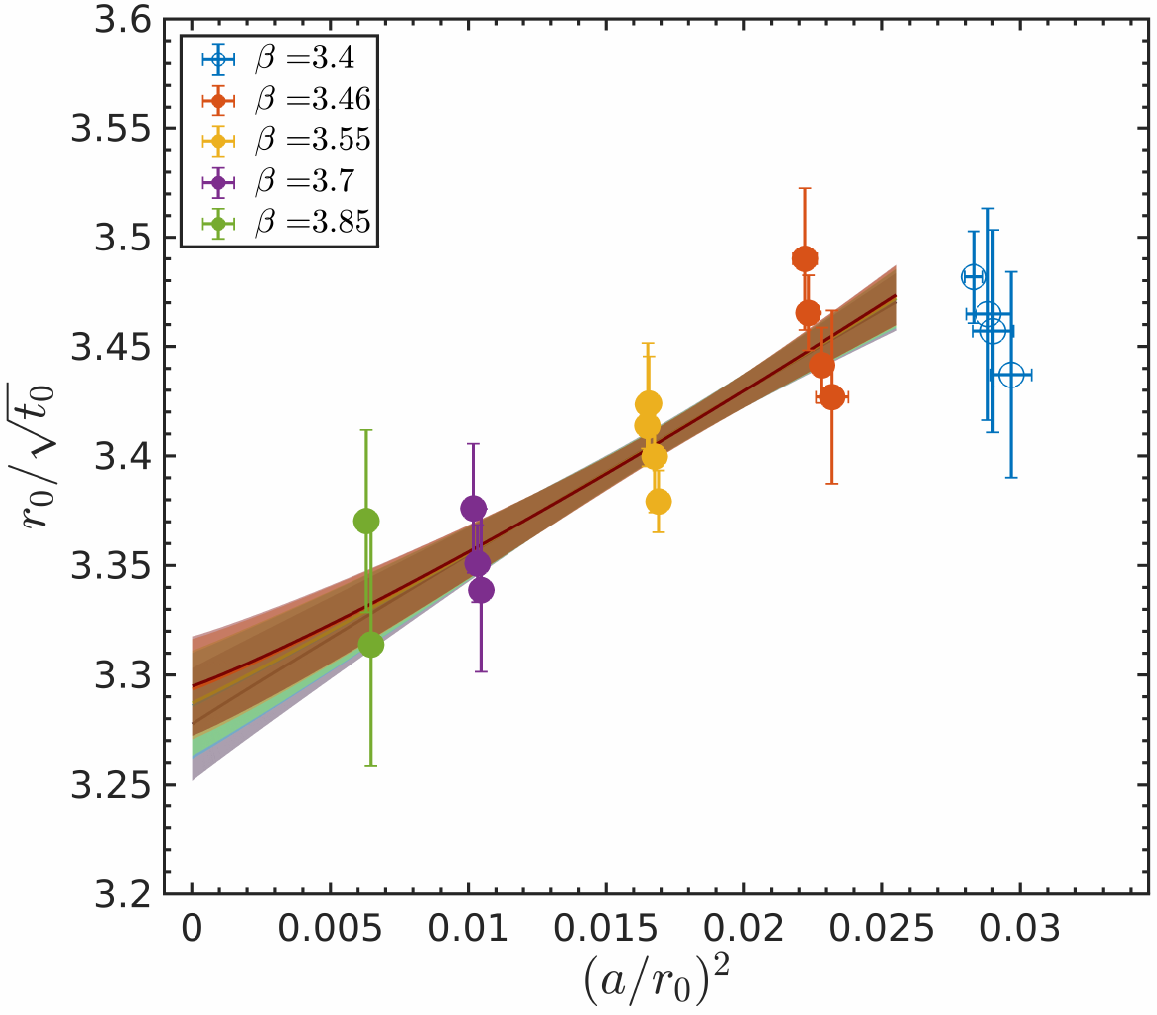}\\
   \caption{Fits with different $(a/r_0^{\rm sym})^2 \ \ln\left[r_0^{\rm sym}/a  \right]^{\hat\Gamma}$ terms.
	To plot all data together, the mass dependent part of the fit function was subtracted from the
	data, i.e. the displayed data is in the chiral limit. The bands correspond to the same $\hat \Gamma$ values as in Fig.~\ref{fig:logcorr}, where the bottom one has $\hat \Gamma = -0.111$ and the top one 0.583.
	}\label{fig:logfits}
\end{figure}

Our continuum extrapolations assume pure $O(a^2)$ lattice artifacts. Symanzik's effective theory~\cite{Symanzik:1979ph} for our action
predicts subleading $O(a^3)$ artifacts, which we neglect. What is potentially more problematic, is that
already the leading scaling violations are not expected to be of pure $O(a^2)$ form, but rather receive logarithmic corrections
$O\left( a^2\ln(a\Lambda)^{-\hat \Gamma}\right)$ ~\cite{Balog:2009yj,Husung:2019ytz,Husung:2021mfl}. 
The possible exponents $\hat \Gamma$ are related to anomalous dimensions 
of the various dimension 6 operators that enter the Symanzik action. They are known in perturbation theory and for our action
have values like $\hat \Gamma \in \{-0.11111, 0.24731, 0.51852, \ldots\}$. More logarithmic corrections are expected that depend
on the specific observables. Even if all possible exponents were known, performing a fit containing all such lattice artifact terms is
simply impossible. To judge the potential impact of these logarithms we re-visit Fit 1, but instead of the pure
$(a/r_0^{\rm sym})^2$ term consider $(a/r_0^{\rm sym})^2 \ \ln\left[r_0^{\rm sym}/a  \right]^{\hat\Gamma}$, in which we
vary the $\hat \Gamma$ within a reasonable range. Fig.~\ref{fig:logcorr} summarizes our findings. 
For the $r_0$ case we also show a few of the modified fits in Fig.~\ref{fig:logfits}.
We conclude, that our 
procedure of investigating various different cuts on the data-set, e.g. neglecting the coarsest ($\beta>3.4)$, 
or the two coarsests ($\beta>3.46$) lattice spacings leads to similar or greater spreads of the continuum values than
allowing different logarithmic factors. For our final extrapolations we thus stay with $\hat \Gamma=0$.

\FloatBarrier

\section{Results and Conclusions}\label{sec:co}
\subsection{The scales $r_0$ and $r_1$}

\begin{figure}[h]
  \centering 
  \includegraphics[width=0.32\textwidth,height=.3\linewidth]{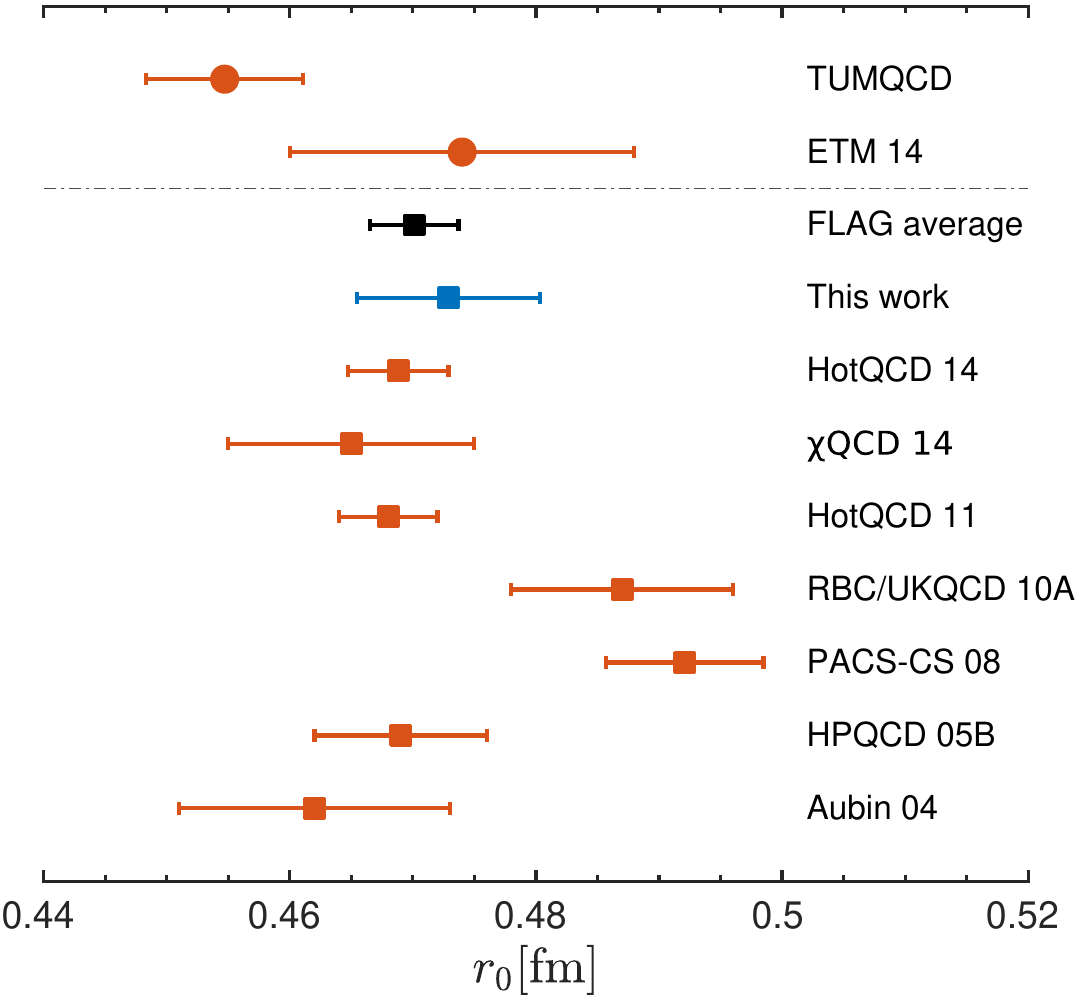}\hfill
  \includegraphics[width=0.32\textwidth,height=.3\linewidth]{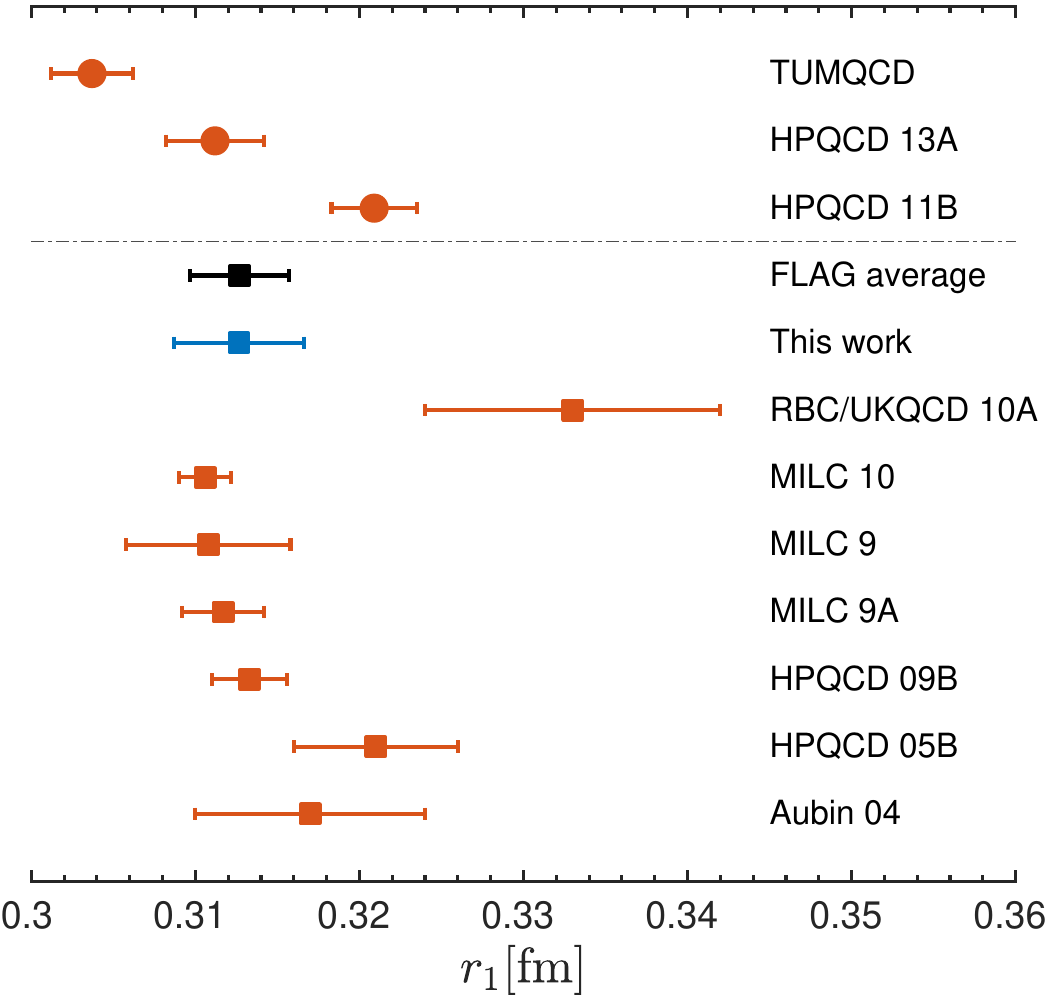}\hfill
  \includegraphics[width=0.32\textwidth,height=.3\linewidth]{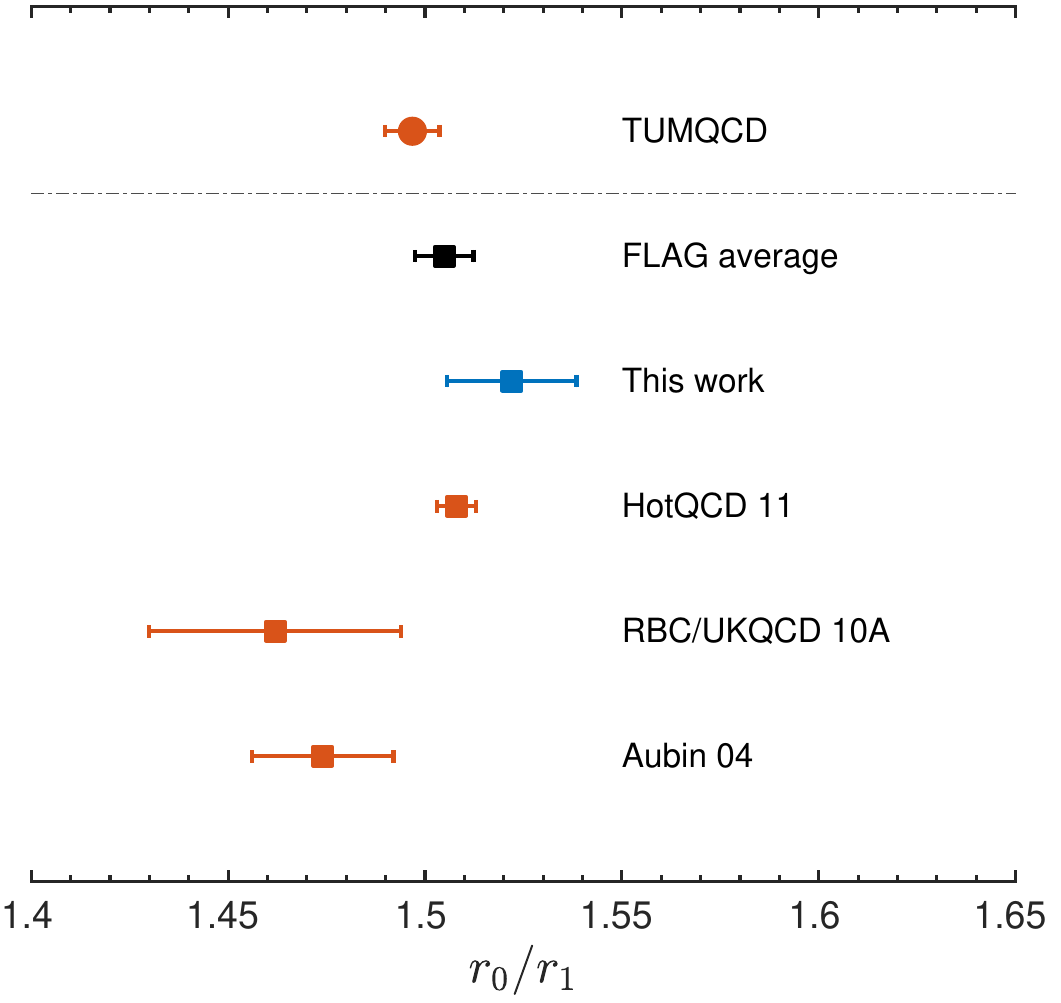}
  \caption{Results of $r_0$ on the left, $r_1$ in the middle, and $\frac{r_0}{r_1}$ to the right. All of them are in comparision with the results from the FLAG review 2021. The upmost points represented by circles are calculated using $\Nf=2+1+1$ QCD while the others are $\Nf=2+1$ QCD. \cite{FlavourLatticeAveragingGroupFLAG:2021npn,EuropeanTwistedMass:2014osg,MILC:2009ltw,Yang:2014sea,Dowdall:2013rya,MILC:2010hzw,HotQCD:2014kol,RBC:2010qam,Gray:2005ur,Aubin:2004wf,Davies:2009tsa,MILC:2009mpl,PACS-CS:2008bkb,HPQCD:2011qwj,Bazavov:2011nk,Brambilla:2022het} }
  \label{fig:scaleresults}
\end{figure}

The main result of this work are the values of the scales $r_0$ and $r_1$ from the static force, as well as their ratio, at the physical 
mass point of $\Nf=2+1$ QCD,

\begin{eqnarray}
                             r_0 &=& 0.4729(57)(48)\,{\rm fm} ,\\ 
                             r_1 &=& 0.3127(24)(32)\,{\rm fm} ,\\ 
{\rm and} \quad \frac{r_0}{r_1}  &=& 1.532(12)\, . 
\end{eqnarray}

In $r_0$ and $r_1$ the first error is from the uncertainty of the dimensionless ratio $r_i/\sqrt{t_0}$ in the continuum limit, 
as listed below,  and the second error due to 
the scale $\sqrt{t_0}$ in fm~\cite{Strassberger:2021tsu}. Both should be combined in quadrature. For the dimensionless ratios $r_i/\sqrt{t_0}$ we obtain

\begin{eqnarray}
                             \frac{r_0}{\sqrt{t_0}} &=& 3.277(39)\, ,\\ 
{\rm and} \quad              \frac{r_1}{\sqrt{t_0}} &=& 2.167(16)\, . 
\end{eqnarray}

Figure \ref{fig:scaleresults} compares these values with previous determinations for $\Nf=2+1$ together with the Flag average \cite{FlavourLatticeAveragingGroupFLAG:2021npn} and $\Nf=2+1+1$. In \cite{Bruno:2014ufa,Knechtli:2017xgy} it was shown that effects of a dynamical charm quark on ratios of low energy scales are typically at permille level. Thus we expect the results for the scales $r_0$, $r_1$ and the ratio $r_0/r_1$ to agree between $\Nf=2+1$ QCD and $\Nf=2+1+1$ QCD at the current level of accuracy.
As a follow up to section \ref{s:E300} an analysis has been performed where E300 has been removed from the calculations altogether. The result is that only the value of $r_1$ changes slighty within the errors. The values with the E300-cut in place are $r_0=0.4728(76)$fm, $r_1=0.3106(37)$fm, and $\frac{r_0}{r_1}= 1.526(17)$.

\FloatBarrier

\subsection{The $\Lambda$-parameter in units of $r_0$ for $\Nf=3$ QCD} \label{Cha:Lambda}


\begin{figure}[h]
  \centering 
  \includegraphics[height=.43\linewidth]{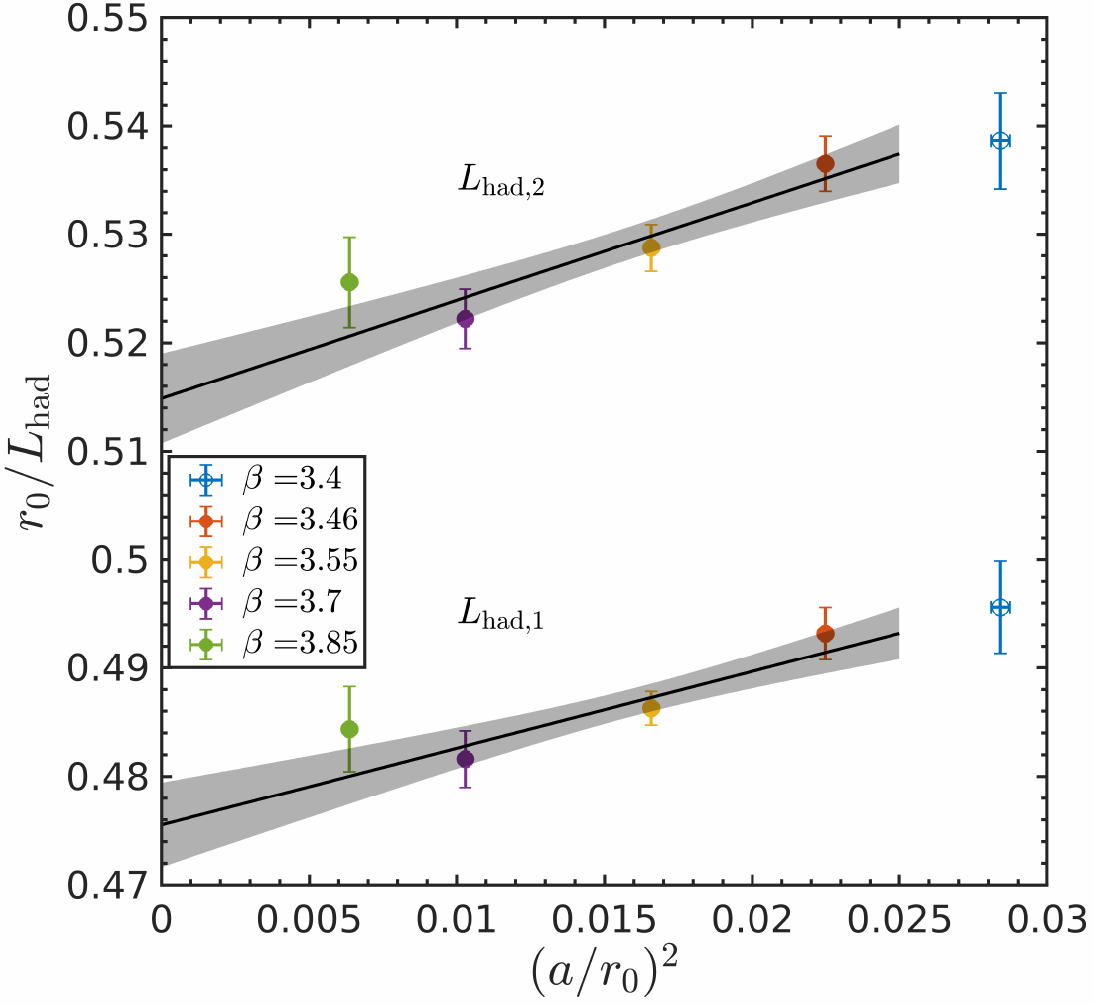}
  \caption{Continuum extrapolations of the ratio of $r_0/a$ over $L_{\rm had}/a$ for two definitions of the scale $L_{\rm had}$ as explained in the text.}
  \label{fig:mu_had}
\end{figure}

The strong coupling ``constant'' is a fundamental parameter of QCD and a precise knowledge of its value of
great importance for many predictions that QCD makes. The value of the coupling at any scale can be computed
from its RG-equation, provided that the dimensionful $\Lambda$-parameter in that scheme is known. It has been 
a great success of lattice QCD to be able to produce the most accurate results for $\Lambda_{\overline{\rm MS}}$
using only very precise low energy experiments as inputs 
(e.g. the masses and decay rates of pseudo-scalar mesons) \cite{DallaBrida:2020pag}. To remove the additional complications of
scale-setting from the results, lattice collaborations often publish dimensionless intermediate results, that
can be compared without setting the scale. These are typically the $\Lambda$-parameter in units of $r_0$,
$\sqrt{t_0}$ or $w_0$. Values for $r_0\Lambda$ and $\sqrt{t_0}\Lambda$ are collected, assessed and averaged
by the FLAG group~\cite{FlavourLatticeAveragingGroupFLAG:2021npn} in addition to the results in MeV.

In recent years the $r_0$ units were largely replaced by scales based on the gradient flow, and e.g. the 
recent result of the ALPHA collaboration~\cite{Bruno:2017gxd} has only used the latter. That particular calculation 
constructs the fully non-perturbative scale evolution of a coupling over a wide range of scales using
a finite-size-scaling method. This procedure yields a value for $L_{\rm had}\Lambda_{\overline{\rm MS}}^{(3)}$ in 
the contiunuum limit of massless three flavor QCD, where 
$L_{\rm had}$ is some constant linear box size. In a last step $L_{\rm had}$ is traded for 
$\sqrt{t_0}$ by extrapolating the ratio of $L_{\rm had}/a$ over $\sqrt{t_0}/a$ to the continuum limit.
For this to be possible, both quantities need to be known using the same lattice action at the same set 
of improved bare couplings $\tilde g_0^2$.

This was all carried out in~\cite{Bruno:2017gxd}, using two different definitions of $L_{\rm had}$, both implicetly defined
by the gradient-flow coupling $\overline{g}_{\rm GF}(L_{\rm had}^{-1})$ having a particular value, namely
\begin{eqnarray}
    \overline{g}^2_{\rm GF}(L_{{\rm had},1}^{-1}) &=  11.31 &\qquad \rightarrow L_{{\rm had},1}\Lambda_{\overline{\rm MS}}^{(3)}=1.729(57) \, ,\\
    \overline{g}^2_{\rm GF}(L_{{\rm had},2}^{-1}) &=  10.20 &\qquad \rightarrow L_{{\rm had},2}\Lambda_{\overline{\rm MS}}^{(3)}=1.593(53) \, .
\end{eqnarray}
The finite lattice spacing results for $L_{{\rm had},i}/a$ were tabulated in Table V of 
the supplementary material of \cite{Bruno:2017gxd}.

To obtain the $\Lambda$ parameter in $r_0$ units, we could go ahead and multiply the original result
for $\sqrt{t_0}\Lambda$ by our ratio $r_0/\sqrt{t_0}$. This would however involve two continuum extrapolations,
namely the one for $L_{\rm had}/\sqrt{t_0}$ and ours from section~\ref{sec:ex}. But because we use the exact same action 
at the same set of improved bare couplings, we are in the fortunate position to be able to completely eliminate the 
flow quantity $t_0$ from the calulation and directly continuum extrapolate the ratio $r_0/a$ over $L_{\rm had}/a$.
This extrapolation turns out to be very well behaved and is shown in \fig{fig:mu_had}, yielding
\begin{eqnarray}
    \frac{r_0}{L_{{\rm had},1}} & = 0.4755(39) & \qquad \rightarrow r_0 \Lambda^{(3)}_{\overline{\rm{MS}}} = 0.822(28) \, ,  \label{e:r0Lambda1}\\
    \frac{r_0}{L_{{\rm had},2}} & = 0.5149(41) & \qquad \rightarrow r_0 \Lambda^{(3)}_{\overline{\rm{MS}}} = 0.820(28) \, ,   \label{e:r0Lambda2}
\end{eqnarray}
both in good agreement with the FLAG average $r_0\Lambda^{(3)}_{\overline{\rm{MS}}} =  0.808(29)$, which includes
the calculations~\cite{Bazavov:2019qoo,Cali:2020hrj,Ayala:2020odx,Bruno:2017gxd,PACS-CS:2009zxm,Maltman:2008bx,McNeile:2010ji}. Both values of the $\Lambda$ parameter in $r_0$ units, \eq{e:r0Lambda1} and \eq{e:r0Lambda2} are equally good and we quote
\begin{equation}
  r_0 \Lambda^{(3)}_{\overline{\rm{MS}}} = 0.820(28) \label{e:r0Lambda}
\end{equation}
as our final result.

\FloatBarrier

\subsection{Shape of the potential}

\begin{figure}[h]
  \centering 
  \includegraphics[height=.43\linewidth]{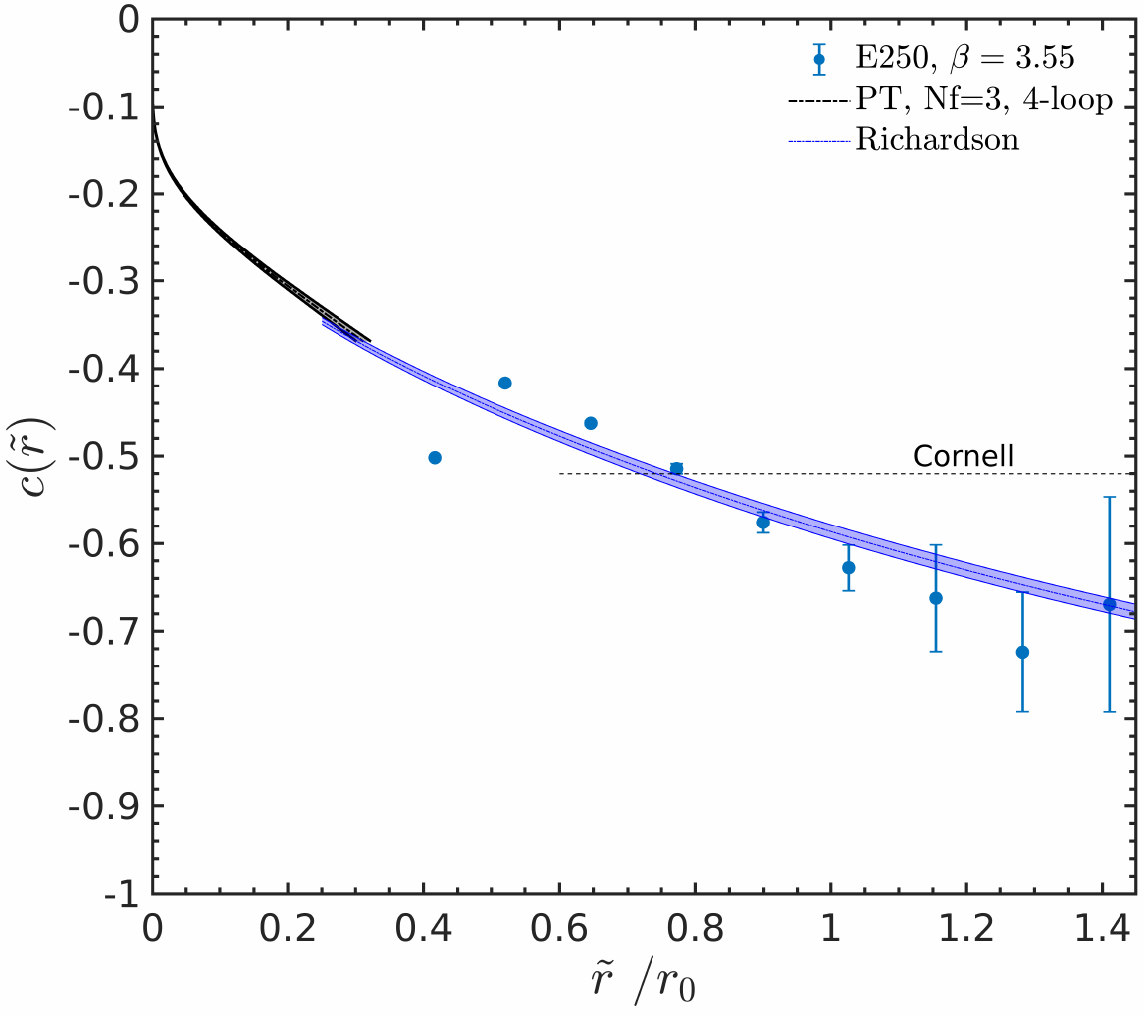}
  \caption{The shape parameter $c(r)$ (\eq{eq:c} and its lattice definition \eq{eq:ImprovedShape}) for the ensemble E250 at the physical point is plotted as a function of the improved distance $\Tilde{r}$ (defined through \eq{eq:cTree}) in units of $r_0$. For comparison we show the perturbative (PT) curve at short distance, the value $c=-0.52$ in the Cornell model \eq{e:Cornell} and the curve derived from the Richardson potential  \eq{e:Richardson}.}
  \label{fig:Shape}
\end{figure}

\FloatBarrier

The renormalized quantity
\begin{equation}\label{eq:c}
  c(r) = \frac{1}{2}r^3V''(r)
\end{equation}
can be used to study the shape of the static potential. At short distances $r\to0$ the dependence of $c(r)$ on the distance $r$ is governed by the renormalization group equation. This is because of the relation to the renormalized coupling $\alpha_c(\mu) = -c(r)/\CF$ given in \eq{e:ccoupling}. At large distance the behaviour of $c(r)$ depends on the flavour content. In pure gauge theory it  approaches asymptotically the value $c(\infty)=\gamma=-\pi(d-2)/24$, see \eq{e:effstring}. In a theory with dynamical fermions string breaking happens and so $c(\infty)=0$ due to the flattening of the ground state static potential. Therefore we expect $c(r)$ to be very sensitive to sea quarks effects.

For these reasons $c(r)$ is an interesting quantity to be studied by lattice simulations. An improved lattice definition of $c(r)$ is given in \eq{eq:ImprovedShape}, in terms of a distance $r=\Tilde{r}$ chosen such that at tree level in perturbation theory $\alpha_{c,{\rm tree}}=g_0^2/(4\pi)$, cf. \eq{eq:cTree}. Precise values of $c(r)$ have been calculated in pure gauge theory in \cite{Luscher:2002qv} using a multi-level sampling algorithm. They have been compared to lattice results from $\Nf=2$ QCD simulations in \cite{Donnellan:2010mx} showing the expected large effects due to the sea quarks. It is also interesting to compare $c(r)$ measured on the lattice to the expectations from phenomenological potentials like the Cornell \eq{e:Cornell} or the Richardson \eq{e:Richardson} potential. In particular the Cornell potential yields a constant $c(r)\equiv-0.52$.

\Fig{fig:Shape} shows our data for $c(\Tilde{r})$ as a function of $\Tilde{r}/r_0$ computed on the ensemble E250 at the physical point. Notice that string breaking happens at $\Tilde{r}/r_0\approx2.6$ \cite{Bulava:2024jpj} so the data shown are for distances well below this threshold. For $\Tilde{r}\to0$ we plot the perturbative curve (black) obtained using the 4-loop $\beta$-function for the coupling $\alpha_c$ as given in Appendix B.1 of \cite{Donnellan:2010mx} using for the $\Lambda$ parameter the value in \eq{e:r0Lambda}. The grey band represents the uncertainty in the $\Lambda$ parameter from \eq{e:r0Lambda}. We see that our data do not reach distances small enough to make contact with the perturbative behaviour. In the region $0.5$--$1.3\,r_0$ the data points follows remarkably close the curve of the Richardson potential (blue) using \eq{e:r0Lambda} for the $\Lambda$ parameter. The blue band represents the uncertainty of the $\Lambda$ parameter from \eq{e:r0Lambda}. We observe a dependence of $c$ on the distance which is clearly different from the constant behaviour assumed in the Cornell model.  The statistical quality of the data degrades fast with increasing distance. Together with the noise-to-signal problem of the Wilson loops there is a factor $r^3$ in the definition of $c$ which amplifies further the problem at large distances $r$. The situation can be improved if multi-level sampling algorithms with dynamical fermions \cite{Ce:2016idq} are applied but this is beyond the scope of the present study.

\section*{Acknowledgements}  We thank Rainer Sommer for valuable discussions. The authors gratefully acknowledge the Gauss Centre for Supercomputing e.V. (www.gauss-centre.eu) for funding this project by providing computing time on the GCS Supercomputer SuperMUC-NG at Leibniz Supercomputing Centre (www.lrz.de). Some computations were carried out on the PLEIADES cluster at the University of Wuppertal, which was supported by the Deutsche Forschungsgemeinschaft (DFG) and the Bundesministerium für Bildung und Forschung (BMBF). The work is supported by the German Research Foundation (DFG) research unit FOR5269 "Future methods for studying confined gluons in QCD". The project is receiving funding from the programme " Netzwerke 2021", an initiative of the Ministry of Culture and Science of the State of Northrhine Westphalia, in the NRW-FAIR network, funding code NW21-024-A (R.H.). The sole responsibility for the content of this publication lies with the authors.
\newpage
\appendix
\section{Appendix}

\subsection{Wilson-loop parameters}\label{s:apploop}

\begin{table}[!h]
   \centering
   \begin{tabular}{c c c c c c}
   \toprule
       $\beta$& id & $T_{\rm max}/a$ & $r_{\rm max}/a$ & smearing &$ N_{\rm meas} $\\
   \midrule
       \multirow{5}{*}{\rotatebox[origin=c]{90}{3.40}}
       &H101 &   24 & 16 & 8, 14 & 2016 \\
       &H102-0 & 24 & 16 & 8, 14 & 997 \\
       &H102-1 & 24 & 16 & 8, 14 & 1008 \\
       &H105 &   24 & 16 & 8, 14 & 1970 \\
       &C101 &   24 & 16 & 8, 14 & 2000 \\
   \midrule
       \multirow{4}{*}{\rotatebox[origin=c]{90}{3.46}}
       &B450 &  24 & 16 & 11, 18 & 1612 \\
       &S400 &  24 & 16 & 11, 18 & 2873 \\
       &D450 &  24 & 16 & 11, 18 & 500 \\
       &D452 &  24 & 16 & 11, 18 & 999 \\
   \midrule
       \multirow{5}{*}{\rotatebox[origin=c]{90}{3.55}}
       &N202 &  24 & 16 & 15, 25 & 1902 \\
       &N203 &  24 & 16 & 15, 25 & 1543 \\
       &N200 &  24 & 16 & 15, 25 & 1712 \\
       &D200 &  24 & 16 & 15, 25 & 2001 \\
       &E250 &  24 & 16 & 15, 25 & 950 \\
   \midrule
       \multirow{4}{*}{\rotatebox[origin=c]{90}{3.70}}
       &N300-0& 28 & 20 & 20, 30 & 507 \\
       &N300-1& 28 & 20 & 20, 30 & 1479 \\
       &J303 &  28 & 20 & 20, 30 & 1073 \\
       &E300 &  28 & 20 & 20, 30, 45, 60 & 1137 \\
   \midrule
       \multirow{2}{*}{\rotatebox[origin=c]{90}{3.85}}
       &J500 &  32 & 24 & 25, 35 & 1444 \\
       &J501 &  32 & 24 & 25, 35 & 3559 \\
   \bottomrule
   \end{tabular}
   \caption{Maximal temporal and spatial sizes of the Wilon loops, smearing levels and number of measurements are listed for each ensemble.}
   \label{tab:smearing}
\end{table}

\FloatBarrier
\newpage
\subsection{Improved distance}\label{s:apprI}

To determine the improved distance we calculate the momentum integral \eq{eq:ftree} numerically as a $L/a \to \infty$ limit of discrete momentum sums. The sums, which we denote by $G(L/a)$ were computed for lattices of sizes $L/a=64$ up to $L/a=512$. To extrapolate to an infinite lattice we use the ansatz $ G(L/a) = G(\infty) + c_2(a/L)^3 + c_3(a/L)^5$, see appendix A of~\cite{Donnellan:2010mx}. Other ans\"atze including the terms $c_4(a/L)^7$ and $c_5(a/L)^9$ were tried as well to estimate the systematics. Using \eq{eq:r2ftree} we find $r_I$. The improved distance for $\Tilde{r}/a$ is determined similarly.

\begin{table}[!h]
   \centering
   \begin{tabular}{c c c}
   \toprule

$r/a$ & $r_I/a$ & $\Tilde{r}/a$ \\ \hline
4   & 3.568   & 4.053         \\
5   & 4.541   & 5.030         \\
6   & 5.523   & 6.015         \\
7   & 6.510   & 7.001         \\
8   & 7.504   & 7.992         \\
9   & 8.499   & 8.987         \\
10  & 9.497   & 9.984         \\
11  & 10.496  & 10.983        \\
12  & 11.495  & 11.982        \\
13  & 12.495  & 12.982        \\
14  & 13.494  & 13.983        \\
15  & 14.494  & 14.983        \\
16  & 15.494  & 15.984        \\
17  & 16.495  & 16.984        \\
18  & 17.495  & 17.985        \\
19  & 18.495  & 18.985        \\
20  & 19.495  & 19.986      \\
   \bottomrule
   \end{tabular}
   \caption{Values of $r_I/a$ used for $V'$ and $\Tilde{r}/a$ for $V''$ for one level of HYP2 smearing in static quark action.}
   \label{tab:r_I}
\end{table}
\FloatBarrier
\newpage
\subsection{Raw results per ensemble}\label{s:appr0}
\begin{table}[!h]
   \centering
   \begin{tabular}{c c c c c c c c c}
   \toprule
       $\beta$& id & $r_0/a$ & $r_1/a$ & $r_0/r_1$\\
   \midrule
       \multirow{4}{*}{\rotatebox[origin=c]{90}{3.40}}
       &H101 &5.81(7)&3.842(13)& 1.511(19)\\
       &H102 &5.87(8)&3.886(14) & 1.511(20)\\
       &H105 &5.89(8)&3.860(11) &  1.526(21)\\
       &C101 &5.94(4)&3.874(8)& 1.534(10) \\
   \midrule
       \multirow{4}{*}{\rotatebox[origin=c]{90}{3.46}}
       &B450 &6.57(8)&4.299(19) & 1.527(20) \\
       &S400 &6.71(7)&4.331(17) &  1.549(15)\\
       &D450 &6.62(3)&4.327(9) & 1.530(9) \\
       &D452 &6.69(3)&4.340(8) & 1.542(8) \\
   \midrule
       \multirow{5}{*}{\rotatebox[origin=c]{90}{3.55}}
       &N202 &7.72(6)&5.066(19) & 1.525(14) \\
       &N203&7.77(5)&5.036(14) & 1.543(10) \\
       &N200 &7.78(6)&5.025(13) & 1.549(12) \\
       &D200 &7.69(3)&5.041(9) & 1.526(6) \\
       &E250 &7.79(2)&5.054(5) & 1.541(5) \\
   \midrule
       \multirow{3}{*}{\rotatebox[origin=c]{90}{3.70}}
       &N300& 9.77(11) & 6.413(31) & 1.525(20) \\
       &J303 &9.84(5)&6.432(13) & 1.530(8) \\
       &E300 &9.84(11)&6.529(24) & 1.506(18) \\
   \midrule
       \multirow{2}{*}{\rotatebox[origin=c]{90}{3.85}}
       &J500 &12.43(9)&8.163(28) & 1.523(12)\\
       &J501 &12.61(15)&8.226(32) & 1.533(19)\\
   \bottomrule
   \end{tabular}
   \caption{The values of $r_0/a$, $r_1/a$, and the ratio $r_0/r_1$ for all ensembles.}
   \label{tab:r0_res}
\end{table}

\newpage
\subsection{Fit results}\label{s:appfit}
\begin{table}[!h]
  \centering
  \scriptsize{
   \begin{tabular}{c c c c c c c c }
   \toprule
       Fit $\#$& Cut Method & $r_0$[fm] & $\chi ^2$ /d.o.f. & $r_1$[fm] & $\chi ^2$ /d.o.f. & $r_0/r_1$ & $\chi ^2$ /d.o.f.  \\
   \midrule

       \multirow{5}{*}{\rotatebox[origin=c]{90}{Fit 1}}
        & No cuts           & 0.4758(57)  & 12.6/15                           & 0.3100(32) & 30.4/15   & 1.5344(95)   & 13.6/15      \\
        & $m_\pi<400$MeV    & 0.4750(57)  & 10.5/10                           & 0.3100(32) & 27.5/10   & 1.5325(94)   & 12.0/10      \\
        & $\beta$>3.4       & 0.4724(61)  & 9.7/11                            & 0.3104(33) & 27.0/11   & 1.5224(127)  & 10.0/11       \\
        & $\beta$>3.46      & 0.4728(73)  & 7.2/7                             & 0.3130(35) & 18.4/7    & 1.5142(199)  & 7.6/7       \\
        & \makecell{$m_\pi<400$MeV\\ $\&$ $\beta$>3.4}  & 0.4716(62) & 8.1/7  & 0.3104(33) & 25.0/7    & 1.5198(123)  & 8.5/7       \\
   \midrule
       \multirow{5}{*}{\rotatebox[origin=c]{90}{Fit 2}}
        & No cuts           & 0.4763(57)    & 11.1/14                         & 0.3099(32) & 30.1/14   & 1.5341(93)    & 13.1/14     \\
        & $m_\pi<400$MeV    &  0.4778(59)   & 4.2/9                           & 0.3102(33) & 23.6/9    & 1.5396(95)    & 8.1/9    \\
        & $\beta$>3.4       & 0.4730(62)    & 8.6/10                          & 0.3104(33) & 26.3/10   & 1.5222(125)   & 9.8/10    \\
        & $\beta$>3.46      & 0.4731(73)    & 6.5/6                           & 0.3127(35) & 18.2/6    & 1.5124(186)   & 7.2/6    \\
        & \makecell{$m_\pi<400$MeV\\ $\&$ $\beta$>3.4}  & 0.4745(64) & 2.3/6  & 0.3107(33) & 21.0/6    & 1.5272(126)   & 5.5/6    \\

   \midrule
       \multirow{5}{*}{\rotatebox[origin=c]{90}{Fit 3}}
        & No cuts           & 0.4762(63)   &  12.4/14                         & 0.3099(33) & 30.7/14   & 1.5350(105)    & 13.4/14    \\
        & $m_\pi<400$MeV    & 0.4755(68)   & 10.4/9                           & 0.3099(33) & 27.7/9    & 1.5351(149)    & 11.5/9     \\
        & $\beta$>3.4       & 0.4724(78)   & 9.8/10                           & 0.3104(35) & 27.2/10   & 1.5225(152)    & 10.0/10     \\
        & $\beta$>3.46      & 0.4724(175)  & 7.2/6                            & 0.3144(49) & 16.4/6    & 1.5114(295)    & 7.5/6       \\
        & \makecell{$m_\pi<400$MeV\\ $\&$ $\beta$>3.4}  & 0.4716(86) & 8.1/6  & 0.3103(36) & 25.0/6    & 1.5213(230)    & 8.5/6     \\

   \midrule
       \multirow{5}{*}{\rotatebox[origin=c]{90}{Fit 4}}
        & No cuts           & 0.4769(61)    & 12.3/14                         & 0.3106(33) & 30.0/14   & 1.5351(109)    & 13.0/14\\
        & $m_\pi<400$MeV    & 0.4762(66)    & 10.3/9                          & 0.3106(33) & 27.1/9    & 1.5350(141)    & 11.4/14\\
        & $\beta$>3.4       & 0.4733(73)    & 9.66/10                         & 0.3108(35) & 26.8/10   & 1.5231(160)    & 10.0/14\\
        & $\beta$>3.46  & 0.4729(166)       & 7.1/6                           & 0.3145(48) & 16.4/6    & 1.5121(334)    & 8.1/14\\
        & \makecell{$m_\pi<400$MeV\\ $\&$ $\beta$>3.4}  & 0.4726(82) & 8.0/6  & 0.3107(36) & 24.6/6    & 1.5220(216)    & 8.3/14\\

   \bottomrule
   \end{tabular}
   }
   \caption{The different fits together with the cuts used to find the scales $r_0$, $r_1$, and the ratio $r_0/r_1$ at physical point used in figures~\ref{fig:Fit} and~\ref{fig:FitRatio}.}
   \label{tab:FitResults}
\end{table}

\FloatBarrier


\bibliographystyle{utphys} 
\bibliography{paper}

\providecommand{\href}[2]{#2}\begingroup\raggedright\begin{thebibliography}{100}

\bibitem{Sommer:1993ce}
R.~Sommer, ``{A New way to set the energy scale in lattice gauge theories and
  its applications to the static force and alpha-s in SU(2) Yang-Mills
  theory},'' \href{http://dx.doi.org/10.1016/0550-3213(94)90473-1}{{\em Nucl.
  Phys. B} {\bfseries 411} (1994) 839--854},
  \href{http://arxiv.org/abs/hep-lat/9310022}{{\ttfamily
  arXiv:hep-lat/9310022}}.

\bibitem{Wilson:1974sk}
K.~G. Wilson, ``{Confinement of Quarks},''
  \href{http://dx.doi.org/10.1103/PhysRevD.10.2445}{{\em Phys. Rev. D}
  {\bfseries 10} (1974) 2445--2459}.

\bibitem{DiGiacomo:1989yp}
A.~Di~Giacomo, M.~Maggiore, and S.~Olejnik, ``{Evidence for Flux Tubes From
  Cooled {QCD} Configurations},''
  \href{http://dx.doi.org/10.1016/0370-2693(90)90828-T}{{\em Phys. Lett. B}
  {\bfseries 236} (1990) 199--202}.

\bibitem{DiGiacomo:1990hc}
A.~Di~Giacomo, M.~Maggiore, and S.~Olejnik, ``{Confinement and Chromoelectric
  Flux Tubes in Lattice {QCD}},''
  \href{http://dx.doi.org/10.1016/0550-3213(90)90567-W}{{\em Nucl. Phys. B}
  {\bfseries 347} (1990) 441--460}.

\bibitem{Singh:1993jj}
V.~Singh, D.~A. Browne, and R.~W. Haymaker, ``{Structure of Abrikosov vortices
  in SU(2) lattice gauge theory},''
  \href{http://dx.doi.org/10.1016/0370-2693(93)91146-E}{{\em Phys. Lett. B}
  {\bfseries 306} (1993) 115--119},
  \href{http://arxiv.org/abs/hep-lat/9301004}{{\ttfamily
  arXiv:hep-lat/9301004}}.

\bibitem{Bali:1994de}
G.~S. Bali, K.~Schilling, and C.~Schlichter, ``{Observing long color flux tubes
  in SU(2) lattice gauge theory},''
  \href{http://dx.doi.org/10.1103/PhysRevD.51.5165}{{\em Phys. Rev. D}
  {\bfseries 51} (1995) 5165--5198},
  \href{http://arxiv.org/abs/hep-lat/9409005}{{\ttfamily
  arXiv:hep-lat/9409005}}.

\bibitem{Bali:2000gf}
G.~S. Bali, ``{QCD forces and heavy quark bound states},''
  \href{http://dx.doi.org/10.1016/S0370-1573(00)00079-X}{{\em Phys. Rept.}
  {\bfseries 343} (2001) 1--136},
  \href{http://arxiv.org/abs/hep-ph/0001312}{{\ttfamily arXiv:hep-ph/0001312}}.

\bibitem{Luscher:2002qv}
M.~L{\"u}scher and P.~Weisz, ``{Quark confinement and the bosonic string},''
  \href{http://dx.doi.org/10.1088/1126-6708/2002/07/049}{{\em JHEP} {\bfseries
  07} (2002) 049}, \href{http://arxiv.org/abs/hep-lat/0207003}{{\ttfamily
  arXiv:hep-lat/0207003}}.

\bibitem{Greensite:2005yu}
J.~Greensite, S.~Olejnik, M.~Polikarpov, S.~Syritsyn, and V.~Zakharov,
  ``{Localized eigenmodes of covariant Laplacians in the Yang-Mills vacuum},''
  \href{http://dx.doi.org/10.1103/PhysRevD.71.114507}{{\em Phys. Rev. D}
  {\bfseries 71} (2005) 114507},
  \href{http://arxiv.org/abs/hep-lat/0504008}{{\ttfamily
  arXiv:hep-lat/0504008}}.

\bibitem{Andreev:2020pqy}
O.~Andreev, ``{String Breaking, Baryons, Medium, and Gauge/String Duality},''
  \href{http://dx.doi.org/10.1103/PhysRevD.101.106003}{{\em Phys. Rev. D}
  {\bfseries 101} no.~10, (2020) 106003},
  \href{http://arxiv.org/abs/2003.09880}{{\ttfamily arXiv:2003.09880
  [hep-ph]}}.

\bibitem{Bulava:2019iut}
J.~Bulava, B.~H\"orz, F.~Knechtli, V.~Koch, G.~Moir, C.~Morningstar, and
  M.~Peardon, ``{String breaking by light and strange quarks in QCD},''
  \href{http://dx.doi.org/10.1016/j.physletb.2019.05.018}{{\em Phys. Lett. B}
  {\bfseries 793} (2019) 493--498},
  \href{http://arxiv.org/abs/1902.04006}{{\ttfamily arXiv:1902.04006
  [hep-lat]}}.

\bibitem{Bulava:2024jpj}
J.~Bulava, F.~Knechtli, V.~Koch, C.~Morningstar, and M.~Peardon, ``{The
  quark-mass dependence of the potential energy between static colour sources
  in the QCD vacuum with light and strange quarks},''
  \href{http://dx.doi.org/10.1016/j.physletb.2024.138754}{{\em Phys. Lett. B}
  {\bfseries 854} (2024) 138754},
  \href{http://arxiv.org/abs/2403.00754}{{\ttfamily arXiv:2403.00754
  [hep-lat]}}.

\bibitem{Bernard:2000gd}
C.~W. Bernard, T.~Burch, K.~Orginos, D.~Toussaint, T.~A. DeGrand, C.~E. DeTar,
  S.~A. Gottlieb, U.~M. Heller, J.~E. Hetrick, and B.~Sugar, ``{The Static
  quark potential in three flavor QCD},''
  \href{http://dx.doi.org/10.1103/PhysRevD.62.034503}{{\em Phys. Rev. D}
  {\bfseries 62} (2000) 034503},
  \href{http://arxiv.org/abs/hep-lat/0002028}{{\ttfamily
  arXiv:hep-lat/0002028}}.

\bibitem{Bazavov:2017dsy}
A.~Bazavov, P.~Petreczky, and J.~H. Weber, ``{Equation of State in 2+1 Flavor
  QCD at High Temperatures},''
  \href{http://dx.doi.org/10.1103/PhysRevD.97.014510}{{\em Phys. Rev. D}
  {\bfseries 97} no.~1, (2018) 014510},
  \href{http://arxiv.org/abs/1710.05024}{{\ttfamily arXiv:1710.05024
  [hep-lat]}}.

\bibitem{Bruno:2014jqa}
M.~Bruno {\em et~al.}, ``{Simulation of QCD with N$_{f} =$ 2 $+$ 1 flavors of
  non-perturbatively improved Wilson fermions},''
  \href{http://dx.doi.org/10.1007/JHEP02(2015)043}{{\em JHEP} {\bfseries 02}
  (2015) 043}, \href{http://arxiv.org/abs/1411.3982}{{\ttfamily arXiv:1411.3982
  [hep-lat]}}.

\bibitem{Mohler:2017wnb}
D.~Mohler, S.~Schaefer, and J.~Simeth, ``{CLS 2+1 flavor simulations at
  physical light- and strange-quark masses},''
  \href{http://dx.doi.org/10.1051/epjconf/201817502010}{{\em EPJ Web Conf.}
  {\bfseries 175} (2018) 02010},
  \href{http://arxiv.org/abs/1712.04884}{{\ttfamily arXiv:1712.04884
  [hep-lat]}}.

\bibitem{Luscher:2010iy}
M.~L\"uscher, ``{Properties and uses of the Wilson flow in lattice QCD},''
  \href{http://dx.doi.org/10.1007/JHEP08(2010)071}{{\em JHEP} {\bfseries 08}
  (2010) 071}, \href{http://arxiv.org/abs/1006.4518}{{\ttfamily arXiv:1006.4518
  [hep-lat]}}. [Erratum: JHEP 03, 092 (2014)].

\bibitem{Peter:1997me}
M.~Peter, ``{The Static potential in QCD: A Full two loop calculation},''
  \href{http://dx.doi.org/10.1016/S0550-3213(97)00373-8}{{\em Nucl. Phys. B}
  {\bfseries 501} (1997) 471--494},
  \href{http://arxiv.org/abs/hep-ph/9702245}{{\ttfamily arXiv:hep-ph/9702245}}.

\bibitem{Schroder:1998vy}
Y.~Schr{\"o}der, ``{The Static potential in QCD to two loops},''
  \href{http://dx.doi.org/10.1016/S0370-2693(99)00010-6}{{\em Phys. Lett. B}
  {\bfseries 447} (1999) 321--326},
  \href{http://arxiv.org/abs/hep-ph/9812205}{{\ttfamily arXiv:hep-ph/9812205}}.

\bibitem{Melles:2000dq}
M.~Melles, ``{The Static QCD potential in coordinate space with quark masses
  through two loops},''
  \href{http://dx.doi.org/10.1103/PhysRevD.62.074019}{{\em Phys. Rev. D}
  {\bfseries 62} (2000) 074019},
  \href{http://arxiv.org/abs/hep-ph/0001295}{{\ttfamily arXiv:hep-ph/0001295}}.

\bibitem{Anzai:2009tm}
C.~Anzai, Y.~Kiyo, and Y.~Sumino, ``{Static QCD potential at three-loop
  order},'' \href{http://dx.doi.org/10.1103/PhysRevLett.104.112003}{{\em Phys.
  Rev. Lett.} {\bfseries 104} (2010) 112003},
  \href{http://arxiv.org/abs/0911.4335}{{\ttfamily arXiv:0911.4335 [hep-ph]}}.

\bibitem{Brambilla:2009bi}
N.~Brambilla, A.~Vairo, X.~Garcia~i Tormo, and J.~Soto, ``{The QCD static
  energy at NNNLL},'' \href{http://dx.doi.org/10.1103/PhysRevD.80.034016}{{\em
  Phys. Rev. D} {\bfseries 80} (2009) 034016},
  \href{http://arxiv.org/abs/0906.1390}{{\ttfamily arXiv:0906.1390 [hep-ph]}}.

\bibitem{Brodsky:1999fr}
S.~J. Brodsky, M.~Melles, and J.~Rathsman, ``{The Two loop scale dependence of
  the static QCD potential including quark masses},''
  \href{http://dx.doi.org/10.1103/PhysRevD.60.096006}{{\em Phys. Rev. D}
  {\bfseries 60} (1999) 096006},
  \href{http://arxiv.org/abs/hep-ph/9906324}{{\ttfamily arXiv:hep-ph/9906324}}.

\bibitem{Donnellan:2010mx}
M.~Donnellan, F.~Knechtli, B.~Leder, and R.~Sommer, ``{Determination of the
  Static Potential with Dynamical Fermions},''
  \href{http://dx.doi.org/10.1016/j.nuclphysb.2011.03.013}{{\em Nucl. Phys. B}
  {\bfseries 849} (2011) 45--63},
  \href{http://arxiv.org/abs/1012.3037}{{\ttfamily arXiv:1012.3037 [hep-lat]}}.

\bibitem{vanRitbergen:1997va}
T.~van Ritbergen, J.~A.~M. Vermaseren, and S.~A. Larin, ``{The Four loop beta
  function in quantum chromodynamics},''
  \href{http://dx.doi.org/10.1016/S0370-2693(97)00370-5}{{\em Phys. Lett. B}
  {\bfseries 400} (1997) 379--384},
  \href{http://arxiv.org/abs/hep-ph/9701390}{{\ttfamily arXiv:hep-ph/9701390}}.

\bibitem{Czakon:2004bu}
M.~Czakon, ``{The Four-loop QCD beta-function and anomalous dimensions},''
  \href{http://dx.doi.org/10.1016/j.nuclphysb.2005.01.012}{{\em Nucl. Phys. B}
  {\bfseries 710} (2005) 485--498},
  \href{http://arxiv.org/abs/hep-ph/0411261}{{\ttfamily arXiv:hep-ph/0411261}}.

\bibitem{Nambu:1978bd}
Y.~Nambu, ``{QCD and the String Model},''
  \href{http://dx.doi.org/10.1016/0370-2693(79)91193-6}{{\em Phys. Lett. B}
  {\bfseries 80} (1979) 372--376}.

\bibitem{Eichten:1978tg}
E.~Eichten, K.~Gottfried, T.~Kinoshita, K.~D. Lane, and T.-M. Yan,
  ``{Charmonium: The Model},''
  \href{http://dx.doi.org/10.1103/PhysRevD.17.3090}{{\em Phys. Rev. D}
  {\bfseries 17} (1978) 3090}. [Erratum: Phys.Rev.D 21, 313 (1980)].

\bibitem{Eichten:1979ms}
E.~Eichten, K.~Gottfried, T.~Kinoshita, K.~D. Lane, and T.-M. Yan,
  ``{Charmonium: Comparison with Experiment},''
  \href{http://dx.doi.org/10.1103/PhysRevD.21.203}{{\em Phys. Rev. D}
  {\bfseries 21} (1980) 203}.

\bibitem{Richardson:1978bt}
J.~L. Richardson, ``{The Heavy Quark Potential and the Upsilon, J/psi
  Systems},'' \href{http://dx.doi.org/10.1016/0370-2693(79)90753-6}{{\em Phys.
  Lett. B} {\bfseries 82} (1979) 272--274}.

\bibitem{Bali:2005fu}
{\bfseries SESAM} Collaboration, G.~S. Bali, H.~Neff, T.~Duessel, T.~Lippert,
  and K.~Schilling, ``{Observation of string breaking in QCD},''
  \href{http://dx.doi.org/10.1103/PhysRevD.71.114513}{{\em Phys. Rev. D}
  {\bfseries 71} (2005) 114513},
  \href{http://arxiv.org/abs/hep-lat/0505012}{{\ttfamily
  arXiv:hep-lat/0505012}}.

\bibitem{DellaMorte:2005nwx}
M.~Della~Morte, A.~Shindler, and R.~Sommer, ``{On lattice actions for static
  quarks},'' \href{http://dx.doi.org/10.1088/1126-6708/2005/08/051}{{\em JHEP}
  {\bfseries 08} (2005) 051},
  \href{http://arxiv.org/abs/hep-lat/0506008}{{\ttfamily
  arXiv:hep-lat/0506008}}.

\bibitem{Grimbach:2008uy}
A.~Grimbach, D.~Guazzini, F.~Knechtli, and F.~Palombi, ``{O(a) improvement of
  the HYP static axial and vector currents at one-loop order of perturbation
  theory},'' \href{http://dx.doi.org/10.1088/1126-6708/2008/03/039}{{\em JHEP}
  {\bfseries 03} (2008) 039}, \href{http://arxiv.org/abs/0802.0862}{{\ttfamily
  arXiv:0802.0862 [hep-lat]}}.

\bibitem{FlavourLatticeAveragingGroupFLAG:2021npn}
{\bfseries Flavour Lattice Averaging Group (FLAG)} Collaboration, Y.~Aoki {\em
  et~al.}, ``{FLAG Review 2021},''
  \href{http://dx.doi.org/10.1140/epjc/s10052-022-10536-1}{{\em Eur. Phys. J.
  C} {\bfseries 82} no.~10, (2022) 869},
  \href{http://arxiv.org/abs/2111.09849}{{\ttfamily arXiv:2111.09849
  [hep-lat]}}.

\bibitem{BMW:2012hcm}
{\bfseries BMW} Collaboration, S.~Bors\'anyi, S.~D\"urr, Z.~Fodor,
  C.~Hoelbling, S.~D. Katz, S.~Krieg, T.~Kurth, L.~Lellouch, T.~Lippert, and
  C.~McNeile, ``{High-precision scale setting in lattice QCD},''
  \href{http://dx.doi.org/10.1007/JHEP09(2012)010}{{\em JHEP} {\bfseries 09}
  (2012) 010}, \href{http://arxiv.org/abs/1203.4469}{{\ttfamily arXiv:1203.4469
  [hep-lat]}}.

\bibitem{Luscher:1984xn}
M.~L{\"u}scher and P.~Weisz, ``{On-shell improved lattice gauge theories},''
  \href{http://dx.doi.org/10.1007/BF01205792}{{\em Commun. Math. Phys.}
  {\bfseries 98} no.~3, (1985) 433}. [Erratum: Commun.Math.Phys. 98, 433
  (1985)].

\bibitem{Sheikholeslami:1985ij}
B.~Sheikholeslami and R.~Wohlert, ``{Improved Continuum Limit Lattice Action
  for QCD with Wilson Fermions},''
  \href{http://dx.doi.org/10.1016/0550-3213(85)90002-1}{{\em Nucl. Phys. B}
  {\bfseries 259} (1985) 572}.

\bibitem{Bulava:2013cta}
J.~Bulava and S.~Schaefer, ``{Improvement of $N_f$ = 3 lattice QCD with Wilson
  fermions and tree-level improved gauge action},''
  \href{http://dx.doi.org/10.1016/j.nuclphysb.2013.05.019}{{\em Nucl. Phys. B}
  {\bfseries 874} (2013) 188--197},
  \href{http://arxiv.org/abs/1304.7093}{{\ttfamily arXiv:1304.7093 [hep-lat]}}.

\bibitem{Luscher:2011kk}
M.~L{\"u}scher and S.~Schaefer, ``{Lattice QCD without topology barriers},''
  \href{http://dx.doi.org/10.1007/JHEP07(2011)036}{{\em JHEP} {\bfseries 07}
  (2011) 036}, \href{http://arxiv.org/abs/1105.4749}{{\ttfamily arXiv:1105.4749
  [hep-lat]}}.

\bibitem{Duane:1987de}
S.~Duane, A.~D. Kennedy, B.~J. Pendleton, and D.~Roweth, ``{Hybrid Monte
  Carlo},'' \href{http://dx.doi.org/10.1016/0370-2693(87)91197-X}{{\em Phys.
  Lett. B} {\bfseries 195} (1987) 216--222}.

\bibitem{Hasenbusch:2001ne}
M.~Hasenbusch, ``{Speeding up the hybrid Monte Carlo algorithm for dynamical
  fermions},'' \href{http://dx.doi.org/10.1016/S0370-2693(01)01102-9}{{\em
  Phys. Lett. B} {\bfseries 519} (2001) 177--182},
  \href{http://arxiv.org/abs/hep-lat/0107019}{{\ttfamily
  arXiv:hep-lat/0107019}}.

\bibitem{Kennedy:1998cu}
A.~D. Kennedy, I.~Horvath, and S.~Sint, ``{A New exact method for dynamical
  fermion computations with nonlocal actions},''
  \href{http://dx.doi.org/10.1016/S0920-5632(99)85217-7}{{\em Nucl. Phys. B
  Proc. Suppl.} {\bfseries 73} (1999) 834--836},
  \href{http://arxiv.org/abs/hep-lat/9809092}{{\ttfamily
  arXiv:hep-lat/9809092}}.

\bibitem{Luscher:2012av}
M.~L{\"u}scher and S.~Schaefer, ``{Lattice QCD with open boundary conditions
  and twisted-mass reweighting},''
  \href{http://dx.doi.org/10.1016/j.cpc.2012.10.003}{{\em Comput. Phys.
  Commun.} {\bfseries 184} (2013) 519--528},
  \href{http://arxiv.org/abs/1206.2809}{{\ttfamily arXiv:1206.2809 [hep-lat]}}.

\bibitem{openQCD}
M.~L{\"u}scher, ``{\url{https://luscher.web.cern.ch/luscher/openQCD/}},''.

\bibitem{Luscher:2008tw}
M.~L{\"u}scher and F.~Palombi, ``{Fluctuations and reweighting of the quark
  determinant on large lattices},''
  \href{http://dx.doi.org/10.22323/1.066.0049}{{\em PoS} {\bfseries
  LATTICE2008} (2008) 049}, \href{http://arxiv.org/abs/0810.0946}{{\ttfamily
  arXiv:0810.0946 [hep-lat]}}.

\bibitem{Kuberski:2023zky}
S.~Kuberski, ``{Low-mode deflation for twisted-mass and RHMC reweighting in
  lattice QCD},'' \href{http://dx.doi.org/10.1016/j.cpc.2024.109173}{{\em
  Comput. Phys. Commun.} {\bfseries 300} (2024) 109173},
  \href{http://arxiv.org/abs/2306.02385}{{\ttfamily arXiv:2306.02385
  [hep-lat]}}.

\bibitem{Mohler:2020txx}
D.~Mohler and S.~Schaefer, ``{Remarks on strange-quark simulations with Wilson
  fermions},'' \href{http://dx.doi.org/10.1103/PhysRevD.102.074506}{{\em Phys.
  Rev. D} {\bfseries 102} no.~7, (2020) 074506},
  \href{http://arxiv.org/abs/2003.13359}{{\ttfamily arXiv:2003.13359
  [hep-lat]}}.

\bibitem{Strassberger:2021tsu}
B.~{Stra\ss berger} {\em et~al.}, ``{Scale setting for CLS 2+1 simulations},''
  \href{http://dx.doi.org/10.22323/1.396.0135}{{\em PoS} {\bfseries
  LATTICE2021} (2022) 135}, \href{http://arxiv.org/abs/2112.06696}{{\ttfamily
  arXiv:2112.06696 [hep-lat]}}.

\bibitem{Bruno:2016plf}
M.~Bruno, T.~Korzec, and S.~Schaefer, ``{Setting the scale for the CLS $2 + 1$
  flavor ensembles},'' \href{http://dx.doi.org/10.1103/PhysRevD.95.074504}{{\em
  Phys. Rev. D} {\bfseries 95} no.~7, (2017) 074504},
  \href{http://arxiv.org/abs/1608.08900}{{\ttfamily arXiv:1608.08900
  [hep-lat]}}.

\bibitem{Wolff:2003sm}
{\bfseries ALPHA} Collaboration, U.~Wolff, ``{Monte Carlo errors with less
  errors},'' \href{http://dx.doi.org/10.1016/S0010-4655(03)00467-3}{{\em
  Comput. Phys. Commun.} {\bfseries 156} (2004) 143--153},
  \href{http://arxiv.org/abs/hep-lat/0306017}{{\ttfamily
  arXiv:hep-lat/0306017}}. [Erratum: Comput.Phys.Commun. 176, 383 (2007)].

\bibitem{Schaefer:2010hu}
{\bfseries ALPHA} Collaboration, S.~Schaefer, R.~Sommer, and F.~Virotta,
  ``{Critical slowing down and error analysis in lattice QCD simulations},''
  \href{http://dx.doi.org/10.1016/j.nuclphysb.2010.11.020}{{\em Nucl. Phys. B}
  {\bfseries 845} (2011) 93--119},
  \href{http://arxiv.org/abs/1009.5228}{{\ttfamily arXiv:1009.5228 [hep-lat]}}.

\bibitem{dobs}
H.~Simma, R.~Sommer, and F.~Virotta, ``{General error computation in lattice
  gauge theory},'' {\em Internal notes of the ALPHA collaboration} (2012-2014)
  .

\bibitem{Ramos:2018vgu}
A.~Ramos, ``{Automatic differentiation for error analysis of Monte Carlo
  data},'' \href{http://dx.doi.org/10.1016/j.cpc.2018.12.020}{{\em Comput.
  Phys. Commun.} {\bfseries 238} (2019) 19--35},
  \href{http://arxiv.org/abs/1809.01289}{{\ttfamily arXiv:1809.01289
  [hep-lat]}}.

\bibitem{Joswig:2022qfe}
F.~Joswig, S.~Kuberski, J.~T. Kuhlmann, and J.~Neuendorf, ``{pyerrors: A python
  framework for error analysis of Monte Carlo data},''
  \href{http://dx.doi.org/10.1016/j.cpc.2023.108750}{{\em Comput. Phys.
  Commun.} {\bfseries 288} (2023) 108750},
  \href{http://arxiv.org/abs/2209.14371}{{\ttfamily arXiv:2209.14371
  [hep-lat]}}.

\bibitem{wloop}
{B.~Leder}, ``{Wilson loop computation},'' 2015.
\newblock \url{https://github.com/bjoern-leder/wloop}.

\bibitem{Hasenfratz:2001hp}
A.~Hasenfratz and F.~Knechtli, ``{Flavor symmetry and the static potential with
  hypercubic blocking},''
  \href{http://dx.doi.org/10.1103/PhysRevD.64.034504}{{\em Phys. Rev. D}
  {\bfseries 64} (2001) 034504},
  \href{http://arxiv.org/abs/hep-lat/0103029}{{\ttfamily
  arXiv:hep-lat/0103029}}.

\bibitem{Eichten:1989zv}
E.~Eichten and B.~R. Hill, ``{An Effective Field Theory for the Calculation of
  Matrix Elements Involving Heavy Quarks},''
  \href{http://dx.doi.org/10.1016/0370-2693(90)92049-O}{{\em Phys. Lett. B}
  {\bfseries 234} (1990) 511--516}.

\bibitem{Campbell:1987nv}
N.~A. Campbell, A.~Huntley, and C.~Michael, ``{Heavy Quark Potentials and
  Hybrid Mesons From SU(3) Lattice Gauge Theory},''
  \href{http://dx.doi.org/10.1016/0550-3213(88)90170-8}{{\em Nucl. Phys. B}
  {\bfseries 306} (1988) 51--62}.

\bibitem{Guagnelli:1998ud}
{\bfseries ALPHA} Collaboration, M.~Guagnelli, R.~Sommer, and H.~Wittig,
  ``{Precision computation of a low-energy reference scale in quenched lattice
  QCD},'' \href{http://dx.doi.org/10.1016/S0550-3213(98)00599-9}{{\em Nucl.
  Phys. B} {\bfseries 535} (1998) 389--402},
  \href{http://arxiv.org/abs/hep-lat/9806005}{{\ttfamily
  arXiv:hep-lat/9806005}}.

\bibitem{Luscher:1990ck}
M.~L{\"u}scher and U.~Wolff, ``{How to Calculate the Elastic Scattering Matrix
  in Two-dimensional Quantum Field Theories by Numerical Simulation},''
  \href{http://dx.doi.org/10.1016/0550-3213(90)90540-T}{{\em Nucl. Phys. B}
  {\bfseries 339} (1990) 222--252}.

\bibitem{Blossier:2009kd}
B.~Blossier, M.~Della~Morte, G.~von Hippel, T.~Mendes, and R.~Sommer, ``{On the
  generalized eigenvalue method for energies and matrix elements in lattice
  field theory},'' \href{http://dx.doi.org/10.1088/1126-6708/2009/04/094}{{\em
  JHEP} {\bfseries 04} (2009) 094},
  \href{http://arxiv.org/abs/0902.1265}{{\ttfamily arXiv:0902.1265 [hep-lat]}}.

\bibitem{Fischer_2020}
M.~Fischer, B.~Kostrzewa, J.~Ostmeyer, K.~Ottnad, M.~Ueding, and C.~Urbach,
  ``On the generalised eigenvalue method and its relation to prony and
  generalised pencil of function methods,''
  \href{http://dx.doi.org/10.1140/epja/s10050-020-00205-w}{{\em The European
  Physical Journal A} {\bfseries 56} no.~8, (Aug., 2020) }.
  \url{http://dx.doi.org/10.1140/epja/s10050-020-00205-w}.

\bibitem{Parisi:1983hm}
G.~Parisi, R.~Petronzio, and F.~Rapuano, ``{A Measurement of the String Tension
  Near the Continuum Limit},''
  \href{http://dx.doi.org/10.1016/0370-2693(83)90930-9}{{\em Phys. Lett. B}
  {\bfseries 128} (1983) 418--420}.

\bibitem{Luscher:2001up}
M.~L{\"u}scher and P.~Weisz, ``{Locality and exponential error reduction in
  numerical lattice gauge theory},''
  \href{http://dx.doi.org/10.1088/1126-6708/2001/09/010}{{\em JHEP} {\bfseries
  09} (2001) 010}, \href{http://arxiv.org/abs/hep-lat/0108014}{{\ttfamily
  arXiv:hep-lat/0108014}}.

\bibitem{Alexandrou:2001ip}
C.~Alexandrou, P.~De~Forcrand, and A.~Tsapalis, ``{The Static three quark SU(3)
  and four quark SU(4) potentials},''
  \href{http://dx.doi.org/10.1103/PhysRevD.65.054503}{{\em Phys. Rev. D}
  {\bfseries 65} (2002) 054503},
  \href{http://arxiv.org/abs/hep-lat/0107006}{{\ttfamily
  arXiv:hep-lat/0107006}}.

\bibitem{Hasenfratz:2001tw}
A.~Hasenfratz, R.~Hoffmann, and F.~Knechtli, ``{The Static potential with
  hypercubic blocking},''
  \href{http://dx.doi.org/10.1016/S0920-5632(01)01733-9}{{\em Nucl. Phys. B
  Proc. Suppl.} {\bfseries 106} (2002) 418--420},
  \href{http://arxiv.org/abs/hep-lat/0110168}{{\ttfamily
  arXiv:hep-lat/0110168}}.

\bibitem{thesisHoffmann}
R.~Hoffmann, ``{The Static Potential with Hypercubic Blocking},'' {\em diploma
  thesis} (2002) .

\bibitem{Asmussen:2023pia}
T.~M.~B. Asmussen, R.~H\"ollwieser, F.~Knechtli, and T.~Korzec, ``{The
  determination of $r_0$ and $r_1$ in Nf=2+1 QCD},''
  \href{http://dx.doi.org/10.22323/1.453.0296}{{\em PoS} {\bfseries
  LATTICE2023} (2024) 296}, \href{http://arxiv.org/abs/2312.14726}{{\ttfamily
  arXiv:2312.14726 [hep-lat]}}.

\bibitem{Aubin_2011}
C.~Aubin, K.~Orginos, D.~Armstrong, V.~Burkert, J.-P. Chen, W.~Detmold,
  J.~Dudek, W.~Melnitchouk, and D.~Richards,
  \href{http://dx.doi.org/10.1063/1.3647217}{``A new approach for delta form
  factors,''} in {\em AIP Conference Proceedings}, p.~621–624.
\newblock AIP, 2011.
\newblock \url{http://dx.doi.org/10.1063/1.3647217}.

\bibitem{Aubin:2012NY}
C.~Aubin and K.~Orginos, ``{An improved method for extracting matrix elements
  from lattice three-point functions},''
  \href{http://dx.doi.org/10.22323/1.139.0148}{{\em PoS} {\bfseries Lattice
  2011} (2012) 148}.

\bibitem{PhysRevD.92.034512}
R.~W. Schiel, ``Expanding the interpolator basis in the variational method to
  explicitly account for backward running states,''
  \href{http://dx.doi.org/10.1103/PhysRevD.92.034512}{{\em Phys. Rev. D}
  {\bfseries 92} (Aug, 2015) 034512}.
  \url{https://link.aps.org/doi/10.1103/PhysRevD.92.034512}.

\bibitem{Ottnad_2018}
K.~Ottnad, T.~Harris, H.~Meyer, G.~von Hippel, J.~Wilhelm, and H.~Wittig,
  ``Nucleon average quark momentum fraction with nf = 2+1 wilson fermions,''
  \href{http://dx.doi.org/10.1051/epjconf/201817506026}{{\em EPJ Web of
  Conferences} {\bfseries 175} (2018) 06026}.
  \url{http://dx.doi.org/10.1051/epjconf/201817506026}.

\bibitem{Niedermayer:2000yx}
F.~Niedermayer, P.~Rufenacht, and U.~Wenger, ``{Fixed point gauge actions with
  fat links: Scaling and glueballs},''
  \href{http://dx.doi.org/10.1016/S0550-3213(00)00731-8}{{\em Nucl. Phys. B}
  {\bfseries 597} (2001) 413--450},
  \href{http://arxiv.org/abs/hep-lat/0007007}{{\ttfamily
  arXiv:hep-lat/0007007}}.

\bibitem{1100705}
H.~Akaike, ``A new look at the statistical model identification,''
  \href{http://dx.doi.org/10.1109/TAC.1974.1100705}{{\em IEEE Transactions on
  Automatic Control} {\bfseries 19} no.~6, (1974) 716--723}.

\bibitem{Jay:2020jkz}
W.~I. Jay and E.~T. Neil, ``{Bayesian model averaging for analysis of lattice
  field theory results},''
  \href{http://dx.doi.org/10.1103/PhysRevD.103.114502}{{\em Phys. Rev. D}
  {\bfseries 103} (2021) 114502},
  \href{http://arxiv.org/abs/2008.01069}{{\ttfamily arXiv:2008.01069
  [stat.ME]}}.

\bibitem{Neil:2022joj}
E.~T. Neil and J.~W. Sitison, ``{Improved information criteria for Bayesian
  model averaging in lattice field theory},''
  \href{http://dx.doi.org/10.1103/PhysRevD.109.014510}{{\em Phys. Rev. D}
  {\bfseries 109} no.~1, (2024) 014510},
  \href{http://arxiv.org/abs/2208.14983}{{\ttfamily arXiv:2208.14983
  [stat.ME]}}.

\bibitem{CamposPlasencia:2023cir}
{\bfseries ALPHA} Collaboration, I.~Campos~Plasencia, M.~Dalla~Brida, G.~M.
  de~Divitiis, A.~Lytle, M.~Papinutto, L.~Pirelli, and A.~Vladikas,
  ``{Nonperturbative running of the tensor operator for Nf=3 QCD from the
  chirally rotated Schr\"odinger functional},''
  \href{http://dx.doi.org/10.1103/PhysRevD.109.054511}{{\em Phys. Rev. D}
  {\bfseries 109} no.~5, (2024) 054511},
  \href{http://arxiv.org/abs/2311.15046}{{\ttfamily arXiv:2311.15046
  [hep-lat]}}.

\bibitem{Symanzik:1979ph}
K.~Symanzik, ``{Cutoff dependence in lattice $\mathrm{\phi}_{4}^{4}$ theory},''
  \href{http://dx.doi.org/10.1007/978-1-4684-7571-5_18}{{\em NATO Sci. Ser. B}
  {\bfseries 59} (1980) 313--330}.

\bibitem{Balog:2009yj}
J.~Balog, F.~Niedermayer, and P.~Weisz, ``{Logarithmic corrections to O(a**2)
  lattice artifacts},''
  \href{http://dx.doi.org/10.1016/j.physletb.2009.04.082}{{\em Phys. Lett. B}
  {\bfseries 676} (2009) 188--192},
  \href{http://arxiv.org/abs/0901.4033}{{\ttfamily arXiv:0901.4033 [hep-lat]}}.

\bibitem{Husung:2019ytz}
N.~Husung, P.~Marquard, and R.~Sommer, ``{Asymptotic behavior of cutoff effects
  in Yang\textendash{}Mills theory and in Wilson\textquoteright{}s lattice
  QCD},'' \href{http://dx.doi.org/10.1140/epjc/s10052-020-7685-4}{{\em Eur.
  Phys. J. C} {\bfseries 80} no.~3, (2020) 200},
  \href{http://arxiv.org/abs/1912.08498}{{\ttfamily arXiv:1912.08498
  [hep-lat]}}.

\bibitem{Husung:2021mfl}
N.~Husung, P.~Marquard, and R.~Sommer, ``{The asymptotic approach to the
  continuum of lattice QCD spectral observables},''
  \href{http://dx.doi.org/10.1016/j.physletb.2022.137069}{{\em Phys. Lett. B}
  {\bfseries 829} (2022) 137069},
  \href{http://arxiv.org/abs/2111.02347}{{\ttfamily arXiv:2111.02347
  [hep-lat]}}.

\bibitem{EuropeanTwistedMass:2014osg}
{\bfseries European Twisted Mass} Collaboration, N.~Carrasco {\em et~al.},
  ``{Up, down, strange and charm quark masses with N$_f$ = 2+1+1 twisted mass
  lattice QCD},'' \href{http://dx.doi.org/10.1016/j.nuclphysb.2014.07.025}{{\em
  Nucl. Phys. B} {\bfseries 887} (2014) 19--68},
  \href{http://arxiv.org/abs/1403.4504}{{\ttfamily arXiv:1403.4504 [hep-lat]}}.

\bibitem{MILC:2009ltw}
{\bfseries MILC} Collaboration, A.~Bazavov {\em et~al.}, ``{MILC results for
  light pseudoscalars},'' \href{http://dx.doi.org/10.22323/1.086.0007}{{\em
  PoS} {\bfseries CD09} (2009) 007},
  \href{http://arxiv.org/abs/0910.2966}{{\ttfamily arXiv:0910.2966 [hep-ph]}}.

\bibitem{Yang:2014sea}
Y.-B. Yang {\em et~al.}, ``{Charm and strange quark masses and $f_{D_s}$ from
  overlap fermions},'' \href{http://dx.doi.org/10.1103/PhysRevD.92.034517}{{\em
  Phys. Rev. D} {\bfseries 92} no.~3, (2015) 034517},
  \href{http://arxiv.org/abs/1410.3343}{{\ttfamily arXiv:1410.3343 [hep-lat]}}.

\bibitem{Dowdall:2013rya}
R.~J. Dowdall, C.~T.~H. Davies, G.~P. Lepage, and C.~McNeile, ``{Vus from pi
  and K decay constants in full lattice QCD with physical u, d, s and c
  quarks},'' \href{http://dx.doi.org/10.1103/PhysRevD.88.074504}{{\em Phys.
  Rev. D} {\bfseries 88} (2013) 074504},
  \href{http://arxiv.org/abs/1303.1670}{{\ttfamily arXiv:1303.1670 [hep-lat]}}.

\bibitem{MILC:2010hzw}
{\bfseries MILC} Collaboration, A.~Bazavov {\em et~al.}, ``{Results for light
  pseudoscalar mesons},'' \href{http://dx.doi.org/10.22323/1.105.0074}{{\em
  PoS} {\bfseries LATTICE2010} (2010) 074},
  \href{http://arxiv.org/abs/1012.0868}{{\ttfamily arXiv:1012.0868 [hep-lat]}}.

\bibitem{HotQCD:2014kol}
{\bfseries HotQCD} Collaboration, A.~Bazavov {\em et~al.}, ``{Equation of state
  in ( 2+1 )-flavor QCD},''
  \href{http://dx.doi.org/10.1103/PhysRevD.90.094503}{{\em Phys. Rev. D}
  {\bfseries 90} (2014) 094503},
  \href{http://arxiv.org/abs/1407.6387}{{\ttfamily arXiv:1407.6387 [hep-lat]}}.

\bibitem{RBC:2010qam}
{\bfseries RBC, UKQCD} Collaboration, Y.~Aoki {\em et~al.}, ``{Continuum Limit
  Physics from 2+1 Flavor Domain Wall QCD},''
  \href{http://dx.doi.org/10.1103/PhysRevD.83.074508}{{\em Phys. Rev. D}
  {\bfseries 83} (2011) 074508},
  \href{http://arxiv.org/abs/1011.0892}{{\ttfamily arXiv:1011.0892 [hep-lat]}}.

\bibitem{Gray:2005ur}
A.~Gray, I.~Allison, C.~T.~H. Davies, E.~Dalgic, G.~P. Lepage, J.~Shigemitsu,
  and M.~Wingate, ``{The Upsilon spectrum and m(b) from full lattice QCD},''
  \href{http://dx.doi.org/10.1103/PhysRevD.72.094507}{{\em Phys. Rev. D}
  {\bfseries 72} (2005) 094507},
  \href{http://arxiv.org/abs/hep-lat/0507013}{{\ttfamily
  arXiv:hep-lat/0507013}}.

\bibitem{Aubin:2004wf}
C.~Aubin, C.~Bernard, C.~DeTar, J.~Osborn, S.~Gottlieb, E.~B. Gregory,
  D.~Toussaint, U.~M. Heller, J.~E. Hetrick, and R.~Sugar, ``{Light hadrons
  with improved staggered quarks: Approaching the continuum limit},''
  \href{http://dx.doi.org/10.1103/PhysRevD.70.094505}{{\em Phys. Rev. D}
  {\bfseries 70} (2004) 094505},
  \href{http://arxiv.org/abs/hep-lat/0402030}{{\ttfamily
  arXiv:hep-lat/0402030}}.

\bibitem{Davies:2009tsa}
{\bfseries HPQCD} Collaboration, C.~T.~H. Davies, E.~Follana, I.~D. Kendall,
  G.~P. Lepage, and C.~McNeile, ``{Precise determination of the lattice spacing
  in full lattice QCD},''
  \href{http://dx.doi.org/10.1103/PhysRevD.81.034506}{{\em Phys. Rev. D}
  {\bfseries 81} (2010) 034506},
  \href{http://arxiv.org/abs/0910.1229}{{\ttfamily arXiv:0910.1229 [hep-lat]}}.

\bibitem{MILC:2009mpl}
{\bfseries MILC} Collaboration, A.~Bazavov {\em et~al.}, ``{Nonperturbative QCD
  Simulations with 2+1 Flavors of Improved Staggered Quarks},''
  \href{http://dx.doi.org/10.1103/RevModPhys.82.1349}{{\em Rev. Mod. Phys.}
  {\bfseries 82} (2010) 1349--1417},
  \href{http://arxiv.org/abs/0903.3598}{{\ttfamily arXiv:0903.3598 [hep-lat]}}.

\bibitem{PACS-CS:2008bkb}
{\bfseries PACS-CS} Collaboration, S.~Aoki {\em et~al.}, ``{2+1 Flavor Lattice
  QCD toward the Physical Point},''
  \href{http://dx.doi.org/10.1103/PhysRevD.79.034503}{{\em Phys. Rev. D}
  {\bfseries 79} (2009) 034503},
  \href{http://arxiv.org/abs/0807.1661}{{\ttfamily arXiv:0807.1661 [hep-lat]}}.

\bibitem{HPQCD:2011qwj}
{\bfseries HPQCD} Collaboration, R.~J. Dowdall {\em et~al.}, ``{The Upsilon
  spectrum and the determination of the lattice spacing from lattice QCD
  including charm quarks in the sea},''
  \href{http://dx.doi.org/10.1103/PhysRevD.85.054509}{{\em Phys. Rev. D}
  {\bfseries 85} (2012) 054509},
  \href{http://arxiv.org/abs/1110.6887}{{\ttfamily arXiv:1110.6887 [hep-lat]}}.

\bibitem{Bazavov:2011nk}
A.~Bazavov {\em et~al.}, ``{The chiral and deconfinement aspects of the QCD
  transition},'' \href{http://dx.doi.org/10.1103/PhysRevD.85.054503}{{\em Phys.
  Rev. D} {\bfseries 85} (2012) 054503},
  \href{http://arxiv.org/abs/1111.1710}{{\ttfamily arXiv:1111.1710 [hep-lat]}}.

\bibitem{Brambilla:2022het}
{\bfseries TUMQCD} Collaboration, N.~Brambilla, R.~L. Delgado, A.~S. Kronfeld,
  V.~Leino, P.~Petreczky, S.~Steinbei\ss{}er, A.~Vairo, and J.~H. Weber,
  ``{Static energy in ($2+1+1$)-flavor lattice QCD: Scale setting and charm
  effects},'' \href{http://dx.doi.org/10.1103/PhysRevD.107.074503}{{\em Phys.
  Rev. D} {\bfseries 107} no.~7, (2023) 074503},
  \href{http://arxiv.org/abs/2206.03156}{{\ttfamily arXiv:2206.03156
  [hep-lat]}}.

\bibitem{Bruno:2014ufa}
{\bfseries ALPHA} Collaboration, M.~Bruno, J.~Finkenrath, F.~Knechtli,
  B.~Leder, and R.~Sommer, ``{Effects of Heavy Sea Quarks at Low Energies},''
  \href{http://dx.doi.org/10.1103/PhysRevLett.114.102001}{{\em Phys. Rev.
  Lett.} {\bfseries 114} no.~10, (2015) 102001},
  \href{http://arxiv.org/abs/1410.8374}{{\ttfamily arXiv:1410.8374 [hep-lat]}}.

\bibitem{Knechtli:2017xgy}
{\bfseries ALPHA} Collaboration, F.~Knechtli, T.~Korzec, B.~Leder, and G.~Moir,
  ``{Power corrections from decoupling of the charm quark},''
  \href{http://dx.doi.org/10.1016/j.physletb.2017.10.025}{{\em Phys. Lett. B}
  {\bfseries 774} (2017) 649--655},
  \href{http://arxiv.org/abs/1706.04982}{{\ttfamily arXiv:1706.04982
  [hep-lat]}}.

\bibitem{DallaBrida:2020pag}
M.~Dalla~Brida, ``{Past, present, and future of precision determinations of the
  QCD parameters from lattice QCD},''
  \href{http://dx.doi.org/10.1140/epja/s10050-021-00381-3}{{\em Eur. Phys. J.
  A} {\bfseries 57} no.~2, (2021) 66},
  \href{http://arxiv.org/abs/2012.01232}{{\ttfamily arXiv:2012.01232
  [hep-lat]}}.

\bibitem{Bruno:2017gxd}
{\bfseries ALPHA} Collaboration, M.~Bruno, M.~Dalla~Brida, P.~Fritzsch,
  T.~Korzec, A.~Ramos, S.~Schaefer, H.~Simma, S.~Sint, and R.~Sommer, ``{QCD
  Coupling from a Nonperturbative Determination of the Three-Flavor $\Lambda$
  Parameter},'' \href{http://dx.doi.org/10.1103/PhysRevLett.119.102001}{{\em
  Phys. Rev. Lett.} {\bfseries 119} no.~10, (2017) 102001},
  \href{http://arxiv.org/abs/1706.03821}{{\ttfamily arXiv:1706.03821
  [hep-lat]}}.

\bibitem{Bazavov:2019qoo}
{\bfseries TUMQCD} Collaboration, A.~Bazavov, N.~Brambilla, X.~Garcia~i Tormo,
  P.~Petreczky, J.~Soto, A.~Vairo, and J.~H. Weber, ``{Determination of the QCD
  coupling from the static energy and the free energy},''
  \href{http://dx.doi.org/10.1103/PhysRevD.100.114511}{{\em Phys. Rev. D}
  {\bfseries 100} no.~11, (2019) 114511},
  \href{http://arxiv.org/abs/1907.11747}{{\ttfamily arXiv:1907.11747
  [hep-lat]}}.

\bibitem{Cali:2020hrj}
S.~Cali, K.~Cichy, P.~Korcyl, and J.~Simeth, ``{Running coupling constant from
  position-space current-current correlation functions in three-flavor lattice
  QCD},'' \href{http://dx.doi.org/10.1103/PhysRevLett.125.242002}{{\em Phys.
  Rev. Lett.} {\bfseries 125} (2020) 242002},
  \href{http://arxiv.org/abs/2003.05781}{{\ttfamily arXiv:2003.05781
  [hep-lat]}}.

\bibitem{Ayala:2020odx}
C.~Ayala, X.~Lobregat, and A.~Pineda, ``{Determination of $\alpha(M_z)$ from an
  hyperasymptotic approximation to the energy of a static quark-antiquark
  pair},'' \href{http://dx.doi.org/10.1007/JHEP09(2020)016}{{\em JHEP}
  {\bfseries 09} (2020) 016}, \href{http://arxiv.org/abs/2005.12301}{{\ttfamily
  arXiv:2005.12301 [hep-ph]}}.

\bibitem{PACS-CS:2009zxm}
{\bfseries PACS-CS} Collaboration, S.~Aoki {\em et~al.}, ``{Precise
  determination of the strong coupling constant in $N_f$ = 2+1 lattice QCD with
  the Schrodinger functional scheme},''
  \href{http://dx.doi.org/10.1088/1126-6708/2009/10/053}{{\em JHEP} {\bfseries
  10} (2009) 053}, \href{http://arxiv.org/abs/0906.3906}{{\ttfamily
  arXiv:0906.3906 [hep-lat]}}.

\bibitem{Maltman:2008bx}
K.~Maltman, D.~Leinweber, P.~Moran, and A.~Sternbeck, ``{The Realistic Lattice
  Determination of alpha(s)(M(Z)) Revisited},''
  \href{http://dx.doi.org/10.1103/PhysRevD.78.114504}{{\em Phys. Rev. D}
  {\bfseries 78} (2008) 114504},
  \href{http://arxiv.org/abs/0807.2020}{{\ttfamily arXiv:0807.2020 [hep-lat]}}.

\bibitem{McNeile:2010ji}
C.~McNeile, C.~T.~H. Davies, E.~Follana, K.~Hornbostel, and G.~P. Lepage,
  ``{High-Precision c and b Masses, and QCD Coupling from Current-Current
  Correlators in Lattice and Continuum QCD},''
  \href{http://dx.doi.org/10.1103/PhysRevD.82.034512}{{\em Phys. Rev. D}
  {\bfseries 82} (2010) 034512},
  \href{http://arxiv.org/abs/1004.4285}{{\ttfamily arXiv:1004.4285 [hep-lat]}}.

\bibitem{Ce:2016idq}
M.~C\`e, L.~Giusti, and S.~Schaefer, ``{Domain decomposition, multi-level
  integration and exponential noise reduction in lattice QCD},''
  \href{http://dx.doi.org/10.1103/PhysRevD.93.094507}{{\em Phys. Rev. D}
  {\bfseries 93} no.~9, (2016) 094507},
  \href{http://arxiv.org/abs/1601.04587}{{\ttfamily arXiv:1601.04587
  [hep-lat]}}.

\end{thebibliography}\endgroup

\end{document}